\documentclass[12pt]{article}

\usepackage{axodraw}
\usepackage{epsfig,graphicx}
\usepackage{amsmath,amsfonts,amssymb}

\newcommand{\unit}{\leavevmode\hbox{\small1\kern-3.6pt\normalsize1}}

\def\cosb{\cos\beta}
\def\sinb{\sin\beta}

\def\cw{\cos\theta_W}

\def\mz{{M_Z}}
\def\mw{{M_W}}

\def\sqrtwo{\sqrt{2}}

\def\cfhi{C_{f\higgsi}}
\def\cfhj{C_{f\higgsj}}

\def\chsnsn{C_{\higgsi\tilde\nu\tilde\nu}}
\def\chisnsn{C_{\higgsi\tilde\nu\tilde\nu}}
\def\chjsnsn{C_{\higgsj\tilde\nu\tilde\nu}}
\def\chksnsn{C_{\higgsk\tilde\nu\tilde\nu}}
\def\chlsnsn{C_{\higgsl\tilde\nu\tilde\nu}}

\def\chhh{C_{\higgsi\higgsj\higgsk}}
\def\chhhk{C_{\higgsi\higgsj\higgsk}}
\def\chhhl{C_{\higgsi\higgsj\higgsl}}
\def\chaa{C_{\phiggsi\phiggsj\higgsk}}
\def\chaak{C_{\phiggsi\phiggsj\higgsk}}
\def\chaal{C_{\phiggsi\phiggsj\higgsl}}
\def\chcc{C_{\chiggsp\chiggsm\higgsk}}
\def\chkaiz{C_{\higgsk\phiggsi Z}}
\def\chlaiz{C_{\higgsl\phiggsi Z}}
\def\chkcw{C_{\higgsk\chiggsp W^-}}
\def\chlcw{C_{\higgsl\chiggsp W^-}}

\def\chhsnsn{C_{\higgsi\higgsj\tilde\nu\tilde\nu}}
\def\caasnsn{C_{\phiggsi\phiggsj\tilde\nu\tilde\nu}}
\def\cccsnsn{C_{\chiggsp\chiggsm\tilde\nu\tilde\nu}}

\def\cnnhi{C_{NN\higgsi}}
\def\cnnhk{C_{NN\higgsk}}
\def\cnnhl{C_{NN\higgsl}}

\def\csnnneui{C_{\tilde\nu N\neuti}}
\def\csnnneuj{C_{\tilde\nu N\neutj}}

\def\l{\lambda}
\def\k{\kappa}
\def\vevs{v_s}
\def\al{{A_\lambda}}
\def\ak{{A_\kappa}}

\def\ln{{\lambda_N}}
\def\aln{{A_{\lambda_N}}}
\def\mn{{m_{\tilde{N}}}}

\def\neuti{{{\tilde\chi}_{i}}}
\def\neutj{{{\tilde\chi}_{j}}}

\def\neutmassi{{m_{{\tilde\chi}_{i}}}}
\def\neutmassj{{m_{{\tilde\chi}_{j}}}}
\def\neutmass{{m_{{\tilde\chi}_{1}^0}}}

\def\neutcompb{{N^{\tilde\chi}_{11}}}

\def\neutcompd{{N^{\tilde\chi}_{13}}}
\def\neutcompu{{N^{\tilde\chi}_{14}}}
\def\neutcomps{{N^{\tilde\chi}_{15}}}

\def\neuticompb{{N^{\tilde\chi}_{i1}}}
\def\neuticompw{{N^{\tilde\chi}_{i2}}}
\def\neuticompd{{N^{\tilde\chi}_{i3}}}
\def\neuticompu{{N^{\tilde\chi}_{i4}}}
\def\neuticomps{{N^{\tilde\chi}_{i5}}}

\def\neutcompb{{N^{1}_{\tilde\chi_1}}}

\def\neutcompd{{N^{3}_{\tilde\chi_1}}}
\def\neutcompu{{N^{4}_{\tilde\chi_1}}}
\def\neutcomps{{N^{5}_{\tilde\chi_1}}}

\def\neuticompb{{N^{1}_{\tilde\chi_i}}}
\def\neuticompw{{N^{2}_{\tilde\chi_i}}}
\def\neuticompd{{N^{3}_{\tilde\chi_i}}}
\def\neuticompu{{N^{4}_{\tilde\chi_i}}}
\def\neuticomps{{N^{5}_{\tilde\chi_i}}}

\def\snr{{\tilde N}}

\def\snmassr{{m_{\tilde N_1}}}

\def\sncompl{{N^{\tilde\nu}_{i \tilde\nu_L}}}
\def\sncompr{{N^{\tilde\nu}_{i \tilde N}}}

\def\rhn{{N}}
\def\rhnmass{{M_{N}}}

\def\higgsi{{H_i^0}}
\def\higgsj{{H_j^0}}
\def\higgsk{{H_k^0}}
\def\higgsl{{H_1^0}}

\def\hmassi{m_{H_i^0}}
\def\hmassj{m_{H_j^0}}

\def\hmassl{m_{H_1^0}}
\def\hmassm{m_{H_2^0}}
\def\hmassh{m_{H_3^0}}

\def\hcompls{S_{H_1^0}^3}

\def\hcompid{S_{H_i^0}^1}
\def\hcompiu{S_{H_i^0}^2}
\def\hcompis{S_{H_i^0}^3}

\def\hcompjd{S_{H_j^0}^1}
\def\hcompju{S_{H_j^0}^2}
\def\hcompjs{S_{H_j^0}^3}

\def\phiggsi{{A_a^0}}
\def\phiggsj{{A_b^0}}

\def\phiggsl{{A_1^0}}

\def\phmassi{{m_{A_a^0}}}
\def\phmassl{{m_{A_1^0}}}
\def\phmassh{{m_{A_2^0}}}

\def\phcompid{P_{A_a^0}^1}
\def\phcompiu{P_{A_a^0}^2}
\def\phcompis{P_{A_a^0}^3}

\def\phcompjd{P_{A_b^0}^1}
\def\phcompju{P_{A_b^0}^2}
\def\phcompjs{P_{A_b^0}^3}

\def\chiggsp{{H^+}}
\def\chiggsm{{H^-}}

\def\chmass{{m_{H^+}}}

\def\ahit{{A_{\higgsi}^t}}
\def\ahiu{{A_{\higgsi}^u}}

\def\bhit{{B_{\higgsi}^t}}
\def\bhiu{{B_{\higgsi}^u}}

\def\Ehi{{E_{\higgsi}}}
\def\Ehj{{E_{\higgsj}}}

\def\Fone#1{{{\cal F}_1(#1)}}

\newcommand{\captions}{\sf\caption}

\newcommand{\crosssec}{\sigma_{\snr p}^{\rm SI}}
\newcommand{\nmh}{{\tt NMHDECAY}}
\newcommand{\snrelic}{{\Omega_\snr h^2}}

\newcommand{\bsg}{b\to s\gamma}
\newcommand{\bmumu}{B_S\to\mu^+\mu^-}
\newcommand{\asusy}{a^{\rm SUSY}_\mu}

\def\lsim{\raise0.3ex\hbox{$\;<$\kern-0.75em\raise-1.1ex\hbox{$\sim\;$}}}
\def\gsim{\raise0.3ex\hbox{$\;>$\kern-0.75em\raise-1.1ex\hbox{$\sim\;$}}}

\allowdisplaybreaks

\begin{document}

\thispagestyle{empty}
\begin{flushright}
  FTUAM 09/5\\
  IFT-UAM/CSIC-09-17\\
  FTPI-MINN-09-14\\
  UMN-TH-2742/09

  \vspace*{2.mm}{26 March 2009}
\end{flushright}

\begin{center}
  {\Large \textbf{Right-handed sneutrino
      dark matter\\ in the NMSSM
  } }  
  
  \vspace{0.5cm}
  David G.~Cerde\~no ${}^{1}$,
  Osamu Seto ${}^2$ \\[0.2cm] 
    
  {\textit{ ${}^1$ Departamento de F\'{\i}sica Te\'{o}rica C-XI,
      and 
      Instituto de F\'{\i}sica Te\'{o}rica
      UAM-CSIC, \\[0pt] 
      Universidad Aut\'{o}noma de Madrid, 
      Cantoblanco, E-28049
      Madrid, Spain\\[0pt] 
      ${}^2$ William I. Fine Theoretical Physics Institute,\\[0pt]
      University of Minnesota, Minneapolis, MN 55455, USA
  }}
  
\vspace*{0.7cm}
\begin{abstract}
  We study the properties of the right-handed sneutrino and its
  viability as a WIMP dark matter candidate in an extended
  version of the NMSSM in which a right-handed neutrino 
  superfield is included 
  with a coupling to the singlet Higgs in order to
  provide non-vanishing Majorana neutrino masses. 
  We perform a systematic study of the parameter space,
  including LEP constraints and experimental bounds on low-energy
  observables. 
  We investigate the conditions under which the right-handed sneutrino
  has the correct relic abundance and the dominant annihilation
  channels. Next we calculate the theoretical predictions for the
  sneutrino-proton elastic scattering cross section and compare it with
  present and future experimental sensitivities.
  We find that sneutrinos with a mass in the range of 5-200 GeV can
  reproduce the observed dark matter relic density without being
  excluded by direct dark matter searches and for natural values of
  the input parameters. 
  Interestingly, the predicted scattering cross section 
  is generally within
  the reach of future experiments.
  Finally, we comment on the possible implications for collider
  physics.
\end{abstract}
\end{center}

	\newpage
	


\section{Introduction}

A number of astrophysical and cosmological observations have provided 
substantial evidence supporting the existence of dark matter, 
an abundant component of our Universe made of some new,
yet undiscovered, particles.
A generic weakly interacting massive particle (WIMP) is one of the most
promising and attractive candidates for cold dark matter in our  
Universe~\cite{Bertone:2004pz}, since it would be thermally produced
in sufficient amount to account for the observed
dark matter density. 
Interestingly, 
WIMPs appear in many well motivated extensions of the standard model 
providing new physics at the TeV scale, such as supersymmetry (SUSY).

In supersymmetric models a discrete symmetry (R-parity) is often
imposed in order to forbid lepton and baryon violating
processes which could lead, for instance, to rapid proton decay.
A remarkable
phenomenological consequence is that supersymmetric particles
can only be produced or annihilate in pairs, thus rendering the lightest
SUSY particle (LSP) stable.
If it is neutral and weakly interacting, the LSP can therefore be an
excellent WIMP candidate for dark matter.
Two particles exist with these properties in the
minimal supersymmetric standard model (MSSM), namely
the lightest
neutralino~\cite{Goldberg:1983nd,Jungman:1995df} and the 
sneutrino~\cite{Ibanez}.
The former is the mixture of neutral gauginos and Higgsinos, and one
of the most popular and well studied 
dark matter candidates~\cite{propaganda}. 
In contrast, the (left-handed) sneutrino in the MSSM turns out not to
be viable as dark matter. Being a left-handed field, the sneutrino
has a sizable coupling with the $Z$ boson, which entails a too
large annihilation cross section and therefore a too small relic
abundance. 
Furthermore, its scattering cross section off nuclei, also mediated by
$Z$ boson exchange, is so large that direct detection experiments for
dark matter would have already observed it \cite{Falk:1994es}. 
It should be noted that the inclusion of a
lepton number violating operator can reduce the detection rate
\cite{Hall:1997ah}.
A re-analysis of the sneutrino LSP in the MSSM in light of an updated
data can be found in Ref.\,\cite{Arina:2007tm}.

There are  motivations to consider extensions of the MSSM. 
One of them is the observation of neutrino oscillation phenomena, 
suggested by the solar and atmospheric neutrino anomalies and 
confirmed by various experiments
~\cite{Mohapatra:2005wg}, which imply non-vanishing (albeit very
small) neutrino masses.
However, in the MSSM as well as the SM, neutrinos are massless
and the tiny neutrino mass is usually explained by a see-saw mechanism
in which 
right-handed Majorana neutrinos, $N$, are introduced~\cite{seesaw}
with a large Majorana mass (usually taken to be around the grand
unification scale, but which can also be of order of the electroweak
scale or even lower).

In supersymmetric models, at same time, one has to introduce 
right-handed sneutrinos, which would then mix with the left-handed
sneutrinos. Interestingly, a mixed left- and right-handed
sneutrino has a reduced coupling to the $Z$ boson thereby alleviating
the abovementioned problems and making it viable as dark matter
\cite{ArkaniHamed:2000bq,Hooper:2004dc}.
However, a significant left-right mixture is only possible
in some particular supersymmetry
breaking schemes with a large trilinear
term~\cite{ArkaniHamed:2000bq,Borzumati:2000mc}.  
Such a mechanism is not available in the standard
supergravity mediated supersymmetry breaking, where trilinear terms
are  proportional to the small neutrino Yukawa couplings.
Recently, another realization of large mixing was pointed
out~\cite{valle} by abandoning the canonical see-saw formula
\cite{seesaw} for neutrino masses. 
Another possibility is having a pure right-handed sneutrino 
~\cite{Asaka:2005cn,Gopalakrishna:2006kr,McDonald:2006if,Page:2007sh,
Lee:2007mt,Allahverdi:2007wt,pilaftsis,Cerdeno:2008ep,Deppisch:2008bp}. 
These cannot be thermal relics, since their
coupling to ordinary matter is extremely reduced by the neutrino Yukawa 
coupling
\cite{Asaka:2005cn,Gopalakrishna:2006kr,McDonald:2006if,Page:2007sh},
unless they are somehow coupled to the observable
sector, for example via an extension of the gauge
\cite{Lee:2007mt,Allahverdi:2007wt} or Higgs
\cite{pilaftsis,Cerdeno:2008ep,Deppisch:2008bp} 
sectors\footnote{Recently, 
  non-LSP right-handed sneutrino dark matter model 
  was also proposed~\cite{Kaplan:2009ag}.
}.

The other motivation to extend the MSSM 
is the so-called ``$\mu$ problem'' \cite{Kim:1983dt}.
The superpotential in the MSSM contains a bilinear
term $\mu H_1 H_2$ which
gives supersymmetric Higgs/Higgsino mass terms.
Successful radiative electroweak symmetry breaking
(REWSB)~\cite{Ibanez:1982fr}  
requires $\mu$ of the order of the electroweak scale. The next-to
minimal supersymmetric standard model (NMSSM) 
addresses this problem in a very
natural way. A new singlet superfield, $S$, is introduced
\cite{Nilles:1982dy}
and the bilinear term (forbidden by a global $Z_3$ symmetry) 
is promoted to a trilinear coupling  $\l S H_1
H_2$. Once REWSB takes place the singlet field acquires a vacuum
expectation value (VEV), $\vevs=\langle S\rangle$, thereby generating 
an effective $\mu$ parameter, $\mu=\lambda \vevs$.
The NMSSM also alleviates the ``little hierarchy problem'' of the
Higgs sector in  
the MSSM~\cite{BasteroGil:2000bw}.
Due to the significant changes in the Higgs and neutralino sectors, 
the NMSSM also has an attractive phenomenology \cite{Djouadi:2008uw} 
and interesting
consequences for neutralino dark matter 
\cite{Cerdeno:2004xw,Hugonie:2007vd,Djouadi:2008uj}.
Notice finally that although the $Z_3$ symmetry
of the NMSSM may give rise to a cosmological domain wall problem, 
this can be avoided with the inclusion of
non-renormalisable operators \cite{Abel:1995wk,Menon:2004wv}.

Motivated by above two issues, in Ref.\,\cite{Cerdeno:2008ep} an
extension of the MSSM was proposed  
in which two new singlet superfields were included, as in
Refs.\,\cite{ko99,pilaftsis}. An extra singlet superfield $S$
addresses the $\mu$ problem in the same way as in the NMSSM 
and provides extra Higgs and neutralino states, while an extra singlet
superfield $N$ accounts for right-handed neutrino and sneutrino
states. 
This model possesses two interesting new consequences.
The right-handed neutrino mass is generated spontaneously (or
dynamically) with the electroweak symmetry breaking.
Due to the non-vanishing VEV of the singlet Higgs, an effective 
Majorana mass for the right-handed neutrino appears 
through the new coupling $S N N$ which is of order of the electroweak
scale, in the same way as
the effective $\mu$ term is generated in the NMSSM.
Moreover, the singlet $S$, which couples to the MSSM Higgs sector,
provides electroweak scale interactions of the
right-handed sneutrino with the rest of the MSSM fields. 
Thus, the purely right-handed sneutrino LSP has the properties of a
WIMP.  
In Ref.\,\cite{Cerdeno:2008ep} it was shown that 
right-handed sneutrinos in this construction 
can not only be thermally produced 
in sufficient amount to account for the dark matter in the Universe 
but also that their elastic scattering cross section with nuclei 
is large enough to detect them through Higgs exchange processes.

In this work we perform a systematic analysis of the properties of the
right-handed sneutrino as a dark matter candidate in the model of
Ref.\,\cite{Cerdeno:2008ep}.   
In Section\,\ref{sec:model} the theoretical details of this
construction are introduced, explaining in detail the properties of
the sneutrino spectrum. In Section\,\ref{sec:relic} the thermal
production of right-handed sneutrinos is analysed for several examples
in the parameter space. The right-handed sneutrino relic abundance is
calculated and the importance of the different annihilation channels
is investigated in detail.  
In Section\,\ref{sec:cross} the direct detection properties of the
right-handed sneutrino are studied, comparing the theoretical
predictions of its elastic scattering cross section 
off nuclei with present and
future experimental sensitivities. 
In Section\,\ref{sec:collider}
we briefly comment on the possible implications of this model for
collider physics, describing the characteristic signals that could be
expected. 
In Section\,\ref{sec:conclusions}
we summarize our conclusions. The technical details are relegated to
the Appendices. In Appendix\,\ref{sec:feynman} we include the Feynman
rules for the new diagrams of the model, and in
Appendix\,\ref{sec:wtilde} we detail the calculation of the sneutrino
relic abundance and give the explicit expressions for the different
annihilation channels.

\section{The Model}
\label{sec:model}

The model consists of an extended version of the 
NMSSM with a new singlet right-handed 
neutrino superfield $N$.
The superpotential is therefore given by
\begin{eqnarray}
  W &=& W_{\rm NMSSM} + \lambda_N S N N + y_N L \cdot H_2 N,  \\
  W_{\rm NMSSM} &=& Y_u H_2 \cdot Q u + Y_d H_1 \cdot Q d 
  + Y_e H_1 \cdot L e
  -\lambda S H_1 \cdot H_2 + \frac{1}{3}\kappa S^3 ,
  \label{superpotential}
\end{eqnarray}
where flavour indices are omitted and the dot denotes the $SU(2)_L$
antisymmetric product. As in the NMSSM, a global $Z_3$ symmetry is
imposed for each superfield, so that there are no supersymmetric mass
terms in the superpotential.  Note that the term $NNN$ and $SSN$ are gauge
invariant but not consistent with R-parity and thus are not included. 
Notice also that $N$
does not have a bare Majorana mass but 
acquires a mass through the
non-vanishing singlet Higgs VEV, $\vevs$.

The supersymmetric scalar potential for squarks, sleptons, Higgses and
the right-handed sneutrino, $\tilde N$, is given as $V = V_F+V_D$ with 
\begin{eqnarray}
V_F &=& |Y_u H_2 \tilde{u} + Y_d H_1 \tilde{d}|^2 + |Y_u H_2 \tilde{Q}|^2
 + |Y_d H_1 \tilde{Q}|^2 
 + |Y_e H_1 \tilde{e}+ y_N H_2 \tilde{N}|^2 + |Y_e H_1 \tilde{L}|^2 \nonumber \\
 && + |Y_d \tilde{Q}\tilde{d} + y_N \tilde{L}\tilde{e} -\lambda S H_2|^2
 + |Y_u \tilde{Q}\tilde{u} -\lambda S H_1 + y_N \tilde{L}\tilde{N}|^2 \nonumber\\
 &&
  + |- \lambda H_1 H_2 + \kappa S^2 + \lambda_N \tilde{N}^2 |^2
 + |2\lambda_N S \tilde{N} + y_N \tilde{L} H_2|^2 ,
\end{eqnarray}
and
\begin{eqnarray}
V_D &=&
\frac{g_1^2}{2} \left(H_1^{\dagger}\frac{-1}{2}H_1 + H_2^{\dagger}\frac{1}{2}H_2
  + \tilde{Q}^{\dagger}\frac{1}{6}\tilde{Q}
  + \tilde{u}^{\dagger}\frac{-1}{3}\tilde{u} 
  + \tilde{d}^{\dagger}\frac{1}{3}\tilde{d} 
  + \tilde{L}^{\dagger}\frac{-1}{2}\tilde{L} 
  + \tilde{e}^{\dagger}\tilde{e} \right)^2 \nonumber \\
&& + \frac{g_2^2}{2}\sum_a \left(H_1^{\dagger}\frac{\tau^a}{2}H_1 
+ H_2^{\dagger}\frac{\tau^a}{2}H_2 +\tilde{Q}^{\dagger}\frac{\tau^a}{2}\tilde{Q}
+ \tilde{L}^{\dagger}\frac{\tau^a}{2}\tilde{L} \right)^2  .
\end{eqnarray}
The soft SUSY breaking terms are
\begin{eqnarray}
  -{\cal L}_{\rm scalar \, mass}
  &=& m_{\tilde{Q}}^2 |\tilde{Q}|^2 + m_{\tilde{u}}^2 |\tilde{u}|^2 
  + m_{\tilde{d}}^2 |\tilde{d}|^2 + m_{\tilde{L}}^2 |\tilde{L}|^2 
  + m_{\tilde{e}}^2 |\tilde{e}|^2 \nonumber \\
  && + m_{H_1}^2 |H_1|^2 + m_{H_2}^2 |H_2|^2 + m_S^2 |S|^2 
  + m_{\tilde{N}}^2 |\tilde{N}|  ,
  \label{lagrangian_masses}
\end{eqnarray}
where the new soft scalar masses $\mn$ and $m_S$ are included, and
\begin{eqnarray}
-{\cal L}_{\rm A-terms} &=&  
\left(A_u Y_u H_2 \tilde{Q}\tilde{u} + A_d Y_d H_1 \tilde{Q} \tilde{d} 
  + A_e Y_e H_1 \tilde{L}\tilde{e} + {\rm H.c.} \right) \nonumber \\
&& + \left(-\lambda A_{\lambda} S H_1 H_2 + \frac{1}{3}\kappa A_{\kappa} S^3
  + {\rm H.c.} \right) \nonumber \\
&& + \left( \lambda_N A_{\lambda_N} S \tilde{N}^2 
  + y_N A_{y_N} \tilde{L} H_2 \tilde{N}+ {\rm H.c.}  \right) ,
  \label{lagrangian_couplings}
\end{eqnarray}
which contains the new trilinear soft terms $\aln$ and $A_{y_N}$. 
The sum of the supersymmetric and soft SUSY breaking terms give 
the total scalar potential.

\subsection{Neutrino mass}

As stated above, in this construction, 
right-handed neutrino masses are generated 
by the non-vanishing VEV of the singlet Higgs as 
\begin{equation}
 \rhnmass = 2 \ln v_s\ ,
\label{RighthandedNeutrinoMass}
\end{equation}
being therefore of order of the electroweak scale.
Then, in order to reproduce the small masses of the left-handed
neutrinos, which are given as 
\begin{equation}
 m_{\nu_L} = \frac{y_N^2 v_2^2}{\rhnmass},
\label{NeutrinoMass}
\end{equation}
the low scale seesaw mechanism implies small 
Yukawa couplings of ${\cal O} (10^{-6})$ or less.
Here, $v_{1, 2} = \langle H_{1, 2} \rangle$ denote the VEV of the
Higgs doublet. 
To reproduce light neutrino masses and mixing for neutrino oscillation 
data we would need to introduce the generation structure 
in the right-handed neutrino sector.
However, as we will see, these small neutrino Yukawa couplings are 
completely irrelevant for dark matter physics.
Hence, for simplicity, we consider one generation case, but one may regard
that the considered sneutrino corresponds to 
the lightest one among multi-generations.

\subsection{Sneutrino masses}

The sneutrino mass matrix can be read from the quadratic terms with
respect to $\tilde{L}$ and $\tilde{N}$  
\begin{eqnarray}
V(\tilde{L}, \tilde{N})
 &\subset& |y_N H_2 \tilde{N}|^2 + |2\lambda_N S \tilde{N}|^2 
 + |-\lambda S H_1 + y_N \tilde{L}\tilde{N}|^2 \nonumber \\
&& + |-\lambda H_1 H_2 + \kappa S^2 + \lambda_N \tilde{N}^2 |^2
 + {\rm D - term} \nonumber \\
&& + m_{\tilde{L}}^2 |\tilde{L}|^2  + m_{\tilde{N}}^2 |\tilde{N}| 
  + \left( \lambda_N A_{\lambda_N} S \tilde{N}^2 
  + y_N A_{y_N} \tilde{L} H_2 \tilde{N}+ {\rm H.c.}  \right)\,.
\end{eqnarray}
Decomposing the left-handed sneutrino $\tilde{\nu}_L$ and right-handed sneutrino $\tilde{N}$ as 
\begin{equation}
  \tilde{\nu}_L \equiv \frac{1}{\sqrt{2}}(\tilde{\nu}_{L1} + i
  \tilde{\nu}_{L2}) ,  
  \quad\quad 
  \tilde{N} \equiv \frac{1}{\sqrt{2}}(\tilde{N}_1 + i \tilde{N}_2) ,
\end{equation}
the sneutrino quadratic term can be written as
\begin{equation}
  \frac{1}{2}
  (\tilde{\nu}_{L1}, \tilde{N}_1, \tilde{\nu}_{L2}, \tilde{N}_2)
{\cal M}^2_{\rm Sneutrino} 
\left(
\begin{array}{c}
  \tilde{\nu}_{L1}  \\
  \tilde{N}_1 \\
  \tilde{\nu}_{L2}  \\
  \tilde{N}_2 \\
\end{array}
\right) ,
\label{SneutrinoMassMatrix}
\end{equation}
with 
\begin{eqnarray}
 && {\cal M}^2_{\rm Sneutrino} \nonumber \\
 = &&
\left(
\begin{array}{cccc}
m_{L\bar{L}}^2          
  & \frac{m_{LR}^2+m_{L\bar{R}}^2 + {\rm c.c} }{2}
  &  0  
  & i \frac{m_{LR}^2-m_{L\bar{R}}^2 - {\rm c.c} }{2}  \\
\frac{m_{LR}^2+m_{L\bar{R}}^2 + {\rm c.c} }{2}
  & m_{R\bar{R}}^2 + m_{RR}^2+m_{RR}^{2*}
  & i \frac{m_{LR}^2-m_{L\bar{R}}^2 - {\rm c.c} }{2} 
  & i (m_{RR}^2 - m_{RR}^{2*})  \\
0   
  & i \frac{m_{LR}^2-m_{L\bar{R}}^2 - {\rm c.c} }{2} 
  & m_{L\bar{L}}^2
  & \frac{-m_{LR}^2+m_{L\bar{R}}^2 + {\rm c.c} }{2} \\
i \frac{m_{LR}^2-m_{L\bar{R}}^2 - {\rm c.c} }{2} 
  & i (m_{RR}^2 - m_{RR}^{2*}) 
  & \frac{-m_{LR}^2+m_{L\bar{R}}^2 + {\rm c.c} }{2}
  & m_{R\bar{R}}^2 - m_{RR}^2 - m_{RR}^{2*} 
  \\
\end{array}
\right) .
 \nonumber \\
&& 
\end{eqnarray}
Here, we defined
\begin{eqnarray}
  m_{L\bar{L}}^2
  &\equiv& m_{\tilde{L}}^2 + |y_N v_2|^2 + {\rm D-term} , \nonumber \\
  m_{LR}^2
 &\equiv& y_N\left(-\lambda \vevs v_1 \right)^{\dagger}
  + y_N A_N v_2 , \nonumber\\
  m_{L\bar{R}}^2
  &\equiv& y_N v_2 \left(-\lambda \vevs \right)^{\dagger} , \nonumber\\
  m_{R\bar{R}}^2 
  &\equiv& m_{\tilde{N}}^2 +|2\lambda_N \vevs|^2 
  + |y_N v_2|^2 , \nonumber\\
  m_{RR}^2
  &\equiv& \lambda_N \left( A_{\lambda_N} \vevs+(
  \kappa \vevs^2-\lambda v_1 v_2 )^{\dagger} \right) .
  \label{sn:mrr}
\end{eqnarray}
If these are real, in other words no CP violation, the real part and
imaginary part of sneutrinos do not mix and its mass matrix
(\ref{SneutrinoMassMatrix}) is simplified as 
\begin{eqnarray}
  {\rm Eq.~(\ref{SneutrinoMassMatrix})}
  &=& \frac{1}{2}(\tilde{\nu}_{L1}, \tilde{N}_1)
  \left(
  \begin{array}{cc}
    m_{L\bar{L}}^2           &  m_{LR}^2+m_{L\bar{R}}^2 \\
    m_{LR}^2+m_{L\bar{R}}^2  &  m_{R\bar{R}}^2 + 2m_{RR}^2 \\
  \end{array}
  \right)
  \left(
  \begin{array}{c}
    \tilde{\nu}_{L1}  \\
    \tilde{N}_1 \\
  \end{array}
  \right) \nonumber \\
  &&  +
  \frac{1}{2} (\tilde{\nu}_{L2}, \tilde{N}_2)
  \left(
  \begin{array}{cc}
    m_{L\bar{L}}^2           &  -m_{LR}^2+m_{L\bar{R}}^2 \\
    -m_{LR}^2+m_{L\bar{R}}^2 &  m_{R\bar{R}}^2 - 2m_{RR}^2 \\
  \end{array}
  \right)
  \left(
  \begin{array}{c}
    \tilde{\nu}_{L2}  \\
    \tilde{N}_2 \\
  \end{array}
  \right) .
  \label{SneutrinoMass}
\end{eqnarray}
Note that the mixing between left-handed and right-handed sneutrinos
is induced by $m_{LR}^2$ and $m_{L\bar{R}}^2$, both of which 
are proportional to the neutrino Yukawa coupling $y_N$. 
As mentioned in the previous subsection, 
$y_N$ must be as small as ${\cal O}(10^{-(6-7)})$ to generate sub-eV
left-handed neutrino mass via low scale seesaw mechanism according to 
Eq.\,(\ref{NeutrinoMass}). 
If we rewrite sneutrinos in the mass eigenstates,
\begin{equation}
  \tilde{\nu}_i = \sncompl \tilde{\nu}_L + \sncompr \tilde{N},
\end{equation}
by rotating with an unitary matrix $N^{\tilde{\nu}}$, then, unless
the diagonal elements of the mass matrix (\ref{SneutrinoMass}) 
are extremely degenerated, we find 
\begin{equation}
(\sncompl, \sncompr) = \left\{ 
 \begin{array}{l}
  \left(1 +{\cal O}(y_N), {\cal O}(y_N) \right)  \\
  \left({\cal O}(y_N), 1+ {\cal O}(y_N) \right)
 \end{array}
 \right. .
\end{equation}
Thus, the mixing between left-handed and right-handed sneutrinos is
also of ${\cal O}(10^{-6}-10^{-7)})$ and therefore negligible. 
For all practical purposes in this paper, we can ignore all
off-diagonal elements in Eq.~(\ref{SneutrinoMass}) and regard
sneutrino mass eigenstates as pure left- or right-handed fields. 
One may see that $m_{RR}^2$ splits the masses of $\tilde{N}_1$ and
$\tilde{N}_2$. In particular,
$\tilde{N}_2$ is heavier than $\tilde{N}_1$ for $m_{RR}^2 < 0$ (and
viceversa), and this is the situation we will consider throughout the
rest of this work\footnote{We only do this for simplicity. The
  imaginary component of the right-handed sneutrino is as good a dark
  matter candidate as the real component. In fact they only differ in
  the expression for the annihilation into a pair of right-handed
  neutrinos. 
}.

\begin{figure}
  \hspace*{-0.5cm}
  \epsfig{file=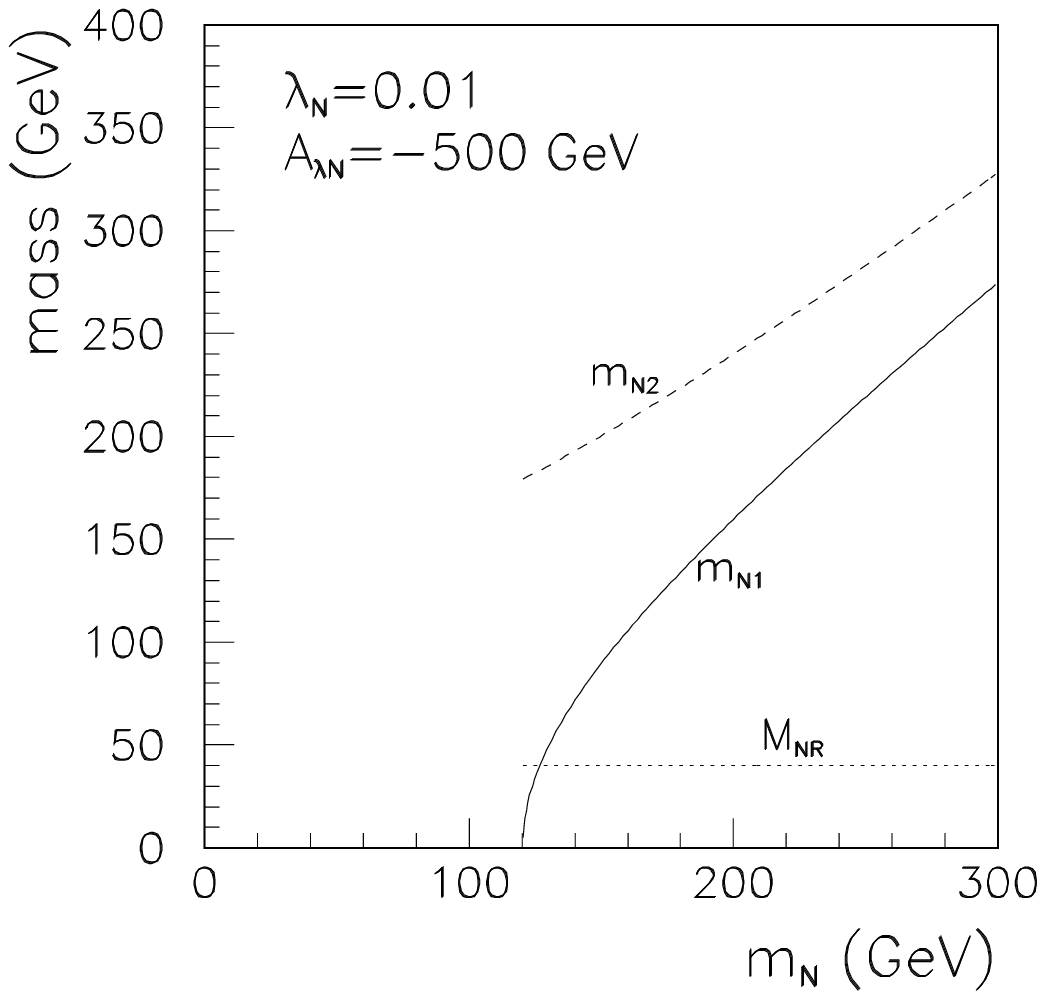,width=9.cm}\hspace*{-1.2cm}
  \epsfig{file=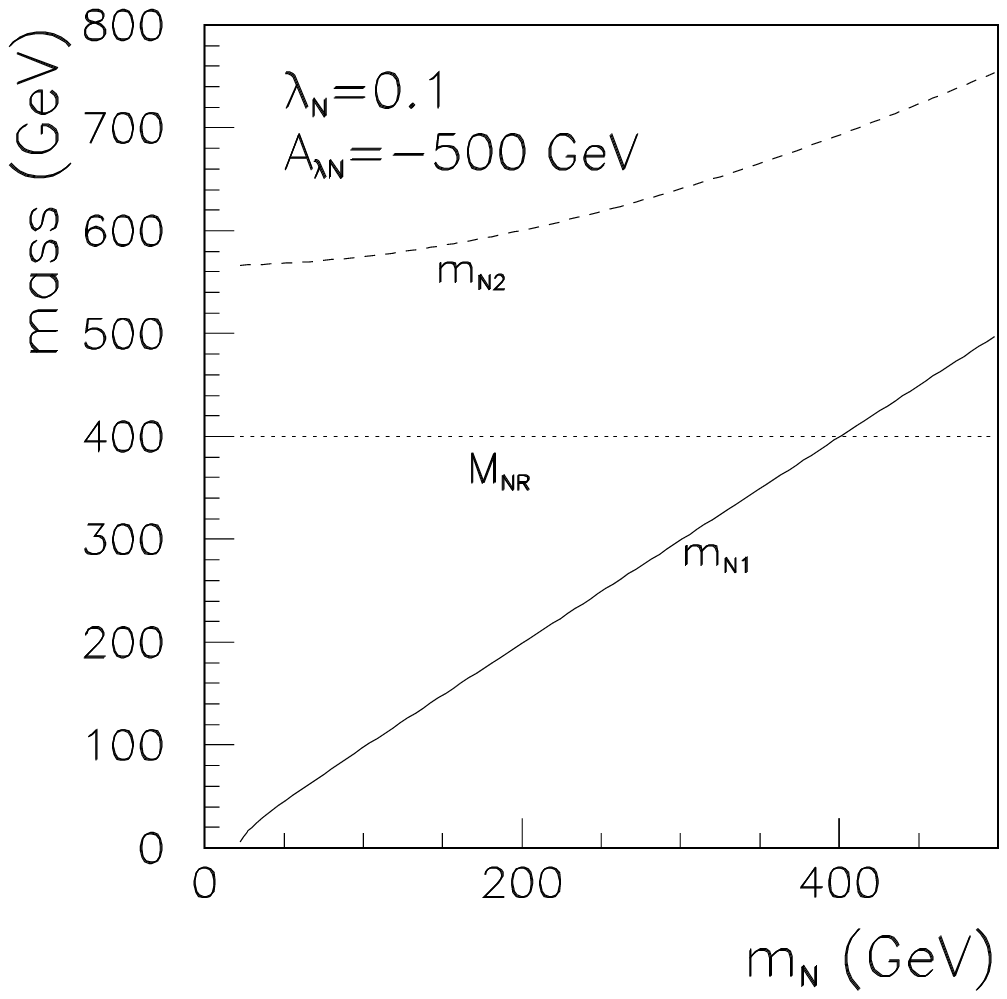,width=9.cm}
  \vspace*{-1cm}
  \captions{Sneutrino spectrum and right-handed neutrino mass,
    $\rhnmass$,  
    as a function of the sneutrino soft
    mass $\mn$ for $\aln=-500$~GeV and $\ln=0.01,\,0.05$. 
    The NMSSM parameters have been chosen as $\lambda=0.1$, 
    $\k=0.05$, $\tan\beta=5$, and $\mu=200\,{\rm GeV}$. 
  }
  \label{fig:spectrum-b}
\end{figure}

Let us briefly illustrate these properties of the sneutrino
spectrum.
As an explicit example, we have taken 
$\mu=200\,{\rm GeV}$
and $\l=0.1$ (which implies $\vevs=2000\,{\rm GeV}$), $\k=0.05$,
together with $\tan\beta=5$, $\al=400$~GeV, and $\ak=0$.
We fix $\aln=-500$~GeV and 
consider two cases, with $\ln=0.01$ and $0.1$. 
From Eq.~(\ref{RighthandedNeutrinoMass}) the right-handed
neutrino mass can be calculated to be 
$\rhnmass=40\,{\rm GeV}$, and $400$~GeV, respectively. 
The sneutrino masses are then
an increasing function of $\mn$. These features are displayed in
Fig.\,\ref{fig:spectrum-b}, where the resulting sneutrino
masses are plotted, together with the right-handed neutrino mass, as a
function of $\mn$.

As commented above, the mass-splitting of the two sneutrino mass
eigenstates is dictated by the $m_{RR}^2$ term in
Eq.~(\ref{sn:mrr}). An increase in $\ln$ implies not only an
enhancement of 
$\rhnmass$ in Eq.~(\ref{RighthandedNeutrinoMass}), but also leads to a
larger mass difference between $\tilde{N}_1$ and $\tilde{N}_2$ through
the increase in $|m_{RR}^2|$. 
This
is clearly evidenced in the two panels of
Fig.\,\ref{fig:spectrum-b}.
Notice that although the neutrino Majorana mass, $\rhnmass$, always
contributes to the sneutrino mass through the $m_{R\bar R}^2$ term in
Eq.~(\ref{sn:mrr}), 
the negativeness of the $m_{RR}^2$ term allows
sneutrinos lighter than right-handed neutrinos.

\begin{figure}
  \hspace*{-0.5cm}
  \epsfig{file=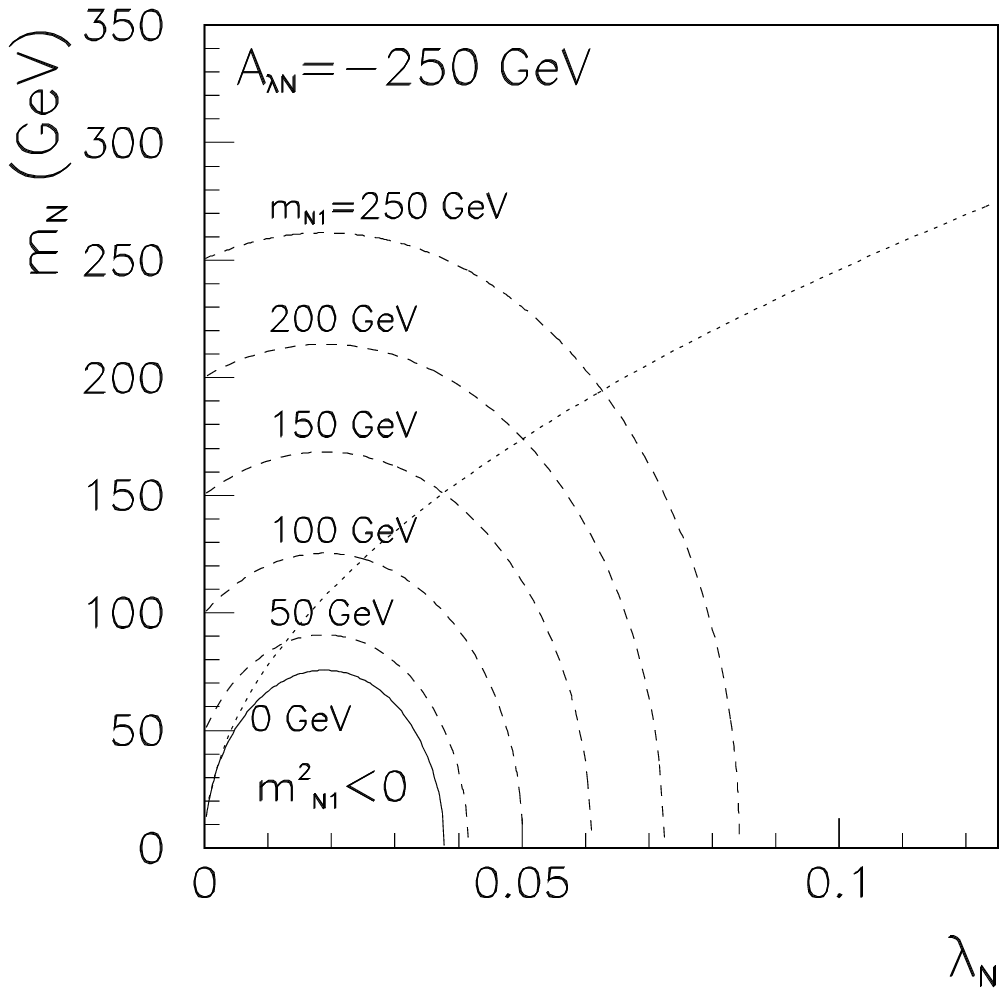,width=9.cm}\hspace*{-1.2cm}
  \epsfig{file=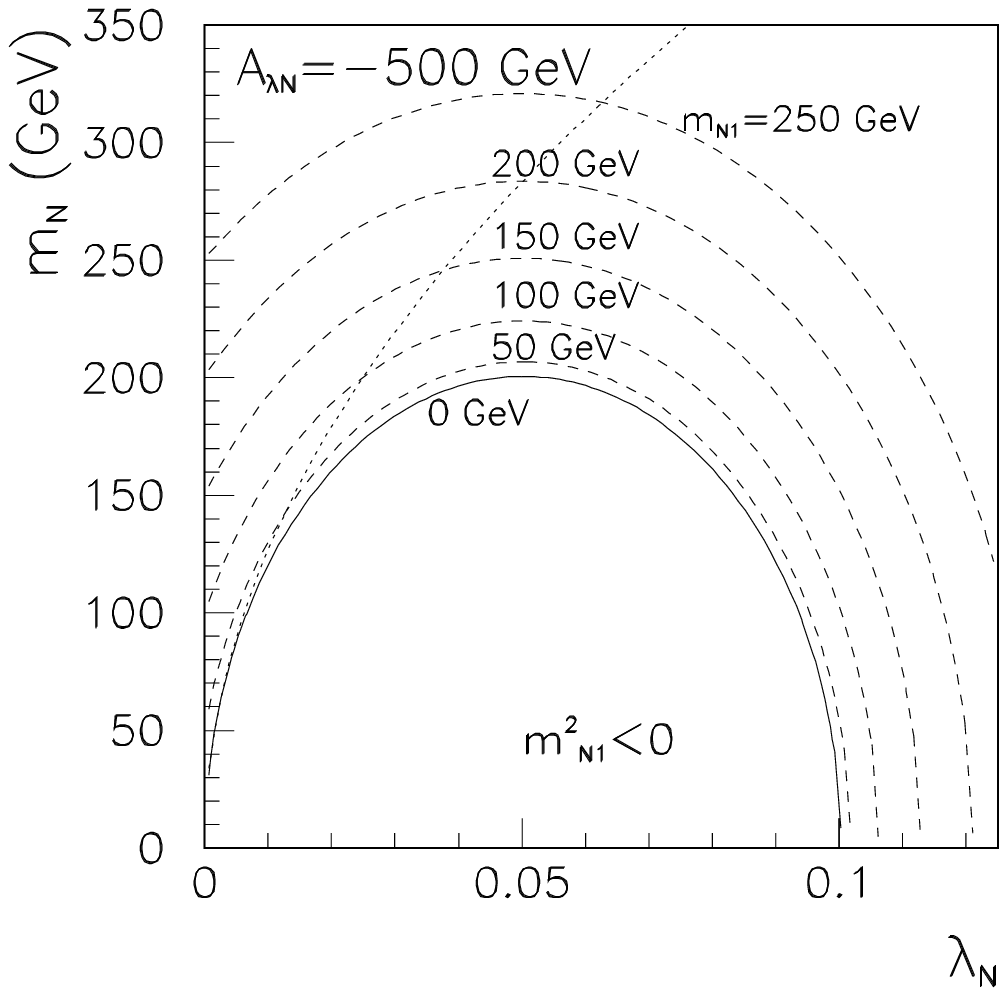,width=9.cm}
  \vspace*{-1cm}
  \captions{$(\mn,\ln)$ plane for the same example as in
     Fig.\,\ref{fig:spectrum-b} but with $\al=250$~GeV (left) and
     500~GeV (right). The trajectories with a fixed sneutrino mass are
     indicated by means of dashed lines. The dotted line represents the
     trajectory along which the sneutrino mass equals the right-handed
     neutrino mass (above that line, $\snmassr>\rhnmass$).
  }
  \label{fig:lnmn-example}
\end{figure}

Variations in $\kappa$ and $\aln$ have no impact on the right-handed
neutrino mass, but affect the mass splitting of sneutrino states
through their contribution to the $m_{RR}^2$ term in
Eq.~(\ref{sn:mrr}). Depending on the relative signs of the various
terms in this expressions, variations in $\kappa$ and $\aln$ can
enhance or decrease the
lightest sneutrino mass, respectively.

In order to illustrate the effect of the $\aln$ parameter, we have
represented in 
Fig.\,\ref{fig:lnmn-example} the trajectories in the $(\mn,\ln)$ plane
with 
a fixed sneutrino mass for $\aln=-250$~GeV and $-500$~GeV. 
For a given value of the sneutrino mass a larger $|\aln|$
allows a wider range of values of $\ln$. We will later use this
flexibility in order to look for regions in the parameter space with
the correct relic abundance.

Variations in $\l$ and $\k$ also alter the allowed range
of values for $\ln$ given a fixed sneutrino mass, since they affect
the $m_{RR}^2$ term. 
In general, an increase (decrease) in $|\l|$ ($|\k|$) leads to a
larger $|m_{RR}^2|$ and therefore to a larger range in $\ln$, with an
effect that mimics the increase in $|\aln|$ discussed above\footnote{
  Remember in this sense that we are concentrating on the case
  $m_{RR}^2<0$, which leads to the real component of the
  sneutrino to be the lightest particle. In the opposite case,
  $m_{RR}^2>0$,  when the lightest component is the imaginary one, the
  range in $\ln$ would be enlarged for a decrease (increase) in $|\l|$
  ($|\k|$), again due to the enhancement of $|m_{RR}^2|$.}.  
Notice also that, for a given $\mu = \l \vevs$,
an increase in $|\l|$ entails a reduction of
the right-handed neutrino mass through the decrease in $\vevs$. This
makes it easier to obtain sneutrinos with a mass larger than the
right-handed neutrino, which might be welcome in order to obtain the
correct relic abundance as we will see in the next section.

\begin{figure}
  \hspace*{-0.5cm}
  \epsfig{file=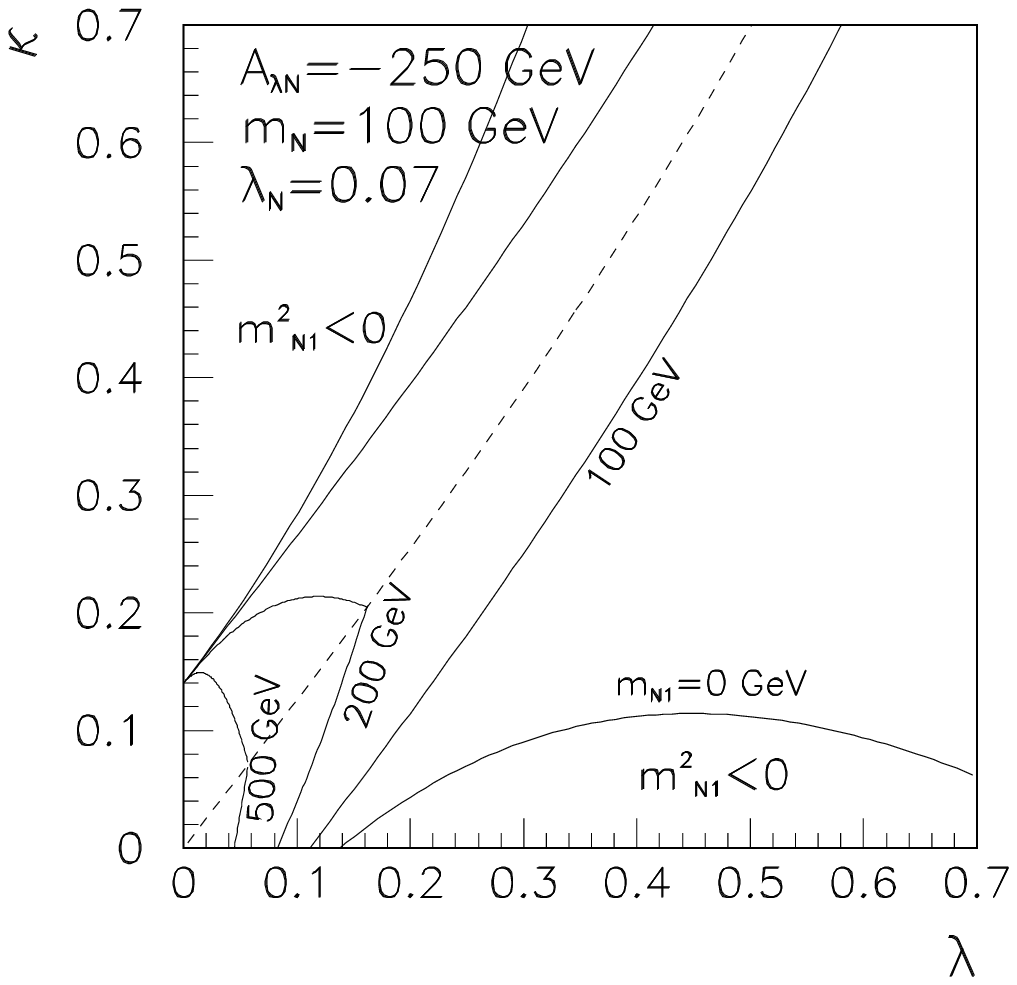,width=9.cm}\hspace*{-1.2cm}
  \epsfig{file=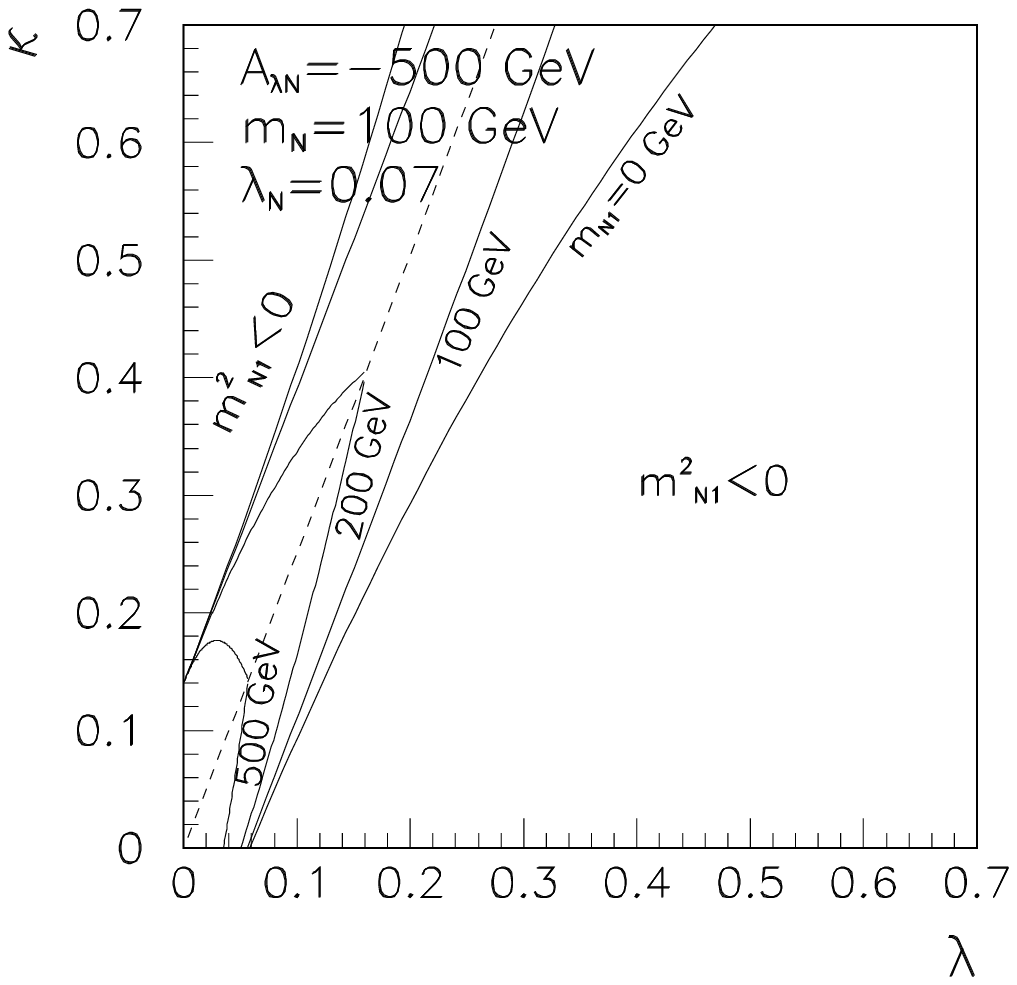,width=9.cm}
  \hspace*{-0.5cm}    
  \epsfig{file=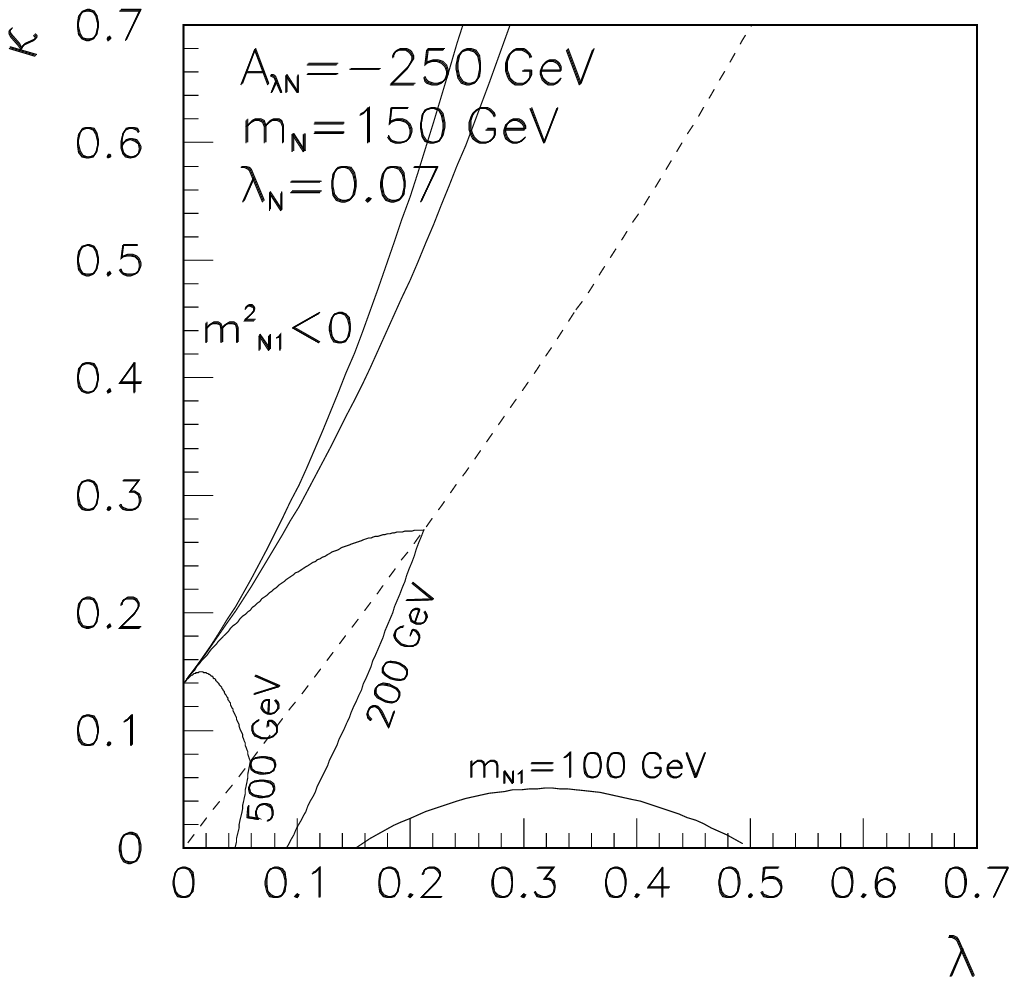,width=9.cm}\hspace*{-1.2cm}
  \epsfig{file=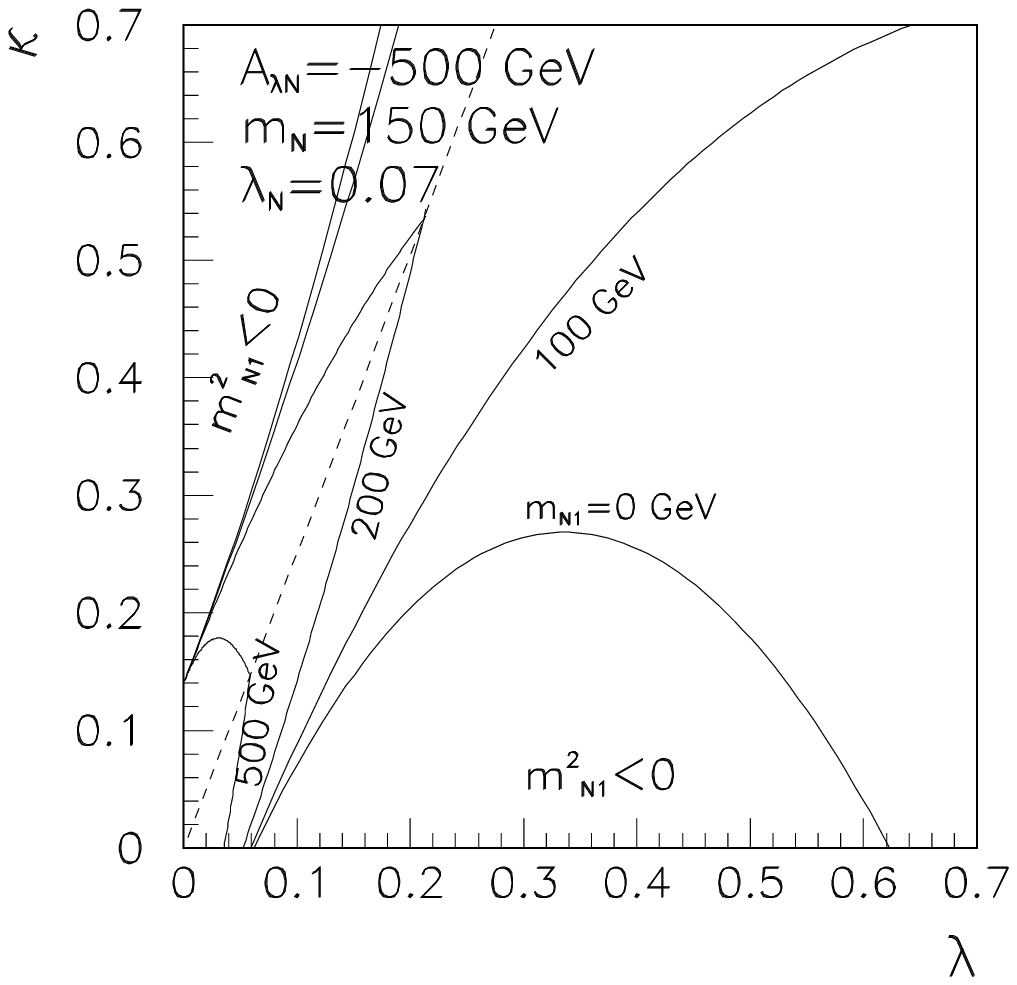,width=9.cm}
  \vspace*{-1cm}
  \captions{Trajectories for constant sneutrino mass in the $(\l,\k)$ plane 
    the same case described in Fig.\,\ref{fig:spectrum-b} and the
    choices $\ln=0.07$, $\mn=100$, $150$~GeV and $\aln=250$, $500$~GeV.
    Below the dotted line, the lightest right-handed sneutrino
    corresponds to the real component, whereas the imaginary component
    becomes lighter in regions above that line. 
  }
  \label{fig:lk-example}
\end{figure}

For completeness, we display in Fig.\,\ref{fig:lk-example} the
contours with a constant sneutrino mass in the $(\l,\k)$ plane for
a specific choice of $\ln$, $\mn$, and $\aln$. As $|\l|$ ($|\k|$)
increases (decreases), the splitting between the two right-handed
sneutrino states (real and imaginary components) decreases and at some
point they become degenerate. This is indicated by means of a dashed
line in the plot. For our choice of parameters the real component of
the right-handed sneutrino is the light state below that line and the
imaginary component becomes lighter in the region above. Notice that
(depending on the values of the input parameters) the sneutrino can
become tachyonic (e.g., for small values of $\k$ or small values of
$\l$). Obviously, the tachyonic regions are larger for a smaller
sneutrino soft mass.

Finally, despite the importance of variations of $\tan\beta$ in the
NMSSM phenomenology, they only affect
the term $v_1v_2$ in Eq.~(\ref{sn:mrr}), 
which is proportional to $\tan\beta/(1+\tan^2\beta)$ 
and have therefore 
little impact on the resulting right-handed sneutrino spectrum.

\subsection{Vacuum and Higgs potential}

So far, we have assumed that $\langle\tilde{N}\rangle=0$ holds in the
vacuum. However, the right-handed sneutrino may have a non vanishing
VEV, depending on the specific choice of 
parameters~\cite{ko99}. We will check here the validity of that
assumption and the implications for the relevant parameters of the
model.

The part of the 
scalar potential related with Higgses and sneutrinos is given by
\begin{eqnarray}
  V &=& |y_N H_2 \tilde{N}|^2+ |\lambda S H_2|^2
  + |\lambda S H_1|^2+ |- \lambda H_1 H_2 + \kappa S^2 + \lambda_N
  \tilde{N}^2 |^2 
  + |2\lambda_N S \tilde{N}|^2  \nonumber\\
  && + V_D(H_1, H_2) \nonumber\\
  && + m_{H_1}^2 |H_1|^2 + m_{H_2}^2 |H_2|^2 + m_S^2 |S|^2 
  + m_{\tilde{N}}^2 |\tilde{N}|^2 \nonumber\\
  && + 
  \left(-\lambda A_{\lambda} S H_1 H_2 + \frac{1}{3}\kappa A_{\kappa}
  S^3 
  + \lambda_N A_{\lambda_N} S \tilde{N}^2 + {\rm H.c.} \right).
\end{eqnarray}
As showed in the previous subsection, the left-handed and right-handed
sneutrinos are almost decoupled from each other.  
Thus, for this discussion, only right-handed sneutrinos are relevant.
From the stationary condition 
\begin{eqnarray}
  \frac{\partial V}{\partial \tilde{N}} &=&
  2 \lambda_N (- \lambda H_1 H_2 + \kappa S^2 )^* \tilde{N}
  +2 |\lambda_N|^2 |\tilde{N}|^2 \tilde{N}^*
  \nonumber\\
  && + \left(|2\lambda_N S|^2+ m_{\tilde{N}}^2 + |y_N H_2|^2\right)
  \tilde{N}^*   
  + 2 \lambda_N A_{\lambda_N} S \tilde{N} \nonumber \\
  &=&  \tilde{N}^*
  \left[ m_{R\bar{R}}^2 + 2 |\lambda_N|^2 |\tilde{N}|^2 
    + \left(2 \lambda_N (- \lambda H_1 H_2 + \kappa S^2 )^*
    + 2 \lambda_N A_{\lambda_N}
    S\right)\left(\frac{\tilde{N}}{\tilde{N}^*}\right) 
    \right]  \nonumber \\
  &=& 0,
\end{eqnarray}
the possible minimum or maximum is given by
\begin{equation}
  |\langle\tilde{N}\rangle|^2  = \left\{
  \begin{array}{c}
    0  \\
    \frac{|2 \lambda_N (- \lambda H_1 H_2 + \kappa S^2 )^*
      + 2 \lambda_N A_{\lambda_N} S | - m_{R\bar{R}}^2 }
         {2|\lambda_N|^2} \quad ( {\rm if} > 0)   \\
  \end{array}
  \right. .
\end{equation}
If $ 2|m_{RR}^2| - m_{R\bar{R}}^2 < 0$,
then, a non vanishing VEV is absent and the origin $\tilde{N} =0$ 
is the true minimum and vacuum.
Of course, this condition is exactly same as the condition for the
positiveness of the 
mass squared of the lighter right-handed sneutrino in
Eq.~(\ref{SneutrinoMass}).
Then, the scalar potential is reduced to the Higgs potential in the
NMSSM.

As stated in Introduction, the domain wall problem in this simplest
Higgs potential can be solved by some 
modifications~\cite{Abel:1995wk,Menon:2004wv}, which have no impact on
the sneutrino physics we are interested in.

\subsection{Higgs and neutralino sectors}

The inclusion of the $S$ superfield implies two new
Higgs states, one CP-even and one CP-odd, which mix with the MSSM
Higgs states. 
We assume 
that there is
no CP-violation in the Higgs sector and therefore 
CP-even and CP-odd states do not mix. Thus  
the corresponding mass matrices, which have exactly the same
expressions as in the NMSSM, can be written in the respective
basis, and the mass eigenstates expressed as a linear superposition
of the corresponding gauge eigenstates. For CP-even and CP-odd
Higgses, we have
\begin{equation}
  H_i^0 = \hcompid\, H^0_{uR}+\hcompiu\, H^0_{dR} + \hcompis\, S_R,
\end{equation}
and
\begin{equation}
  A_a^0 = \phcompid\, H^0_{uI}+\phcompiu\, H^0_{dI} + \phcompis\, S_I,
\end{equation}
for $i=1,2,3$, $a=1,2$, and where the subscripts $R$ and $I$ denote
the real and
imaginary components of the
corresponding fields, respectively.
The minimisation of the scalar potential and subsequent calculation of
the Higgs masses is carried out with the code 
\nmh~2.0 code~\cite{Ellwanger:2005dv}.

The superpartner of the singlet Higgs, the singlino $\tilde S$, 
has the same quantum numbers as the bino, wino and Higgsinos and
therefore mixes with them, giving rise to a fifth
neutralino state. This is exactly the same situation as in the
NMSSM~\cite{Menon:2004wv,Greene:1986th,Cerdeno:2004xw} and the
neutralino mass matrix has the same expression.
The mass matrix is diagonalized as usual by an unitary 
matrix $N^{\tilde{\chi}}$.
The neutralino mass eigenstates are therefore a linear superposition
of bino, wino, Higgsinos and singlino which we express as
\begin{equation}
  \tilde{\chi}^0_i
  = \neuticompb \tilde{B} + \neuticompw \tilde{W^3} + \neuticompd
  \tilde{H_1^0} + \neuticompu \tilde{H_2^0} + \neuticomps \tilde{S} .
\end{equation}

\subsection{Right-handed sneutrino LSP}

The phenomenology of this construction is largely dependent on which
particle plays the role of the lightest supersymmetric particle. 
In the ordinary NMSSM the
neutralino is the LSP in extensive regions of the parameter space,
thereby providing an interesting candidate for dark matter, which can
reproduce the correct relic abundance and be within the sensitivity of
some of the future direct detection experiments
\cite{Cerdeno:2004xw}.

In this extended model another possibility arises, since the
right-handed sneutrino might also be the LSP. 
The mass of the lightest right-handed sneutrino, is given by 
$ m_{R\bar{R}}^2 - 2|m_{RR}^2| $ (\ref{sn:mrr}), 
and is therefore dependent on the set of new parameters. In
particular, as we have shown in Fig.\,\ref{fig:lnmn-example},
below a certain value of $\ln$ it is always possible 
to choose a value of the soft mass
parameter, $\mn$, for which the right-handed sneutrino mass will
be small enough so as to guarantee that it becomes the LSP. 
Being a neutral particle
with electroweak scale interactions, the right-handed sneutrino would
therefore constitute a dark matter candidate within the category of
WIMPs. 
We will explore this possibility by analyzing the sneutrino relic
abundance and direct detection cross section in the following
Sections.

\section{Thermal relic density}
\label{sec:relic}

A crucial feature of this scenario is the existence of direct
couplings of the right-handed sneutrino to Higgses and
neutralinos. These emerge through the term $\ln SNN$ 
in the superpotential (\ref{superpotential}), 
since as we stated above, the singlet
and singlino components of $S$ mix with the CP-even Higgs bosons
and neutralinos,
respectively. 
The strength of the interaction is therefore 
dependent on the value of $\ln$ and 
$\aln$. This provides tree-level interactions with ordinary matter
which are specified in 
Appendix~\ref{sec:feynman}. 
For adequate values of $\ln$ (and $\aln$), 
these couplings would be of electroweak
scale, 
thereby making the right-handed sneutrino a potential WIMP 
candidate\footnote{
  This is contrary to the case of right-handed sneutrinos with
  the MSSM, which do not present (unsuppressed) tree-level couplings
  to ordinary matter and therefore have to be generated via
  non-thermal processes (e.g., late decays of the NLSP)
  ~\cite{Asaka:2005cn,Gopalakrishna:2006kr,McDonald:2006if,Page:2007sh}.}.
Still, in order to determine whether or not the right-handed sneutrino
is a viable WIMP, its relic density has to be evaluated
and compared with the recent measurement by the WMAP satellite, 
$0.1037\le\Omega h^2\le 0.1161$.
\cite{Dunkley:2008ie}.

The possible annihilation products for right-handed sneutrinos in this
construction include the following channels
\begin{description}
\item {(i)}
  $W^+\, W^-$, $Z\, Z$, and $f \bar{f}$
  via $s$-channel Higgs exchange;
\item {(ii)}
  $H_i^0\, H_j^0$, 
  via $s$-channel Higgs exchange, $t$-
  and $u$-channel sneutrino exchange, and a scalar quartic coupling;
\item {(iii)}
  $A_a^0\, A_b^0$, and $H_i^+\, H_j^-$,  
  via  $s$-channel Higgs exchange, and a scalar quartic coupling;
\item {(iv)}
  $Z\,A_a^0$ and $W^\pm\,H^\mp$
  via $s$-channel Higgs exchange;
\item {(v)}
  $NN$, 
  via $s$-channel Higgs exchange and via $t$- and $u$-channel
  neutralinos exchange.
\end{description}
The processes suppressed by the neutrino Yukawa $y_N$ (such as
$s$-channel sneutrino annihilation mediated by the $Z$ boson) 
have not been
included, since these are negligible as shown in the previous
section. The corresponding expressions for the amplitudes of each
channel are listed in 
Appendix\,\ref{sec:wtilde}.

Since all the above annihilations involve s-channel Higgs exchange
processes,  
resonant sneutrino annihilation will take place near the pole, where
$\snmassr\approx 2\hmassi$. 
This implies that the partial-wave expansion method is not
sufficiently accurate. 
Thus, we integrate it directly, following the same procedure detailed
in Ref.\,\cite{nrr02} for the case of the neutralino. 
In our calculation, we assume the absence of a degenerate particle 
and do not include possible associated co-annihilation effects. 
Notice that this could in principle happen 
between the real and imaginary component of sneutrinos 
and between the lightest sneutrino and the lightest neutralino.
The former situation only arises if $m_{RR}^2$ in
(\ref{SneutrinoMass}) 
is very small, which implies some accidental cancellation
in (\ref{sn:mrr}) and only occurs for specific choices if input
parameters which we will explicitly avoid. 
The latter case would only be relevant near those regions in which the
sneutrino mass is close to the NLSP mass. 
These regions of the parameter space can be
easily identified and we will comment on this later.

As we explained above, and is evidenced in the different annihilation
channels previously detailed, the main feature of this construction is
the direct coupling of the sneutrino to the Higgs fields. Hence, the
sneutrino annihilation cross section is extremely dependent on the
structure of the Higgs sector. Although this introduces a strong
dependence of our results on the NMSSM parameter space, there are some
general features which are easy to identify and understand.

\begin{itemize}
\item The coupling 
  $\lambda_N$ determines the overall scale of the annihilation
  cross section. A larger $\lambda_N$ implies more effective
  sneutrino annihilation, and in turn a smaller relic abundance, and
  viceversa. Regarding the supersymmetric spectrum,
  notice that $\ln$ only affects the right-handed neutrino
  and sneutrino masses and does not alter the rest of the NMSSM
  spectrum. Thus, having chosen a set of viable 
  NMSSM input parameters, the
  value of $\ln$ (as well as $\mn$ and $\aln$) 
  can be freely varied in order to reproduce the
  correct sneutrino relic abundance, being only constrained by
  perturbativity (we will impose $\ln<1$), the occurrence of tachyons
  in the sneutrino sector or the upper bound on the sneutrino mass in
  order for it to be the LSP.

\item Annihilation into $H_i^0\, H_j^0$ and  $A_a^0\, A_b^0$
  will turn
  out to be among the most effective 
  channels, as we will later see. Whether these
  are kinematically allowed or not depends on the Higgs
  masses. Interestingly, within the framework of the NMSSM, very light
  CP-even and CP-odd 
  Higgses are possible (as long as they have a significant singlet
  component), making these channels available 
  for a wide range of sneutrino masses. In the next Section we will
  show how this can, for instance,
  make it possible for light sneutrinos to reproduce the correct relic
  abundance.

\item Another relevant contribution to the total annihilation cross
  section is due to the annihilation into a pair of
  right-handed neutrinos, $N$. This channel is kinematically allowed
  when $\snmassr>\rhnmass$. 
  As we saw in
  Fig.\,\ref{fig:spectrum-b}, this is
  allowed above a certain value of the soft sneutrino mass, $\mn$,
  which depends on the specific values of $\ln$ and $\aln$ and which
  increases for larger $\ln$.

\item Finally, 
  for all the possible annihilation products there is always a
  contribution coming from $s$-channel CP-even Higgs exchange. This
  implies that all of them are subject to a resonant effect
  when $2\snmassr\approx\hmassi$, for $i=1,2,3$. 
  The occurrence of resonant sneutrino
  annihilation gives rise to a characteristic decrease of the
  relic abundance at the corresponding values of the sneutrino mass.
  Furthermore, given the possibility of light scalar Higgses in the
  NMSSM, this resonant annihilation can be present even for light
  sneutrinos. 

\end{itemize}

Having understood the basic properties of right-handed sneutrino
annihilation in this framework, we will now proceed to explore the
parameter space and provide specific examples which show how the
correct thermal relic abundance can be obtained.

\subsection{Numerical examples}

The input parameters of this scenario are, on the one hand, 
the usual NMSSM parameters, which we define at low-energy,
\begin{equation}
  \lambda, \, \kappa,\, \tan \beta,\, \mu,\, A_\lambda, \, A_\kappa\,. 
  \label{nmssmparam}
\end{equation}
The soft supersymmetry-breaking terms, namely gaugino masses,
$M_{1,2,3}$, flavour independent 
scalar masses, $m_{Q,L,U,D,E}$, and trilinear parameters,
$A_{U,D,E}$, are also taken as free parameters and specified at
low scale.  
More specifically, we have assumed they gaugino masses mimic, at
low-energy 
the values obtained from a hypothetical GUT unification. Thus we set
$M_1:M_2:M_3 = 1:2:6$. 
The present lower bound on a possible supersymmetric contribution to
the muon anomalous magnetic moment, 
sets stringent upper constraints on
the mass of sleptons. As it was shown in \cite{Cerdeno:2004xw}, for
small values of $\tan\beta$ the
experimental result can be reproduced
with the choice
$m_{L,E}= 150$ GeV and $A_E=-2500$~GeV and a small value for the
gaugino mass
\footnote{It should be
  stressed that if one does not wish to impose the bound on the muon
  anomalous magnetic moment (motivated e.g. by tau data, which lead to
  a better agreement with the SM result), large slepton masses, equal
  to squark masses can be chose. 
  This would, however, have no consequence on the calculation of
  neither the relic abundance of right-handed sneutrinos, nor in their
  detection cross section.}, 
$M_1\lsim160$~GeV. Larger values of $\tan\beta$ are, however less
constrained \cite{Domingo:2008bb}.
In order to satisfy the experimental constraint on the branching ratio
of the $b\to s\gamma$ rare decay a careful choice has to be made of
parameters in (\ref{nmssmparam}). More details on the conditions
under which these bounds can be fulfilled can be found in
\cite{Cerdeno:2004xw,Domingo:2007dx}.

The analysis of the low-energy NMSSM phenomenology has been performed
with the  \nmh~2.0 code~\cite{Ellwanger:2005dv}, which 
minimises the scalar potential, dismissing the presence of
tachyons and/or false minima, and computes the Higgs boson masses
including 1- and 2-loop radiative corrections, as well as the rest of
the supersymmetric masses.
Based on this code, we have built a set of routines which numerically
calculate the right-handed sneutrino spectrum and 
relic density as described in the previous
section. 

More specifically, we impose the experimental
bound on the branching ratio of the rare $\bsg$ decay,
$2.85\times10^{-4}\le\,{\rm BR}(\bsg)\le 4.25\times10^{-4}$ at
$2\sigma$ level, obtained
from the experimental world average 
reported by the Heavy Flavour Averaging Group \cite{bsgHFAG07},
and the theoretical calculation in the Standard Model
\cite{bsg-misiak}, 
with errors combined in quadrature.  
We also take into account the upper constraint on
the $\bmumu$ branching ratio obtained by CDF,
BR$(\bmumu)<5.8\times10^{-8}$ at $95\%$ c.l. \cite{bmumuCDF07}
(which improves the previous one from D0 \cite{bmumuD007}).

Regarding the muon anomalous magnetic moment, a constraint on the
supersymmetric contribution to this observable, $\asusy$, can be
extracted by comparing the experimental result
\cite{g-2}, with the most recent theoretical evaluations of the
Standard Model contributions \cite{g-2_SM,newg2,kino}. 
When $e^+e^-$ data are used the experimental excess in
$a_\mu\equiv(g_\mu-2)/2$ would constrain a possible supersymmetric
contribution to be $\asusy=(27.6\,\pm\,8)\times10^{-10}$, where
theoretical and experimental errors have been combined in
quadrature. However, when tau data are used, a smaller discrepancy
with the experimental measurement is found. Due to this reason, in our
analysis we will not impose this constraint, but only indicate whether
the points are compatible with it at the $2\sigma$ level, for which we
assume the range
$11.6\times10^{-10}\le\asusy\le43.6\times10^{-10}$.

The inclusion of the new superfield $N$ and the corresponding terms in
the superpotential and Lagrangian
leaves three new parameters to be fixed. Following the discussion in
the previous section, these can be chosen as
\begin{equation}
  \ln,\, \mn,\, \aln.
  \label{rhnmssmparam}
\end{equation}
These, together with the NMSSM parameters (\ref{nmssmparam}) fully
specify the model. Our task is now to determine whether the correct
sneutrino relic density can be obtained with reasonable choices of the
above set of parameters.

\begin{table}
  \begin{center}
    \begin{tabular}{|c|r|r|r|r|}
      \hline
      & A)        & B1)     & B2)      & C) \\
      \hline
      $\tan\beta$ 
      & 5                   & 5        & 5       & 3.5   \\
      $\al$     
      & 550 GeV             & 400 GeV  & 400 GeV & 480 GeV\\ 
      $\ak$
      &-200 GeV             &   0 GeV  &  0 GeV  & -50 GeV\\  
      $\mu$     
      & 130 GeV             & 200 GeV  & 200 GeV & 230 GeV\\
      $(\l,\,\k)$ 
      &(0.2, 0.1)           & (0.1, 0.05)& (0.3, 0.2)  &(0.5, 0.3)\\
      \hline
      $M_1$
      & 200 GeV   & 150 GeV & 150 GeV   & 300 GeV \\
      $m_{L,E}$   &  250 GeV  &  250 GeV   & 250 GeV & 250 GeV   \\
      $m_{Q,U,D}$ & 1000 GeV  & 1000 GeV  & 1000 GeV  & 1000 GeV  \\
      $A_{E}$     & -2500 GeV & -2500 GeV & -2500 GeV & -2500 GeV \\
      $A_{U,D}$     &  1500 GeV & 1500 GeV  & 1500 GeV  & 1000 GeV   \\
      \hline
      \hline
      $\hmassl$   &  62.4 GeV &  117.7 GeV & 116.2 GeV & 115.1 GeV\\      
      $\hmassm$   & 119.4 GeV &  200.8 GeV & 267.7 GeV & 263.5 GeV\\      
      $\hmassh$   & 634.1 GeV &  706.4 GeV & 730.6 GeV & 725.4 GeV\\
      $\phmassl$  & 199.6 GeV &   14.2 GeV &  47.6 GeV & 169.7 GeV \\      
      $\phmassh$  & 632.5 GeV &  705.1 GeV & 727.9 GeV & 719.5 GeV\\
      $\neutmass$ &  95.8 GeV &  126.3 GeV  & 124.3 GeV & 190.1 GeV \\      
      \hline
      BR($\bsg$) & $4.15\times 10^{-4}$& $4.02\times 10^{-4}$  &
      $4.02\times 10^{-4}$ & $4.03\times 10^{-4}$ \\ 
      $\asusy$ & $4.08\times 10^{-10}$&$5.58\times 10^{-10}$ &
      $8.05\times 10^{-10}$& $2.96\times 10^{-10}$     \\      
      \hline
    \end{tabular}
  \end{center}
  \captions{
    Set of inputs corresponding to 
    the examples used in the analysis,
    together with the predicted values of BR($\bsg$) and $\asusy$ and
    the relevant part of the resulting supersymmetric spectrum.} 
  \label{tab:examples-real}
\end{table}

Let us now illustrate with some examples the 
theoretical predictions for the relic
abundance of right-handed sneutrinos and the various possibilities
which allow to reproduce the correct relic density. The sets of
parameters for each example are detailed in
Table\,\ref{tab:examples-real}, together with the predicted 
values\footnote{Some of these examples are inspired in the scenarios
  analysed in  \cite{Cerdeno:2004xw} (e.g., cases B1) and B2) are
  similar to the case studied in Fig.\,7 there), and we refer the
  reader to that work for a more detailed study of the NMSSM parameter
  space and associated experimental constraints.} 
of BR($\bsg$) and $\asusy$.

\subsubsection{Predominant annihilation into scalar Higgses}
\label{sec:higgs}

As already mentioned, sneutrino annihilations into Higgs bosons
are generally among the dominant channels, when they are
kinematically allowed. This is also the case in other models for
thermal right-handed sneutrino \cite{Deppisch:2008bp}. However, in our
case, 
the flexibility and interesting properties of the Higgs
sector of the NMSSM make this possibility much very versatile.

As an specific example we will choose the NMSSM input parameters as in
case A) of 
Table\,\ref{tab:examples-real}. 
This point is very characteristic of the 
NMSSM since light CP-even Higgses are possible. In this case the
lightest Higgs has a mass  
$\hmassl\approx62$~GeV, which is consistent with the present LEP
constraints due to its large 
singlet component ($(\hcompls)^2=0.997$). 
The second lightest Higgs is MSSM-like and has a mass of
$\hmassm\approx120$~GeV. 
The lightest neutralino is a mixed bino-Higgsino-singlino state 
($(\neutcompb)^2\approx0.1$, $(\neutcompd)^2+(\neutcompu)^2\approx0.7$
and $(\neutcomps)^2\approx0.2$). 
Its mass
is $\neutmass\approx96$~GeV, and
this sets the upper bound for the sneutrino mass if the latter is to
be the LSP. 
Regarding the right-handed sneutrino sector, its mass is a function of
the parameters $\ln$, $\mn$ and $\aln$, as explained in the previous
section. As a first example we have fixed $\aln=-250$~GeV and performed
a scan in the allowed range for the $\ln$, $\mn$ parameters.

\begin{figure}[!t]
  \hspace*{-0.5cm}
  \epsfig{file=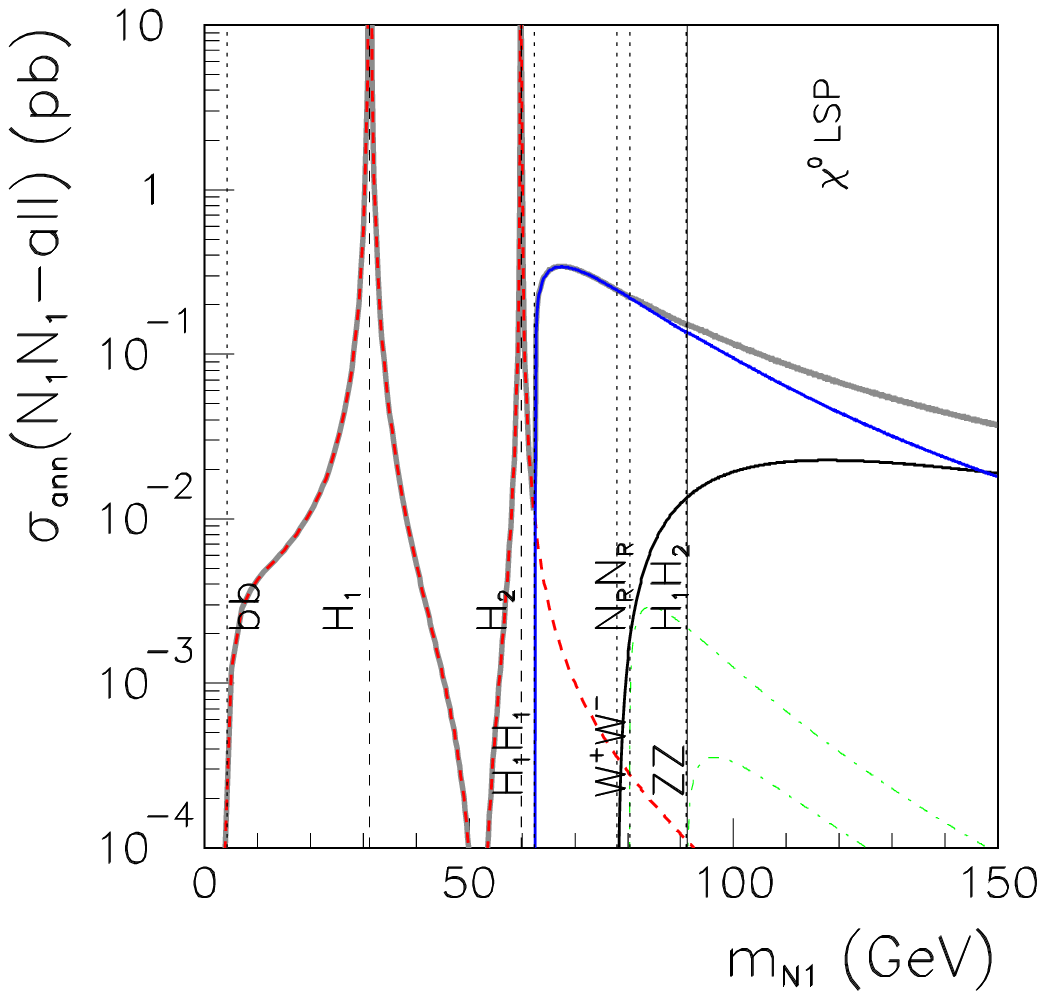,width=9.cm}\hspace*{-1.2cm}
  \epsfig{file=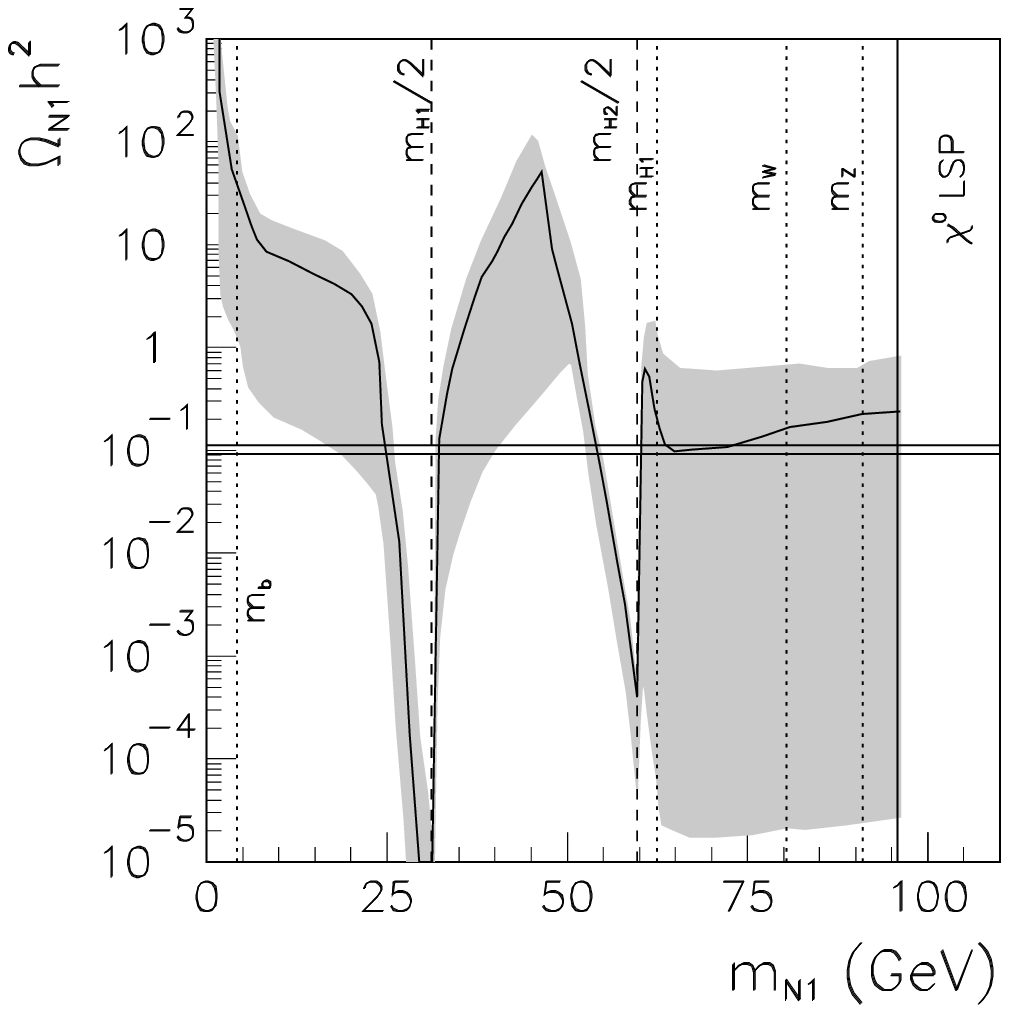,width=9.cm}
  \vspace*{-1cm}
  \captions{Left)
    Total sneutrino annihilation cross-section as a function of the
    sneutrino mass (grey thick solid line) for example A) of
    Table,\ref{tab:examples-real} with $(\l,\k)=(0.2,0.1)$, and
    $\ln=0.06$.     
    The contributions from the different annihilation channels are 
    indicated as follows, dashed red line for $f\bar f$, 
    dot-dashed green lines
    for $ZZ$ and $W^+W^-$, solid blue line for $\higgsi\higgsj$, and
    solid black line for $NN$.
    Vertical dashed lines indicate the location of the resonances in
    the $s$-channels when $2\snmassr\approx \hmassi$, whereas vertical
    dotted
    lines indicate when the different annihilation channels become
    kinematically accessible. The vertical solid line indicates the
    point at which the neutralino becomes the LSP.
    Right) Theoretical predictions for the sneutrino relic density
    as a function of the sneutrino mass for
    the same example, but for a scan with $\ln\in[0.05,0.1]$ and
    and $\mn\in[0,150]$~GeV. 
    The solid line indicates the result for
    $\ln=0.06$. 
  } 
  \label{fig:relic-c-real}
\end{figure}

The resulting sneutrino annihilation cross section is shown in
Fig.\,\ref{fig:relic-c-real}, together with the contribution for each
individual channel, as a function of the sneutrino mass for 
$\ln=0.06$ (the variation in the sneutrino mass is achieved with a
scan in $\mn$).
A first thing to notice is that the various annihilation channels
display a resonant enhancement due to the two lightest CP-even Higgses
when the sneutrino mass is approximately one half of the Higgs mass
for the two lightest scalar Higgses, i.e.,
$\snmassr\approx30$~GeV and $60$~GeV.
The resonance due to the third scalar Higgs state is not
observed since it would occur for large masses
($\snmassr\approx317$~GeV), for which
the right-handed sneutrino is no longer the LSP.

In this example, annihilation into a pair of lightest CP-even Higgses
dominates the total annihilation cross section once this channel
becomes kinematically allowed (for
$\snmassr>\hmassl\approx62$~GeV). Lighter sneutrinos can only
annihilate into a pair fermion-antifermion, being $b\bar b$ the
leading contribution. As observed in the plot, for this choice of
parameters this channel is less effective and only provides a
sufficiently large annihilation cross section close to the resonances
of the scalar Higgses\footnote{
  We remind the reader that for a WIMP of mass 100~GeV the necessary
  annihilation cross-section in order to reproduce the WMAP result is
  around $\sigma_{ann}\sim 0.1$~pb.}.
Notice also that although annihilation into a pair of
right-handed neutrinos is also possible for sneutrinos heavier than
$80$~GeV, this channel only becomes comparable to the Higgs
contribution for sneutrino masses of order $150$~GeV, for which the
sneutrino is no longer the LSP.

The resulting theoretical predictions for the relic density are
depicted on the right hand-side of Fig.\,\ref{fig:relic-c-real} for a
scan of $\ln$ in the range $\ln\in[0.05,0.1]$ and a scan in the
allowed values for $\mn$. The solid line
indicates the result for $\ln=0.06$. We can see how light sneutrinos
with masses as small as approximately $25$~GeV can be obtained with
the correct relic abundance due to the resonant effects in the Higgs
diagrams. Once annihilation into scalar Higgses is allowed,
the correct relic abundance can be obtained in the whole
range\footnote{Since 
  co-annihilation effects have not been included, our result for the 
  relic abundance is not exact when $\snmassr\approx\neutmass$. } 
$\snmassr\approx60-90$~GeV
with $\ln\approx0.06$.

\begin{figure}[!t]
  \begin{center}
    \epsfig{file=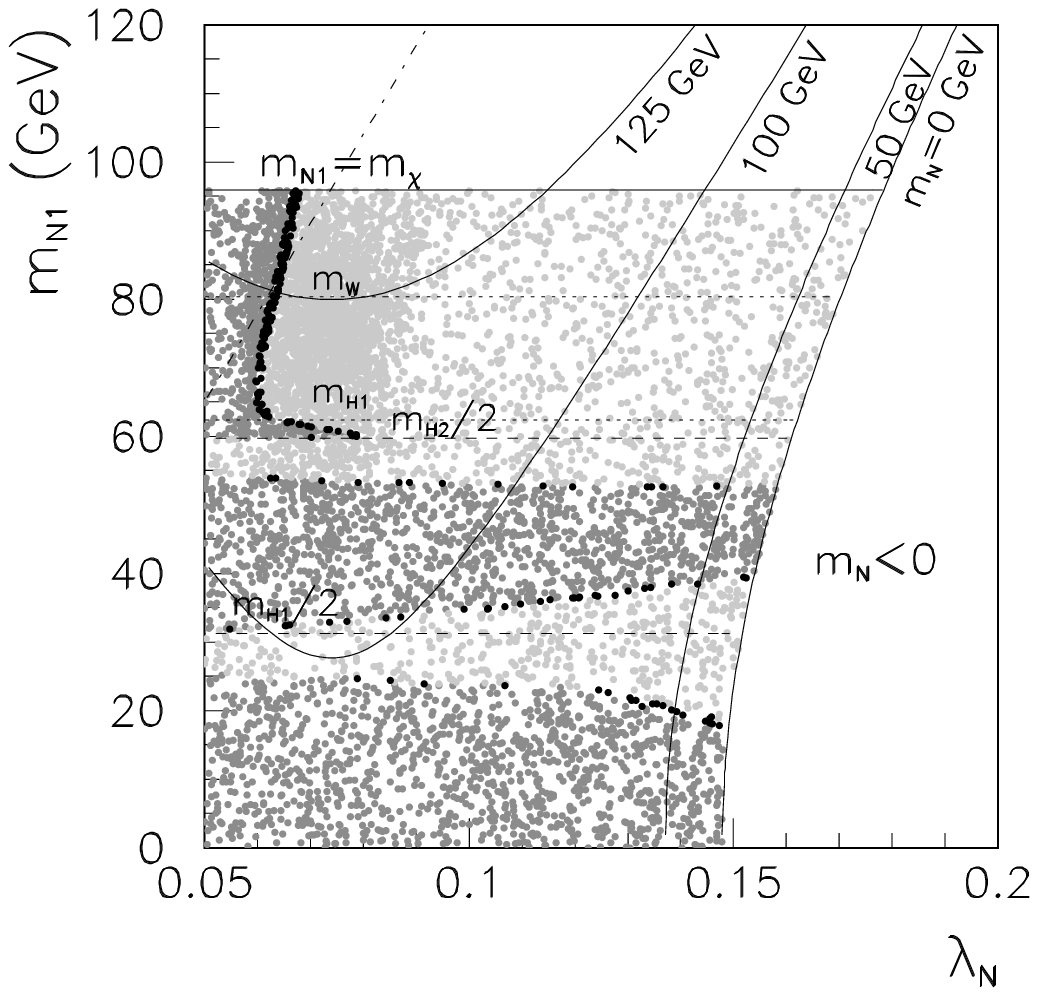,width=9.cm}\hspace*{-1.2cm}
  \end{center}
  \vspace*{-1cm}
  \captions{Effect of the relic density constraint on the
    $(\ln,\snmassr)$ plane for case A) in
    Table\,\ref{tab:examples-real} with $\aln=-250$~GeV.  
    Black dots represent the regions where the
    sneutrino relic abundance is in agreement with the WMAP
    constraint, whereas light(dark) grey dots represent those where
    the sneutrino relic abundance is smaller(larger). The curved 
    solid lines indicate the trajectories with a fixed value
    of the sneutrino soft mass $\mn$.
  } 
  \label{fig:lnmn-c}
\end{figure}

For a better understanding of the effect of the relic density
constraint on the sneutrino parameters, we represent in
Fig.\,\ref{fig:lnmn-c} the ($\ln,\snmassr$) plane, indicating with
black dots the
regions where the sneutrino relic abundance is in agreement with the
WMAP result. Dark grey dots represent the areas in which the
sneutrino relic abundance exceeds the WMAP constraint and light grey
dots are those in which the sneutrino relic density is smaller than
the WMAP bound.
The contours for a fixed sneutrino soft mass, $\mn$, are
indicated by means of solid lines. 
As already explained, when 
annihilation into a pair of Higgs bosons is allowed, 
a value of $\ln\approx0.06$ is favoured in this example. 
As we see, this corresponds to values of the sneutrino soft mass in
the range $100-150$~GeV.
Otherwise the
relic density is too large except along the resonances (the fine-tuned
near horizontal strips), for which the value of $\ln$ can be as large
as $0.15$.

\subsubsection{Predominant annihilation into $b\bar b$ and very
  light sneutrinos}  
\label{sec:bbar}

Obtaining the correct relic density with only annihilation into $b\bar
b$ would require a larger value of $\ln$. However, as we can see from
Fig.\,\ref{fig:lnmn-c}, if we restrict ourselves to positive values of
$\mn$ then for this
choice of parameters there is an upper bound $\ln\lsim0.15-0.18$.
Remember however from the previous section that an increase in
$|\aln|$ allows larger values of $\ln$. With this in mind, we have
reanalysed the previous example, case A) in
Table,\,\ref{tab:examples-real}, but now using $\aln=-500$~GeV, for
which values of $\ln$ up to 0.35 can be used. This considerably
increases the annihilation cross section.

\begin{figure}[!t]
  \hspace*{-0.5cm}
  \epsfig{file=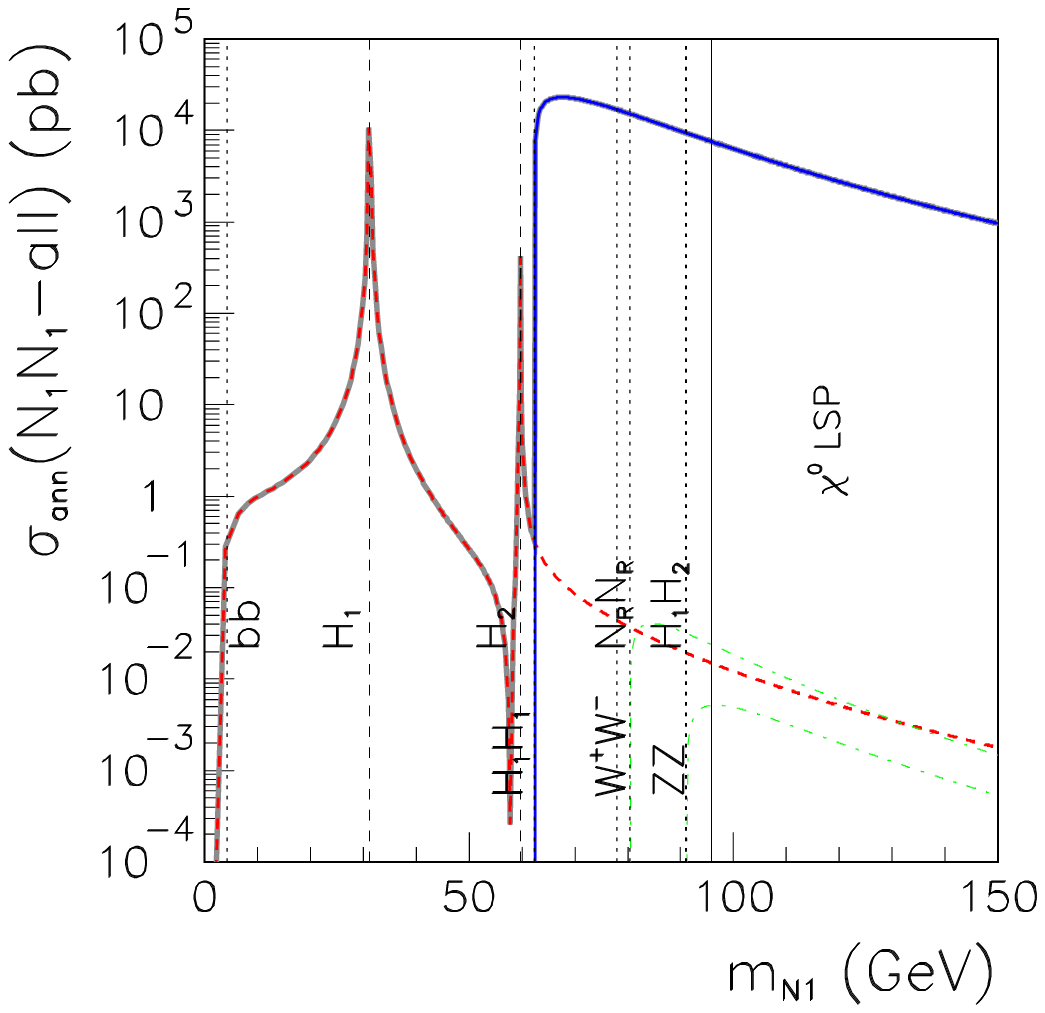,width=9.cm}\hspace*{-1.2cm}
  \epsfig{file=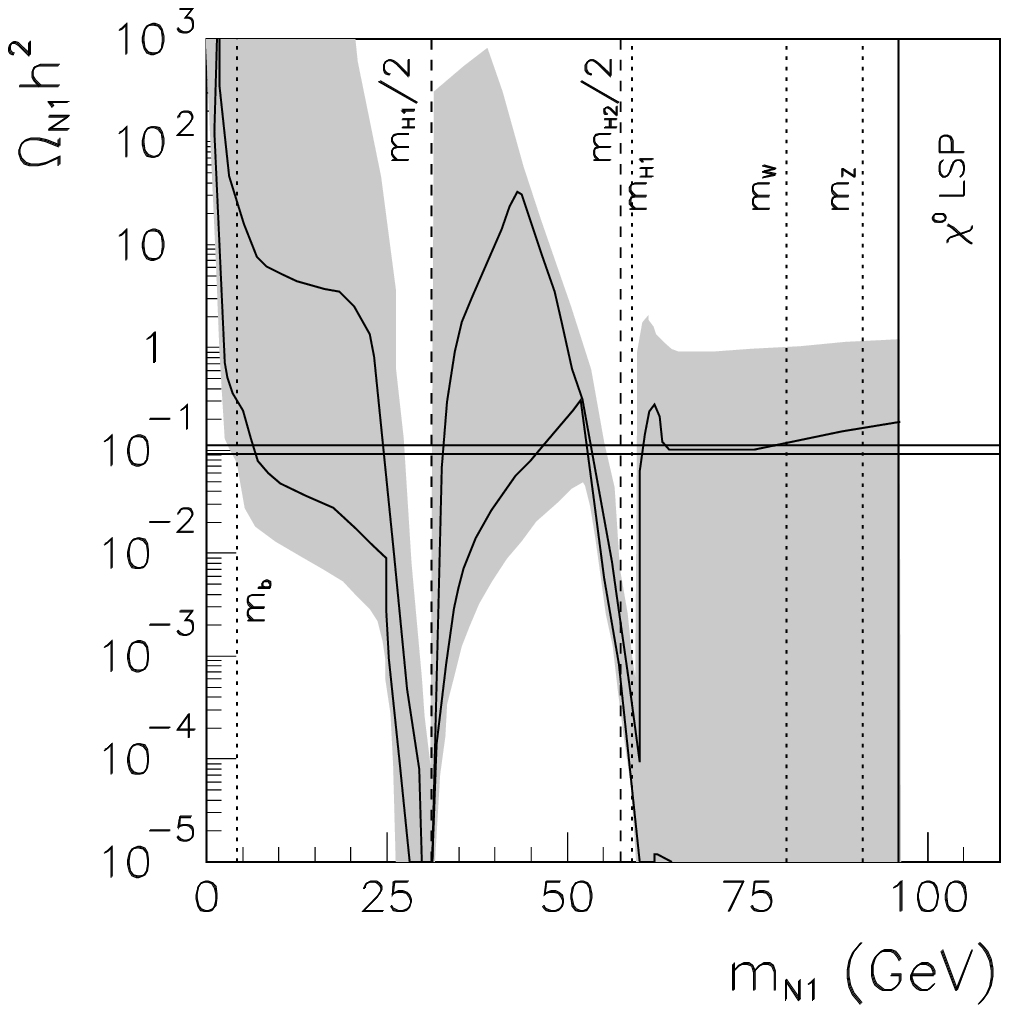,width=9.cm}
  \vspace*{-1cm}
  \captions{Left) The same as in Fig.\ref{fig:relic-b-real} 
    (case A) in Table\,\ref{tab:examples-real}) but with 
    $\aln=-500$~GeV and $\ln=0.25$. 
    Right)  Theoretical predictions for the sneutrino relic density
    as a function of the sneutrino mass for
    the same example, but for a scan with $\ln\in[0.05,0.35]$ and
    and $\mn\in[0,250]$~GeV. 
    The solid lines indicate, from top to bottom, the result for
    $\ln=0.11,\,0.25$. }
  \label{fig:relic-c2-real}
\end{figure}

For this example, 
Fig.\,\ref{fig:relic-c2-real} represents the annihilation
cross section for $\ln=0.25$ and a scan in the allowed values of $\mn$. 
The $b\bar b$ contribution is now
significantly larger, reaching values of order of $1$~pb without the
need of resonant effects. Similarly, the Higgs
contribution also 
increases considerably, thereby leading to a too large annihilation
cross section (notice the different scale on the plot). 
On the right-hand side of
Fig.\,\ref{fig:relic-c2-real} 
the resulting sneutrino relic abundance is plotted for a scan with 
$\ln\in[0.05,0.35]$ and and $\mn\in[0,250]$~GeV, where the solid lines
indicate, from top to bottom, the particular cases $\ln=0.11$ (which
yields similar results as the example studied in
Section\,\ref{sec:higgs}) and $0.25$.
Interestingly, as we can see, for such large values of $\ln$ light
sneutrinos with masses $\snmassr\gsim5$~GeV can reproduce the correct
relic abundance.

\subsubsection{Predominant annihilation into pseudoscalar Higgses}
\label{sec:pseudoscalar}

Another interesting possibility is sneutrino annihilation into a pair of
pseudoscalar Higgses. Although the pseudoscalar is rather heavy in the
MSSM (and experimentally 
constrained to be above $93.4$~GeV \cite{Amsler:2008zzb}),
in the NMSSM (where an extra CP-odd state is present) 
it can be much lighter in some regions of the parameter space
and
experimentally allowed if it has a large
singlet component
\cite{Dobrescu:2000jt,Gunion:2005rw,Dermisek:2005ar}. 
This therefore implies that this channel can be
kinematically allowed for lighter sneutrinos.

In order to illustrate this possibility, we have taken the NMSSM input
parameters as indicated in example B1) of 
Table\,\ref{tab:examples-real}.
This point in the NMSSM parameter space is a good
example of how light pseudoscalars might be phenomenologically
viable. In this case the lightest pseudoscalar has a mass
$\phmassl=14.2$~GeV and it is singlet-like.
The lightest scalar Higgs is MSSM-like with a mass
$\hmassl\approx116$~GeV, whereas the second lightest Higgs, with
$\hmassm\approx200$~GeV is mostly singlet.
The lightest neutralino is a mixed bino-Higgsino state
($(\neutcompb)^2\approx0.66$ and 
$(\neutcompd)^2+(\neutcompu)^2\approx0.25$)
with a mass $\neutmass=126$~GeV. It would be the
lightest supersymmetric particle in the NMSSM for this choice of
parameters and therefore sets once more 
the upper limit for allowed values of
the right-handed sneutrino mass.
Regarding the right-handed sneutrino sector, we have fixed
$\aln=-500$~GeV and performed a scan in the $\ln$, $\mn$ parameters. 
This is precisely the same example that was displayed on 
Fig.\,\ref{fig:lnmn-example}. As one can extract from there, 
a maximal value of
$\ln\sim0.11$ can be used, with the soft sneutrino mass ranging from
$\mn=0$ to approximately $\mn=235$~GeV.

\begin{figure}[!t]
  \hspace*{-0.5cm}
  \epsfig{file=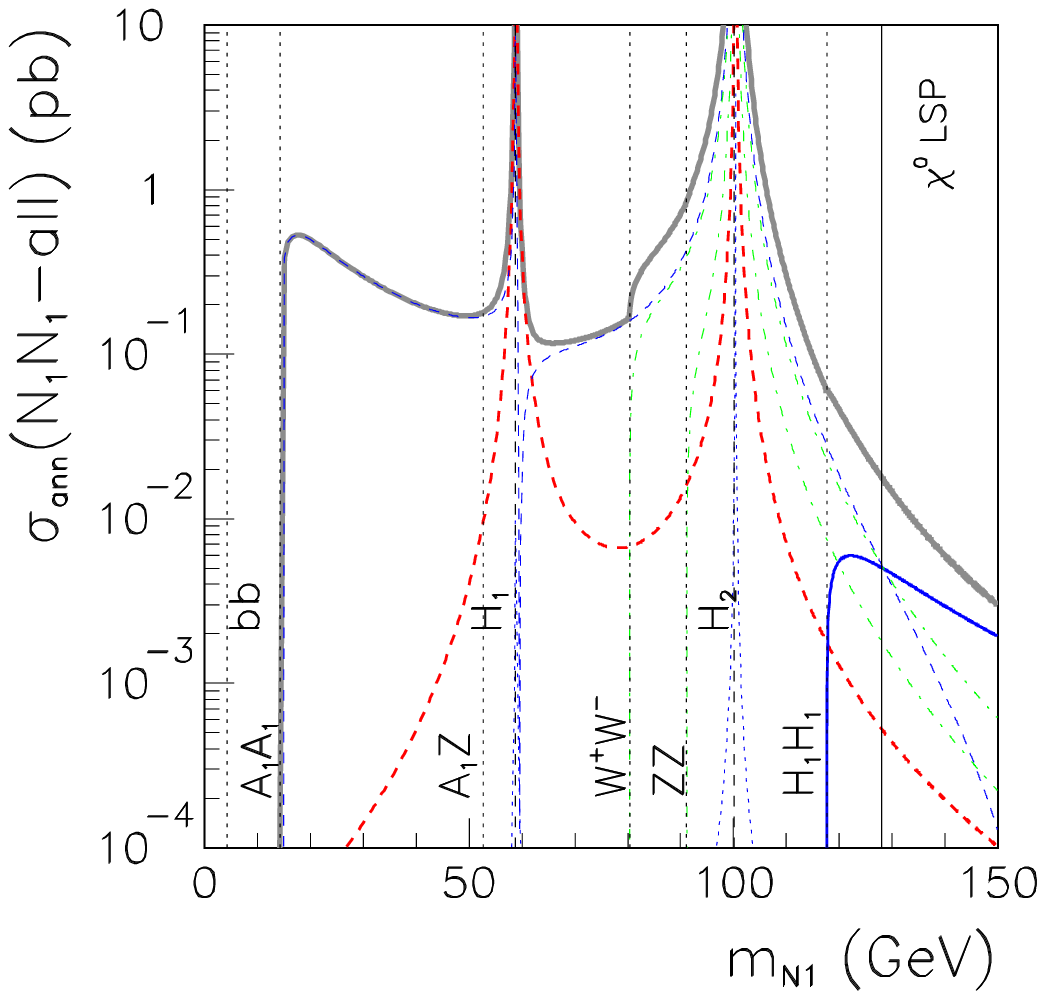,width=9.cm}\hspace*{-1.2cm}
  \epsfig{file=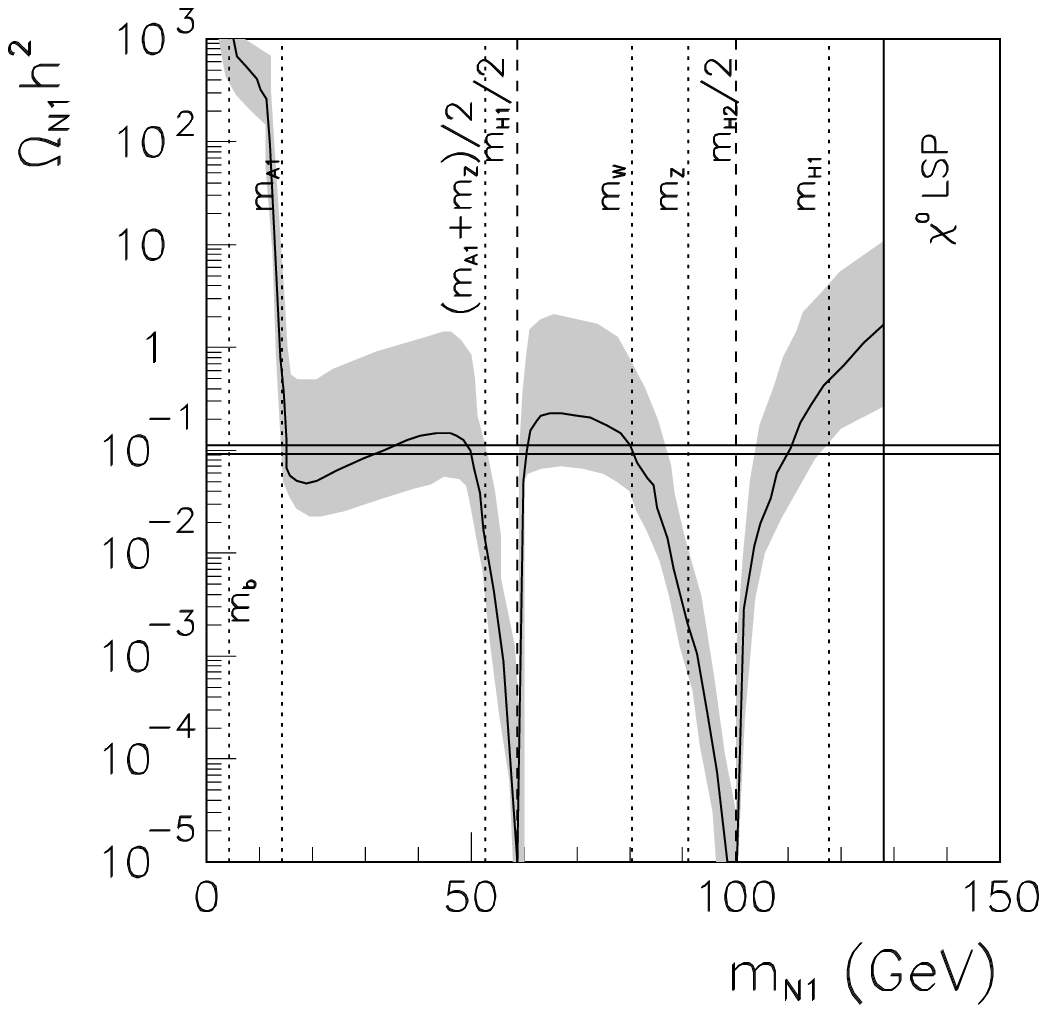,width=9.cm}
  \vspace*{-1cm}
  \captions{Left)
    The same as in Fig.\,\ref{fig:relic-c-real} but for example B1) of
    Table,\ref{tab:examples-real} with $\l=0.1$, $\k=0.05$, and
    $\ln=0.07$.     
    Here the 
    dashed blue line denotes the contribution of the
    $\phiggsi\phiggsj$ channel and the dotted blue line (only visible
    in the Higgs resonances)
    the contribution of 
    $\phiggsi Z$.
    Right) Theoretical predictions for the sneutrino relic density
    as a function of the sneutrino mass for
    the same example, but for a scan with $\ln\in[0.05,0.1]$ and
    and $\mn\in[0,250]$~GeV. 
    The solid line indicates the result for
    $\ln=0.07$. 
  } 
  \label{fig:relic-b-real}
\end{figure}

The predicted sneutrino annihilation cross section is represented on
the left hand-side of Fig.\,\ref{fig:relic-b-real} as a function of the
sneutrino mass, together with the contributions from each individual
annihilation channel. As we can see, for this example the annihilation
into a pair of CP-odd Higgses largely dominates over the other
contributions. This is particularly true for light sneutrinos since,
due to the smallness of $\phmassl$, this channel is kinematically
available for $\snmassr\gsim 14$~GeV. Note that for lighter sneutrinos
the only possibility is annihilation into a pair of $f\bar f$, which 
is less effective. 
The abrupt enhancements at the
resonances of the Higgs-exchanging $s$-channels when
$2\snmassr\sim\hmassl,\hmassm$ are also observed for every channel at
$\snmassr\approx 58$~GeV and $100$~GeV.

The theoretical predictions for the resulting relic density are
depicted on the right hand-side of Fig.\,\ref{fig:relic-b-real} for the
range $\ln\in[0.05,0.11]$. As we see there, the correct relic density
can be obtained for a wide range of sneutrino masses. In particular,
sneutrinos as light as $\snmassr\approx 15$~GeV are now possible due
to their very effective annihilation into a pair of CP-odd
Higgses. 
This is particularly interesting, since these light sneutrinos have
different properties from those obtained in case A). In case
A) they could only annihilate into $b\bar b$ and thus the coupling
$\ln$ had to be of order $0.25$. However, in the present case the
coupling can be much smaller and $\ln\approx0.07$ suffices to obtain
the correct relic abundance. This will have very interesting
properties for their detection, as we will see in
Section\,\ref{sec:cross}.

\begin{figure}
  \begin{center}
    \epsfig{file=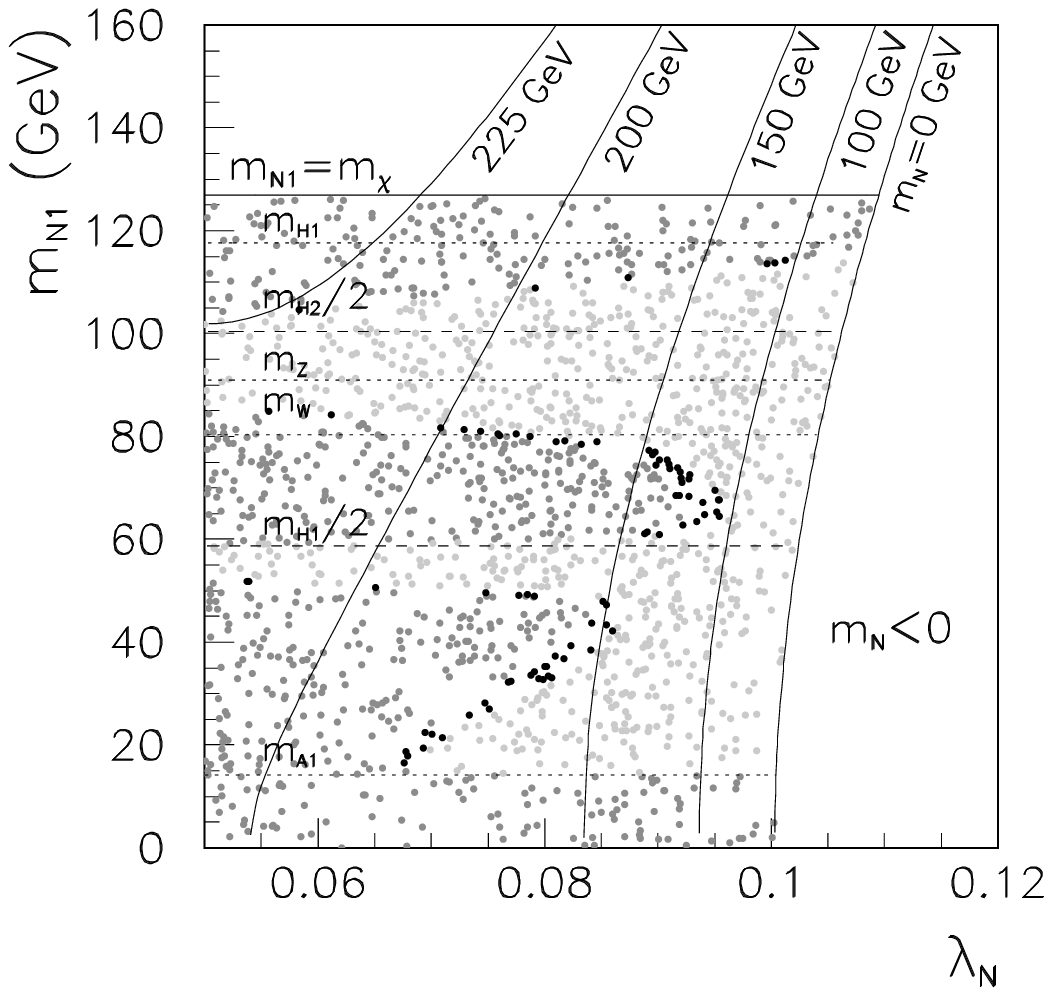,width=9.cm}\hspace*{-1.2cm}
  \end{center}
  \vspace*{-1cm}
  \captions{ The same as in Fig.\,\ref{fig:lnmn-c} but for case B1) of
    Table\,\ref{tab:examples-real} with $\aln=-500$~GeV. 
  } 
  \label{fig:lnmn-b}
\end{figure}

The effect of the relic density constraint on the ($\ln,\snmassr$)
plane is shown in Fig.\,\ref{fig:lnmn-b}, together with the 
contours for a fixed sneutrino soft mass, $\mn$.
As we observe, values of the sneutrino coupling constant
to the Higgs of order $\ln\approx0.06-0.1$ are sufficient to
have viable sneutrino dark matter, whereas values of the soft
sneutrino mass around $\mn\approx100-200$~GeV seem to be preferred. 
Once more, along the Higgs
resonances sneutrino annihilation is more effective and smaller
values of $\ln$ are needed to obtain the correct abundance. 
Remarkably, the region corresponding to viable very light sneutrinos 
(with masses in the $15-40$~GeV range)
can be obtained without requiring a
fine-tuning in the $\ln$ or $\mn$ parameters\footnote{Remember,
  however that there is a certain fine tuning in the NMSSM parameters
  in order to obtain very light viable pseudoscalars. Still, 
  very light right-handed 
  sneutrinos in this construction 
  are much more natural than very light neutralinos
  in the NMSSM, for which the extra condition of a resonant
  annihilation with the very light CP-even Higgs,
  $2\neutmass\approx\phmassl$ 
  is necessary \cite{Gunion:2005rw}.}.

\begin{figure}
  \hspace*{-0.5cm}
  \epsfig{file=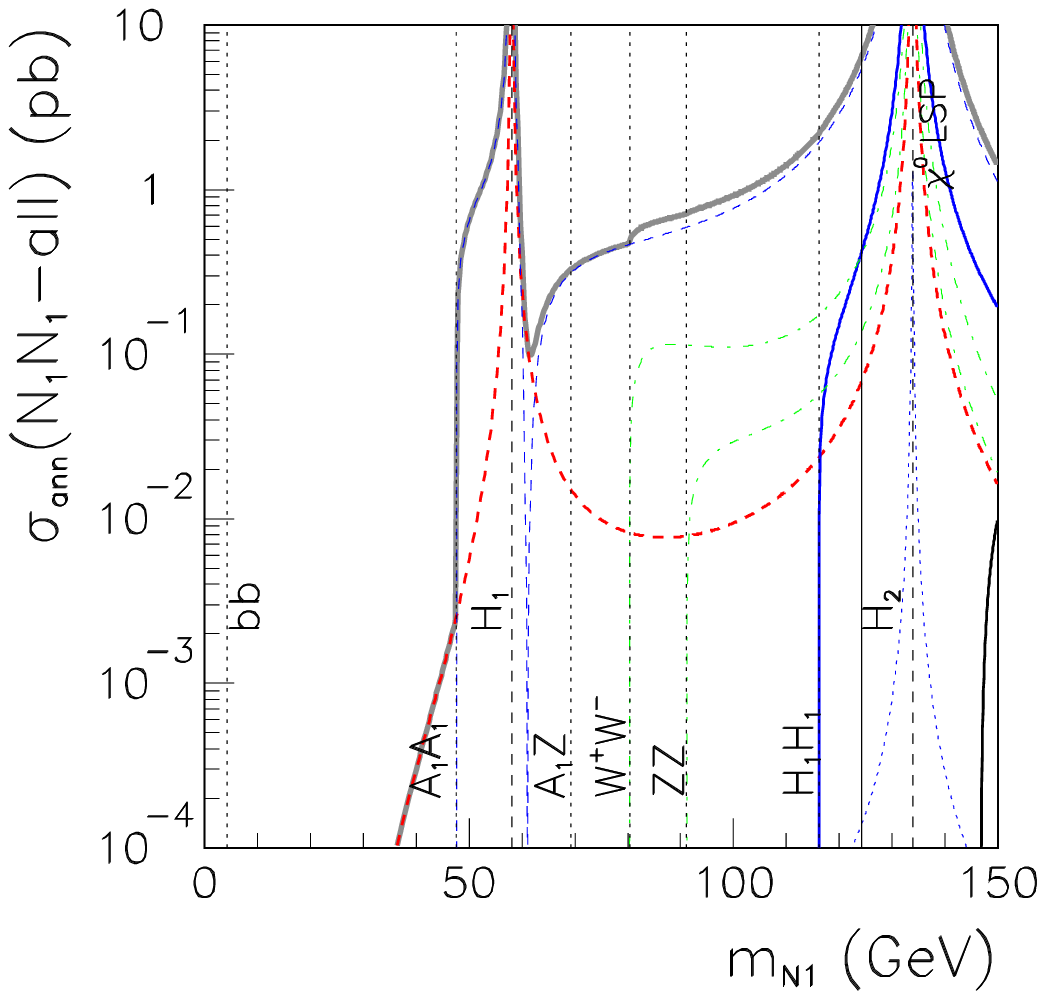,width=9.cm}\hspace*{-1.2cm}
  \epsfig{file=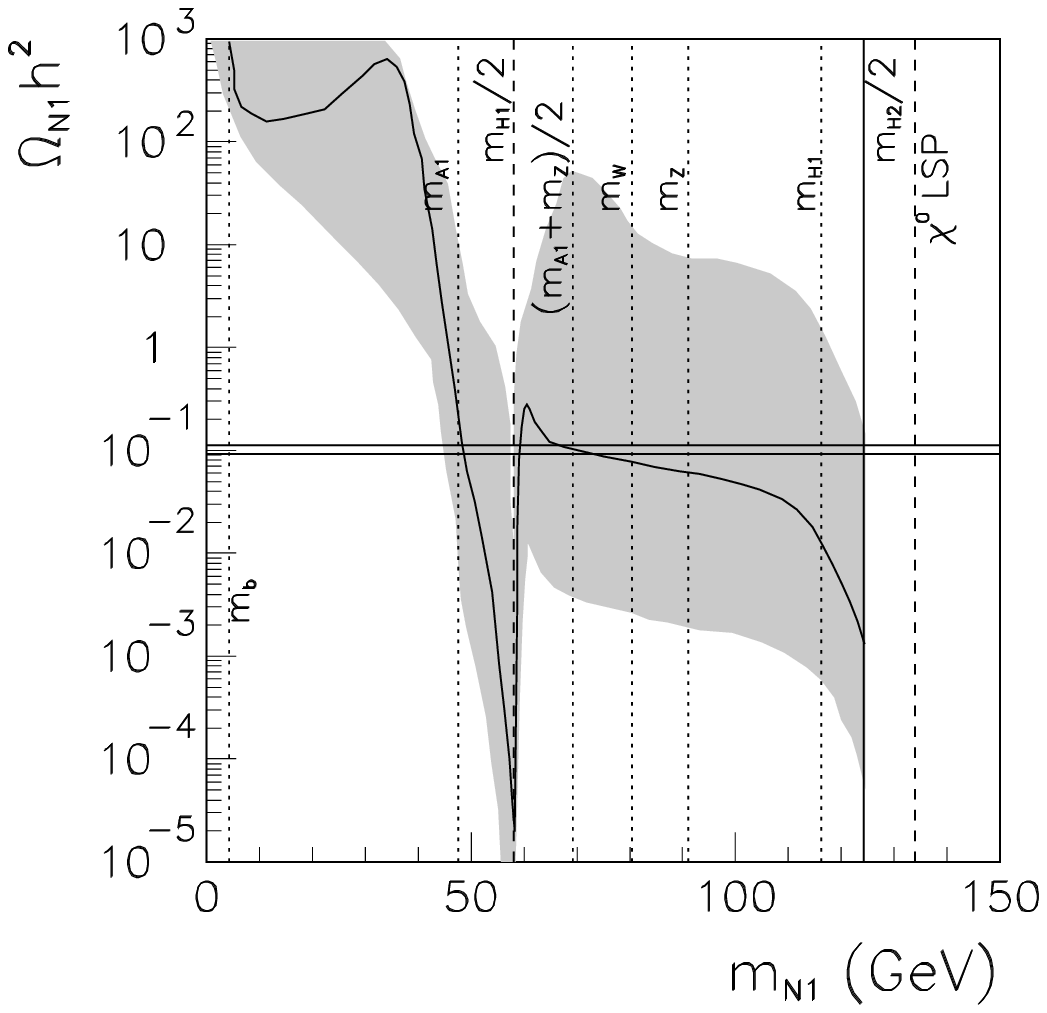,width=9.cm}
  \vspace*{-1cm}
  \captions{Left) The same as in Fig.\ref{fig:relic-b-real} but for
    case B2) of 
    Table\,\ref{tab:examples-real} with $\ln=0.14$. 
    Right) Corresponding relic density as a function of the sneutrino
    mass for  $\ln\in[0.05,0.35]$. The solid line indicates the result
    for $\ln=0.14$.}
  \label{fig:relic-a-real}
\end{figure}

The sneutrino properties are extremely sensitive
to the NMSSM Higgs sector. One should therefore expect that the above
analysis and preferred values for the sneutrino parameters could vary
for a different set of NMSSM inputs. 
In particular, given a different point of the
$(\l,\k)$ plane and due to the changes in the (scalar and pseudoscalar)
Higgs masses and mixings, 
the values of the sneutrino mass at 
which the various channels open and the conditions for
resonant annihilation would differ.

This is indeed the case. For instance, 
for the same choice of parameters as in
example the previous example, 
let us now choose $(\l,\,\k)=(0.3,0.2)$, corresponding to case B2) in
Table\,\ref{tab:examples-real}. This implies an
increase of the pseudoscalar mass (and a mild reduction of its singlet
component) for which $\phmassl\approx48$~GeV, as well as an increase
in the mass of the second lightest scalar Higgs, which remains being
singlet-like (the lightest scalar
Higgs is still MSSM-like and has a mass $\hmassl\approx116$~GeV). 
The rest of the spectrum is
detailed in Table\,\ref{tab:examples-real}.

As a consequence of these changes, 
annihilation into pseudoscalars is no longer
possible for very light sneutrinos, which can now only do it into
$b\bar b$ and their annihilation
cross-section becomes much smaller (for the same value of $\ln$) than
in example A). For sneutrinos with masses 
above $50$~GeV the pseudoscalar channel
is still the dominant one, with a modest contribution from
annihilation into a pair of gauge bosons when these are kinematically
allowed. 
This can be observed on the left-hand side of 
Fig.\,\ref{fig:relic-a-real}, where the corresponding predictions for
the sneutrino annihilation cross section are
shown for $\ln=0.14$.
Notice also that 
due to the change in the CP-even Higgs masses, the
position of the 
resonances change, in particular, the resonance due to the second
heaviest Higgs is not present (it would occur for $\snmassr\approx200$
~GeV, a mass for which the sneutrino is no longer the LSP).

As illustrated on the right-hand side of
Fig.\,\ref{fig:relic-a-real},
the correct relic abundance can still be obtained
for a given range of sneutrino masses. As in the previous examples, we
perform a scan in the allowed values of $\ln$ and $\mn$ for
$\aln=-500$~GeV. As explained in the previous section, due to the
increase in $\aln$, larger values of $\ln$ are viable.

\begin{figure}
  \begin{center}
    \epsfig{file=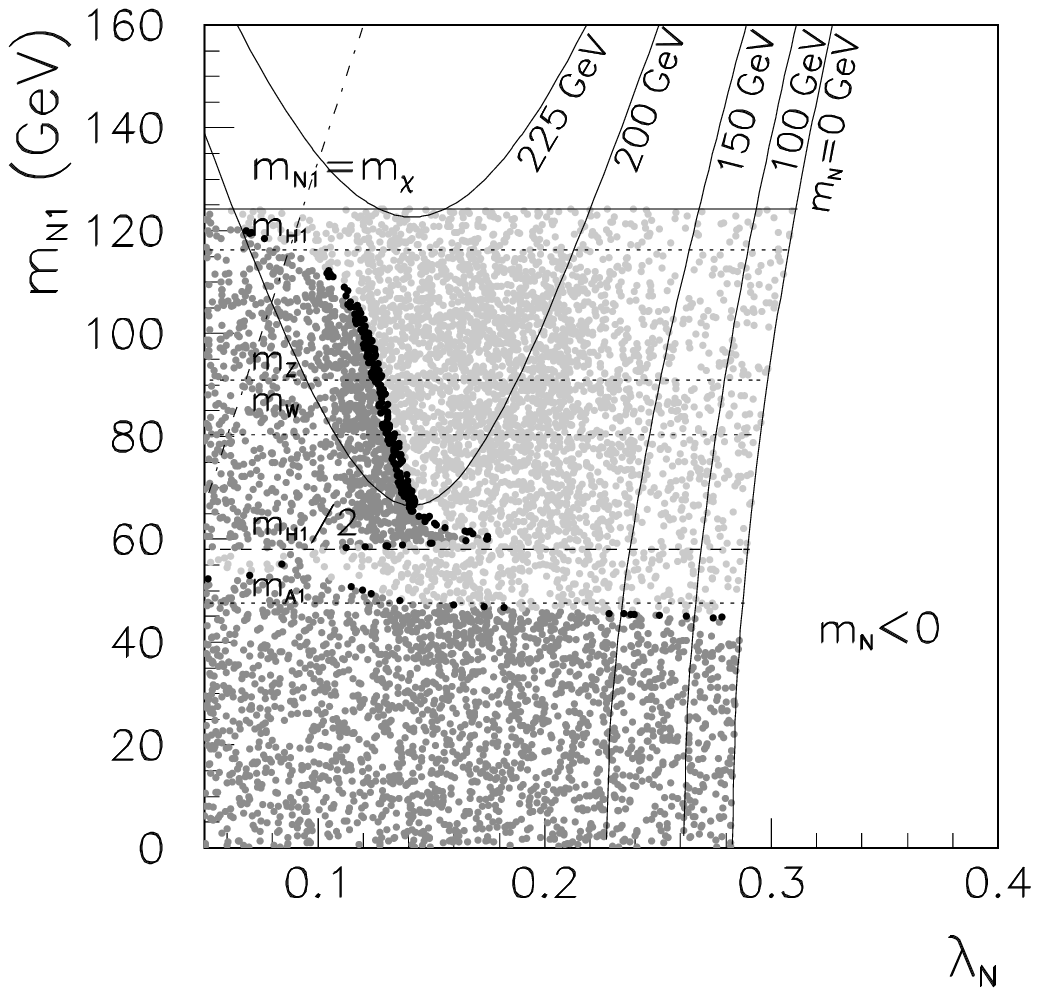,width=9.cm}\hspace*{-1.2cm}
  \end{center}
  \vspace*{-1cm}
  \captions{The same as in Fig.\,\ref{fig:lnmn-c} but for case B2) of
    Table\,\ref{tab:examples-real} with $\aln=-500$~GeV.
  } 
  \label{fig:lnmn-a}
\end{figure}

The effect of the relic density constraint on the ($\ln,\snmassr$)
plane is displayed in Fig.\,\ref{fig:lnmn-a}. The allowed areas for
sneutrinos heavier than $60$~GeV correspond to $\ln\approx0.12-0.14$
with a sneutrino soft mass of order $\mn\approx200-225$~GeV, whereas
lighter sneutrinos are only viable in the thin region corresponding
to resonant annihilation with the lightest scalar Higgs.

As we see from examples B1) and B2), through adequate variations in
the sneutrino parameters (\ref{rhnmssmparam}), it is possible to find
different  
points in the $(\l,\,\k)$ plane where the sneutrino has the correct
relic density. Variations on the $\l$ and $\k$ parameters have a large
impact on the neutralino and Higgs sector and it is therefore not
surprising that as a consequence 
the resulting sneutrino phenomenology is significantly
altered. However, the flexibility in the sneutrino properties
(especially the freedom in choosing an adequate value of its couplings
through $\ln$) makes
it possible to obtain viable sneutrino dark matter for a wide choice
of NMSSM parameters. 

\begin{figure}[!t]
  \hspace*{-0.5cm}
  \epsfig{file=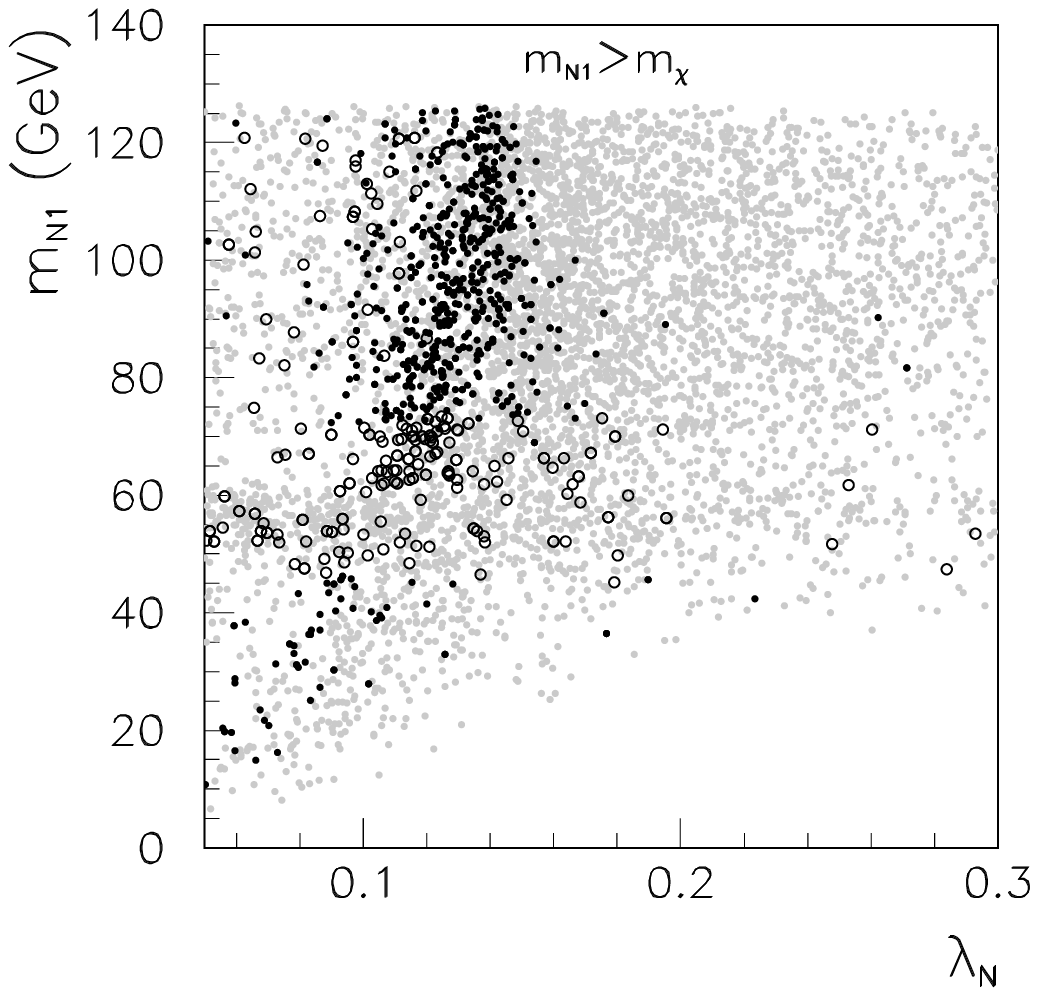,width=9.cm}\hspace*{-1.2cm}
  \epsfig{file=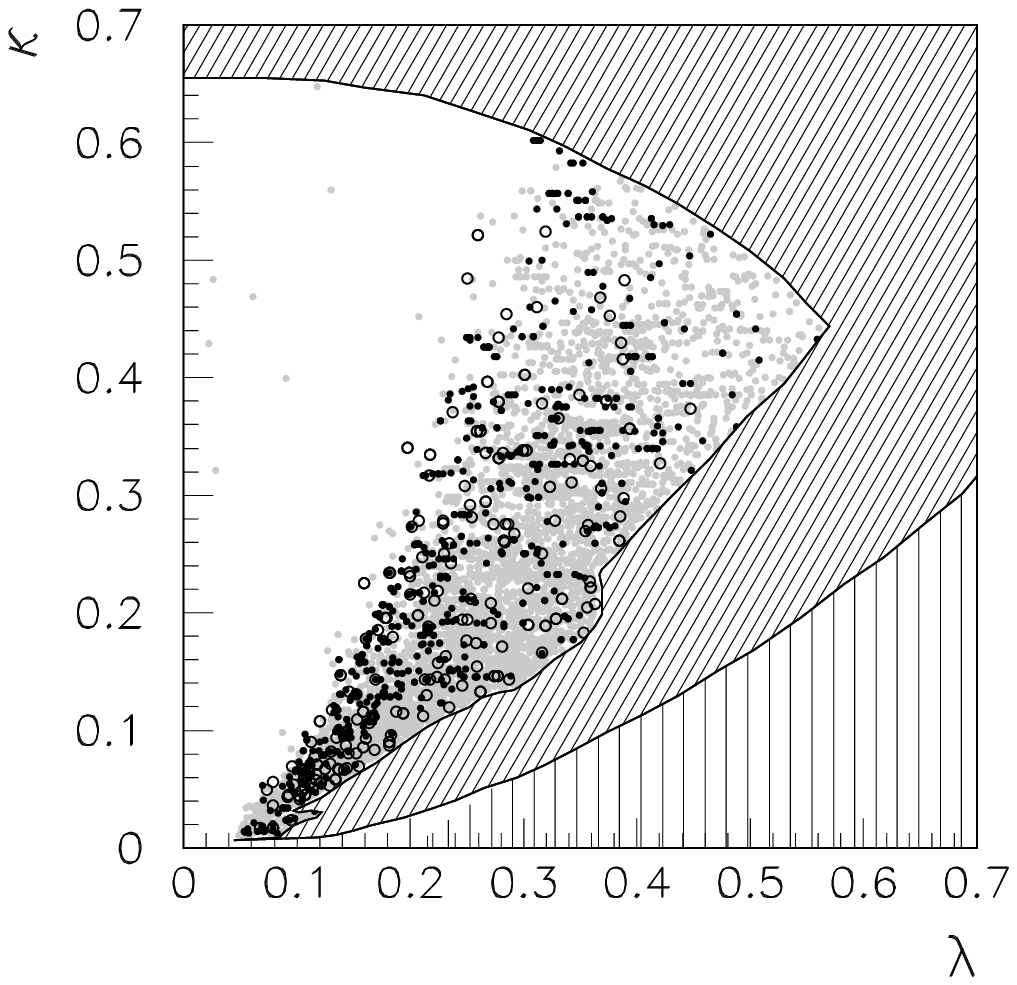,width=9.cm}
  \vspace*{-1cm}
  \captions{Left) 
    The same as in Fig.\,\ref{fig:lnmn-c} but for a scan in
    $(\l,\k)$ as described in the text. Empty circles correspond to
    points with the correct relic abundance due to resonant
    annihilation effects. Right) Regions with the correct sneutrino
    relic abundance (black dots) in the ($\l,\k$) plane. The oblique 
    ruled area
    is excluded due to the occurrence of Landau Poles or because of 
    experimental constraints in the Higgs sector, whereas the
    vertical ruled area is ruled out due to tachyons in the Higgs
    sector.
  } 
  \label{fig:lnmn-scan-ab}
\end{figure}

In order to illustrate this we have combined a scan in the $(\l,\k)$
plane, in the range $\l,\k=0.07-0.7$ with a scan in the sneutrino
parameters taking $\mn=0-250$~GeV and $\ln=0.05-0.3$. 
We have fixed $\aln=-500$~GeV and 
we have taken otherwise the same NMSSM parameters as in example B1)
and B2) of Table\,\ref{tab:examples-real}.
On the left-hand side of 
Fig.\,\ref{fig:lnmn-scan-ab} we represent the resulting sneutrino
mass as a function of the coupling $\ln$. Black dots represent those
with the correct relic abundance and light points correspond to those
where the sneutrino relic density is smaller than the WMAP results
(points with a relic density exceeding the WMAP constraint are not
shown). 
As we can see, the correct relic abundance can be obtained within a
range of $\ln\approx0.06-0.15$ and for a wide range of sneutrino
masses. 
On the right-hand side of Fig.\,\ref{fig:lnmn-scan-ab} the regions
with the correct relic abundance are displayed on the $(\l,\k)$
plane. We clearly see that the sneutrino can be a viable candidate
(for an adequate choice of the $\ln$ and $\mn$ parameters) in
virtually any point in this plane. Notice that the absence of points
in the left-hand side of the plot is due to the fact that the
imaginary component of the sneutrino becomes the lightest one, as
explained in Section\,\ref{sec:model}. This is in principle
no problem and in fact the imaginary sneutrino would have almost the
same properties as the real component\footnote{The only difference
  arises in the expression for its
  annihilation into right-handed neutrinos, 
  as explained in Appendix\,\ref{sec:wtilde}.}, 
however, as we
mentioned earlier,
for simplicity we have not considered such a possibility here.

\subsubsection{Predominant annihilation into 
  right-handed neutrinos or gauge bosons} 
\label{sec:rhn}

\begin{figure}[!t]
  \hspace*{-0.5cm}
  \epsfig{file=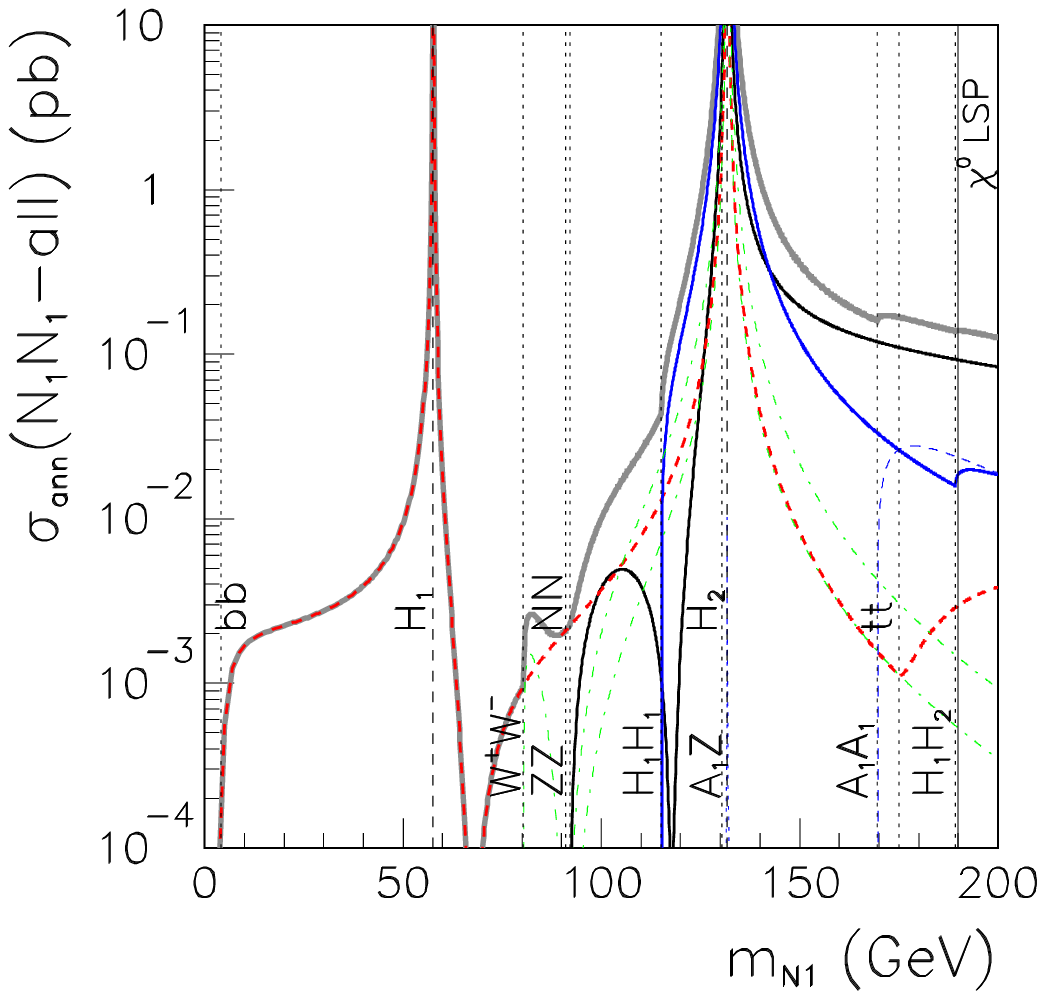,width=9.cm}\hspace*{-1.2cm}
  \epsfig{file=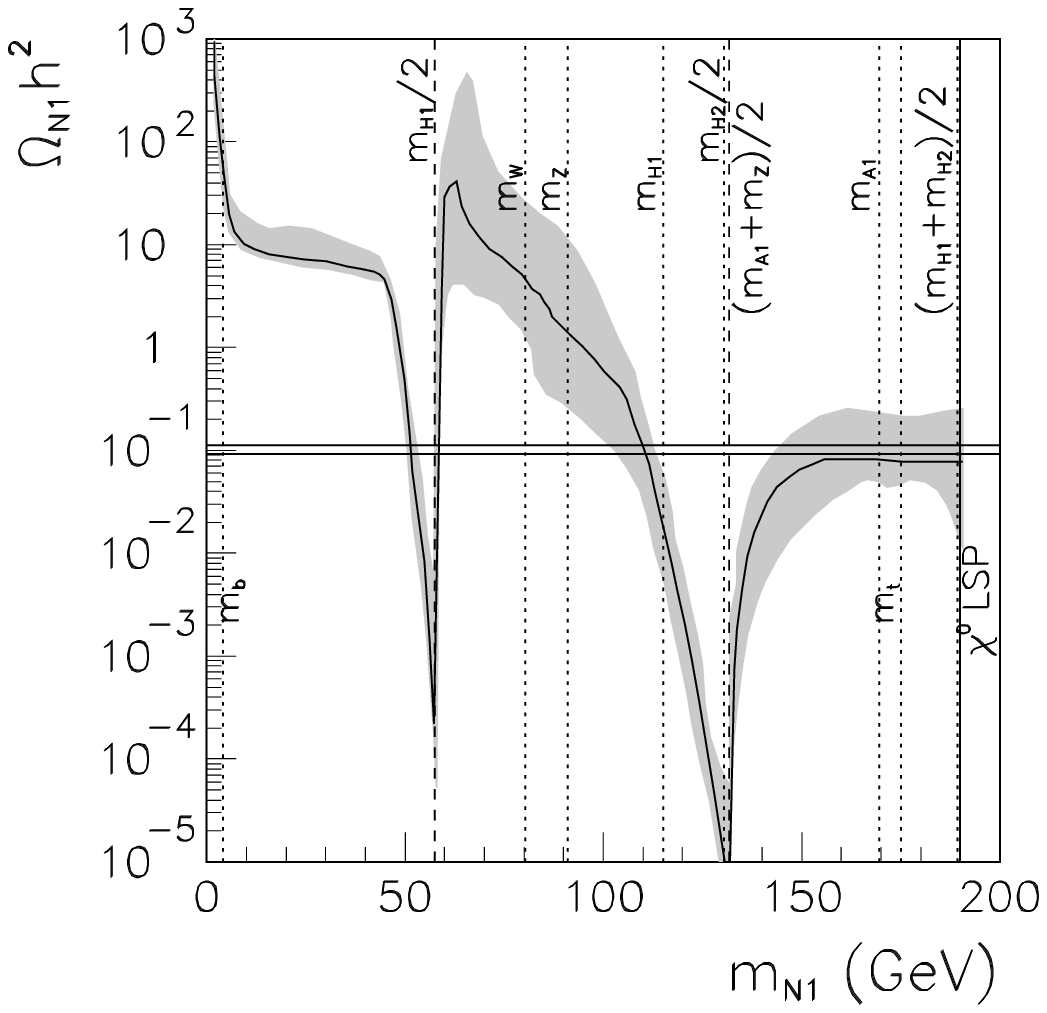,width=9.cm}
  \captions{Left) The same as in Fig.\ref{fig:relic-b-real} but for
    case C) of 
    Table\,\ref{tab:examples-real} with  
    $\ln=0.15$ and $\aln=-250$~GeV. 
    Right) Corresponding relic density as a function of the sneutrino mass for
    for $\ln\in[0.1,0.3]$. The solid line indicates the result
    for $\ln=0.15$.}
  \label{fig:relic-g-real}
\end{figure}

Let us finally illustrate with another example the relevance of 
other annihilation channels, in particular the
annihilation into a pair of right-handed neutrinos. Our set of input
parameters is chosen according to example C) in
Table\,\ref{tab:examples-real}, where we also specify the resulting
spectrum.

We fix $\aln=-250$~GeV and plot the resulting annihilation
cross section for each individual channel on the left-hand side of
Fig.\,\ref{fig:relic-g-real}. Once more,
the 
features of the Higgs sector clearly affect the calculation of the
relic density for the sneutrino. In this example the resonant
annihilation through the exchange of the two lightest Higgses is
visible at $\snmassr\approx58$~GeV and $130$~GeV. 
The lightest Higgs in this example is doublet-like, whereas
the intermediate Higgs, $H_2$, and the lightest pseudoscalar are
almost pure singlets. Sneutrino annihilation into $b\bar b$ is
the only possibility for sneutrino masses below $\mw$, but this
channel is only effective when resonant annihilation through the
lightest Higgs takes place. Annihilation into $ZZ$ or $W^+W^-$ is the 
dominant channel below the Higgs production threshold. Interestingly,
after the second Higgs resonance, annihilation into a pair of
right-handed neutrinos becomes the leading contribution for 
$\snmassr\gsim150$~GeV, although the contributions of the
$\higgsl\higgsl$ and $\phiggsl\phiggsl$ channels are not negligible.

The resulting relic density is shown on the right-hand side of
Fig.\,\ref{fig:relic-g-real} for a scan with  $\ln\in[0.1,0.3]$ and
$\mn\in[0,200]$~GeV. The solid line indicates the result for
$\ln=0.15$. 
We find that in this example the 
correct relic density can be obtained through annihilation into $ZZ$
or $W^+W^-$ (for sneutrinos with a mass around $100$~GeV and
$\ln\approx0.2$) or annihilation into $NN$ and $\higgsl\higgsl$ for
sneutrinos above $\snmassr\gsim 140$~GeV, in which case
$\ln\approx0.1$ is necessary.

\begin{figure}
  \begin{center}
    \epsfig{file=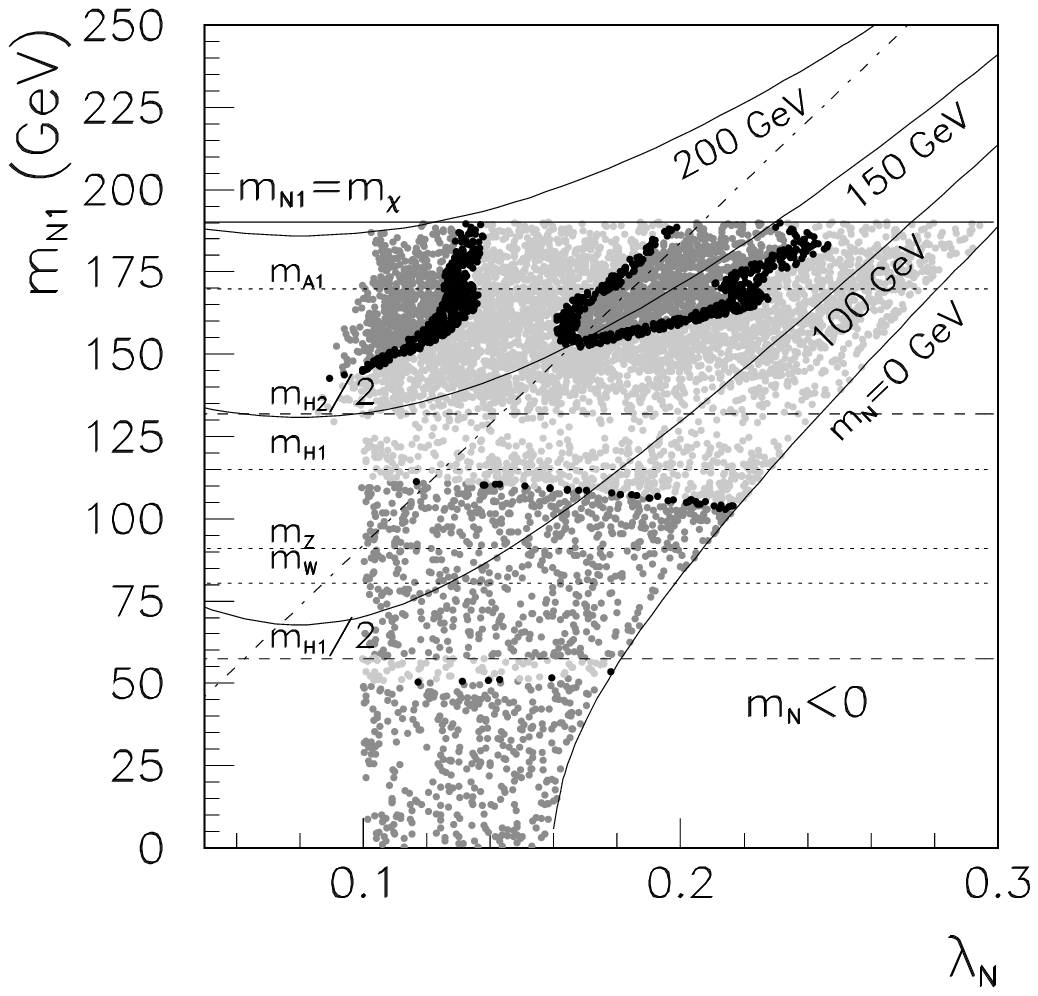,width=9.cm}\hspace*{-1.2cm}
  \end{center}
  \vspace*{-1cm}
  \captions{The same as in Fig.\,\ref{fig:lnmn-c} but for case C) of
    Table\,\ref{tab:examples-real} with $\aln=-250$~GeV.
  } 
  \label{fig:lnmn-g}
\end{figure}

The corresponding $(\ln,\snmassr)$ plane is shown in
Fig.\,\ref{fig:lnmn-g}, indicating the regions consistent with the
WMAP constraint on the relic density. Consistently with the discussion
above, in the region
above the dot-dashed line (which corresponds to the area in which the
sneutrino mass is larger than the right-handed neutrino mass, 
$\snmassr>\rhnmass$) there is a wide region with $\ln\approx0.1$ for
which the WMAP constraint is satisfied. This also corresponds to
values of the soft sneutrino masses in the range
$\mn=150-200$~GeV. Notice also a second region for larger values of
$\ln$ of order $0.15-0.18$.
Below the dot-dashed line, where $\snmassr<\rhnmass$ the WMAP region
corresponds to values of the sneutrino-Higgs coupling in the range
$\ln=0.15-0.24$, and smaller values of the sneutrino soft mass,
$\mn=100-150$~GeV. 
Light sneutrinos in this example have a too large relic abundance and
it is only along the regions with resonant annihilation that the
WMAP constraint is satisfied.

\subsection{Overview}

Throughout this section we have seen how the right-handed sneutrino
can reproduce the correct dark matter abundance under different
conditions and for various points in the NMSSM parameter space. In
this sense, the flexibility to choose adequate values of the
sneutrino parameters $\ln,\,\mn$ and $\aln$ was very useful.
Notice in any case that the preferred values of these parameters are
quite natural. The sneutrino-Higgs coupling $\ln$ can be chosen in the
range $\ln=0.05-0.4$ and the sneutrino soft mass could be of order of
$100-250$~GeV, being of the same order than the slepton soft masses in
our examples.  
As a result, 
sneutrinos in the mass range $\snmassr\approx5-200$~GeV are found that
can reproduce the WMAP constraint on the relic abundance.

It should also be noted that obtaining heavier right-handed sneutrinos
is in principle possible, especially since more channels would be
kinematically  allowed and the relic abundance could be obtained for
moderate values of $\ln$. Notice that, in order to do so, the
mass of the neutralino (and that of the rest of the spectrum) also
must be increased so that the right-handed sneutrino is the
LSP. This can be done through an increase in the gaugino and scalar
mass parameters as well as in the $\mu$ term. However, this entails a
further reduction of the supersymmetric contribution to the muon
anomalous magnetic moment, thereby increasing the tension with its
experimental bound.

\section{Direct detection}
\label{sec:cross}

The direct detection of sneutrinos could take place through their
elastic scattering with nuclei inside a dark matter detector. The
interaction occurs in the non-relativistic regime (given that the
velocity of sneutrinos in the dark matter halo is very small) and
therefore one can easily find a description in terms of an effective
Lagrangian. In our case, there is only one diagram contributing (at
tree level) to this process, namely, the $t$-channel exchange of
neutral Higgses shown in Fig.\,\ref{cross-diag} (the exchange of a $Z$
boson is largely suppressed by the neutrino Yukawa squared and
therefore is completely negligible). The effective
Lagrangian describing the four-field interaction only contains a
scalar coupling which reads
\begin{equation} 
  {\cal L}_{eff}\supset \alpha_{q_i} \tilde N\tilde N
  \bar q_i q_i
\end{equation}
with
\begin{equation}
  \alpha_{q_i}\equiv\sum_{j=1}^3\frac{\chsnsn Y_{q_i}}{m_{H_j^o}^2}
  \label{alphaq}
\end{equation}
where $\chsnsn$ is defined in Appendix \ref{sec:feynman}, $Y_{q_i}$ is the
corresponding quark Yukawa coupling and $i$ labels up-type quarks
($i=1$) and down-type quarks ($i=2$). Notice that the effective
Lagrangian contains no axial-vector coupling since the sneutrino is a
scalar field, therefore implying a vanishing contribution to the
spin-dependent cross section. 
The total spin-independent sneutrino-proton 
scattering cross section yields
\begin{equation}
  \crosssec = \frac1\pi\frac{m_p^2}{(m_p+\snmassr)^2}\,f_p^2\,,
\end{equation}
where $m_p$ is the proton mass and 
\begin{equation}
  \frac{f_p}{m_p}=
  \sum_{q_i=u,d,s}f_{Tq_i}^p\frac{\alpha_{q_i}}{m_{q_i}}+ 
  \frac{2}{27}\ f_{TG}^p\sum_{q_i=c,b,t}\frac{\alpha_{q_i}}{m_{q_i}}\ .
\end{equation}
The quantities $f_{Tq_i}^p$ and $f_{TG}^p$ are the hadronic matrix
elements which parametrize the
quark content of the proton. 
They are subject to considerable uncertainties
\cite{efo-uncertainties,bottino-uncertainties,eoss-uncertainties}
which induce a significant correction to the theoretical predictions
for $\crosssec$.  In our analysis we will consider the most recent
values for these quantities, as explained in 
\cite{eos08}.

\begin{figure}[!t]
  \begin{center}
    \begin{picture}(150,110)(0,-50)
      \DashArrowLine(10,50)(60,20){5}
      \DashArrowLine(60,20)(110,50){5}
      \DashLine(60,20)(60,-20){5}
      \ArrowLine(10,-50)(60,-20)
      \ArrowLine(60,-20)(110,-50)
      \Text(110,35)[c]{$\tilde{N}$}
      \Text(10,35)[c]{$\tilde{N}$}
      \Text(110,-35)[c]{$\bar{q}$}
      \Text(10,-35)[c]{$q$}
      \Text(70,0)[c]{$H_i^0$}
    \end{picture}
    \end{center}
  \captions{Diagram describing the elastic interaction of sneutrinos
    with quarks. }
  \label{cross-diag}
\end{figure}

It is obvious from the previous formulae that the sneutrino detection
cross section is extremely dependent on the features of the Higgs
sector of the model. In particular, $\crosssec$ becomes larger when
the sneutrino-sneutrino-Higgs coupling increases (which according to
its expression in Appendix\,\ref{sec:feynman} can be achieved by 
enhancing $\ln$ or with large values for $|\aln|$). Moreover, from
Eq.\,(\ref{alphaq}) we also see that larger
values $\crosssec$ can be obtained in those regions of the
parameter space where the mass of the lightest Higgs becomes smaller.

We will now calculate the theoretical predictions for the sneutrino
scattering cross section for the various cases studied in
Section\,\ref{sec:relic}. 
Let us begin by addressing case A) in Table\,\ref{tab:examples-real},
for which we have seen in Section\,\ref{sec:higgs} that the
$\higgsl\higgsl$ channel
is the dominant contribution to the annihilation
cross section for moderate values of $\ln$. Choosing $\aln=-250$~GeV
and performing the same scan in $\ln$ and $\mn$ as in
Fig.\,\ref{fig:relic-c-real} we have calculated the resulting
sneutrino scattering cross section off quarks in those cases where
the sneutrino abundance is consistent with the WMAP constraint or
smaller (in which case the sneutrino could be a subdominant dark
matter component).

If the sneutrino only contributes to a fraction of the total dark
matter density, one should expect that it is present in the dark
matter halo in the same proportion as in the Universe, 
contributing only to a fraction of
the local dark matter density. This implies, of course, a reduction of
the detection rate in direct detection experiments. In order to take
this effect into account, it is customary to define the 
sneutrino fractional density $\xi={\rm min}[1,\snrelic/0.1037]$ 
\cite{rescaling} and plot the modulated elastic scattering cross
section, $\xi\crosssec$.

\begin{figure}[!t]
  \hspace*{-0.5cm}
  \epsfig{file=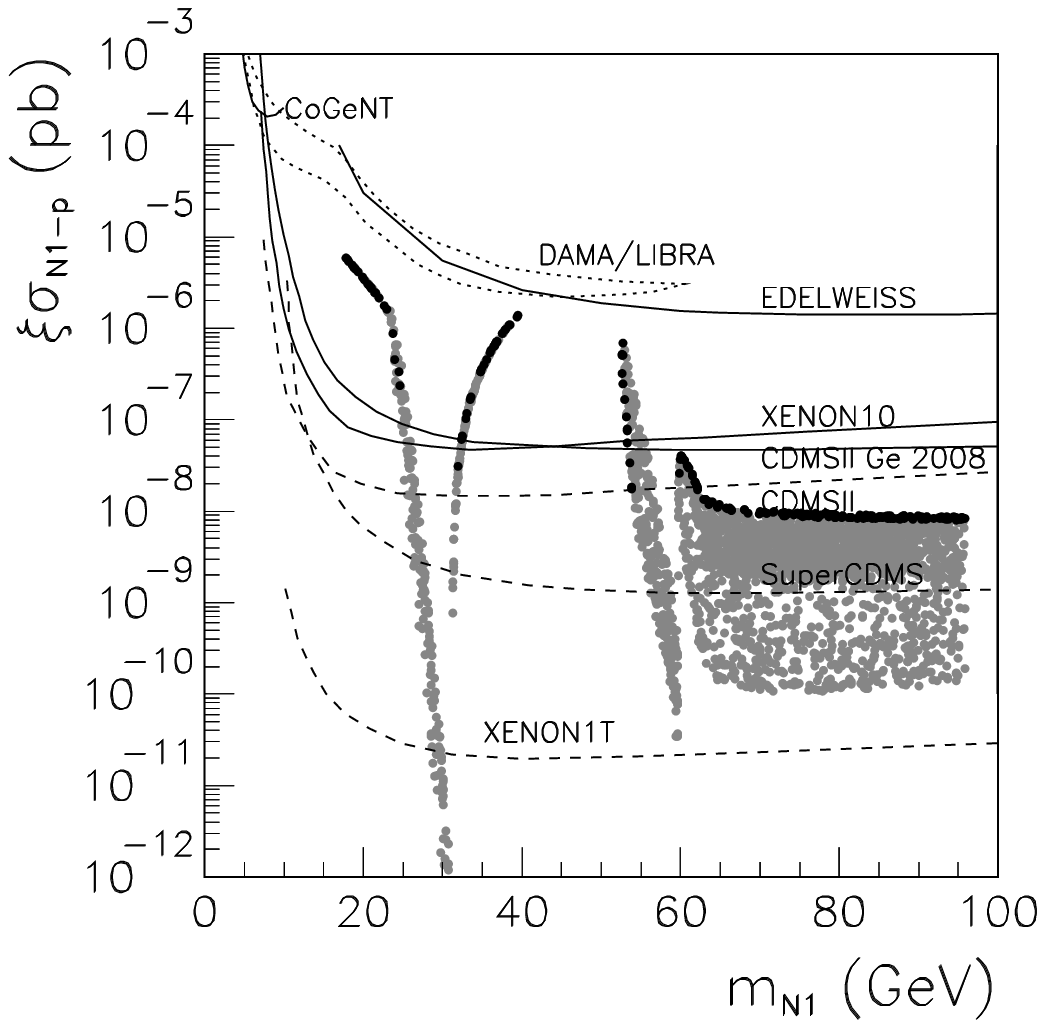,width=9.cm}\hspace*{-1.2cm}
  \epsfig{file=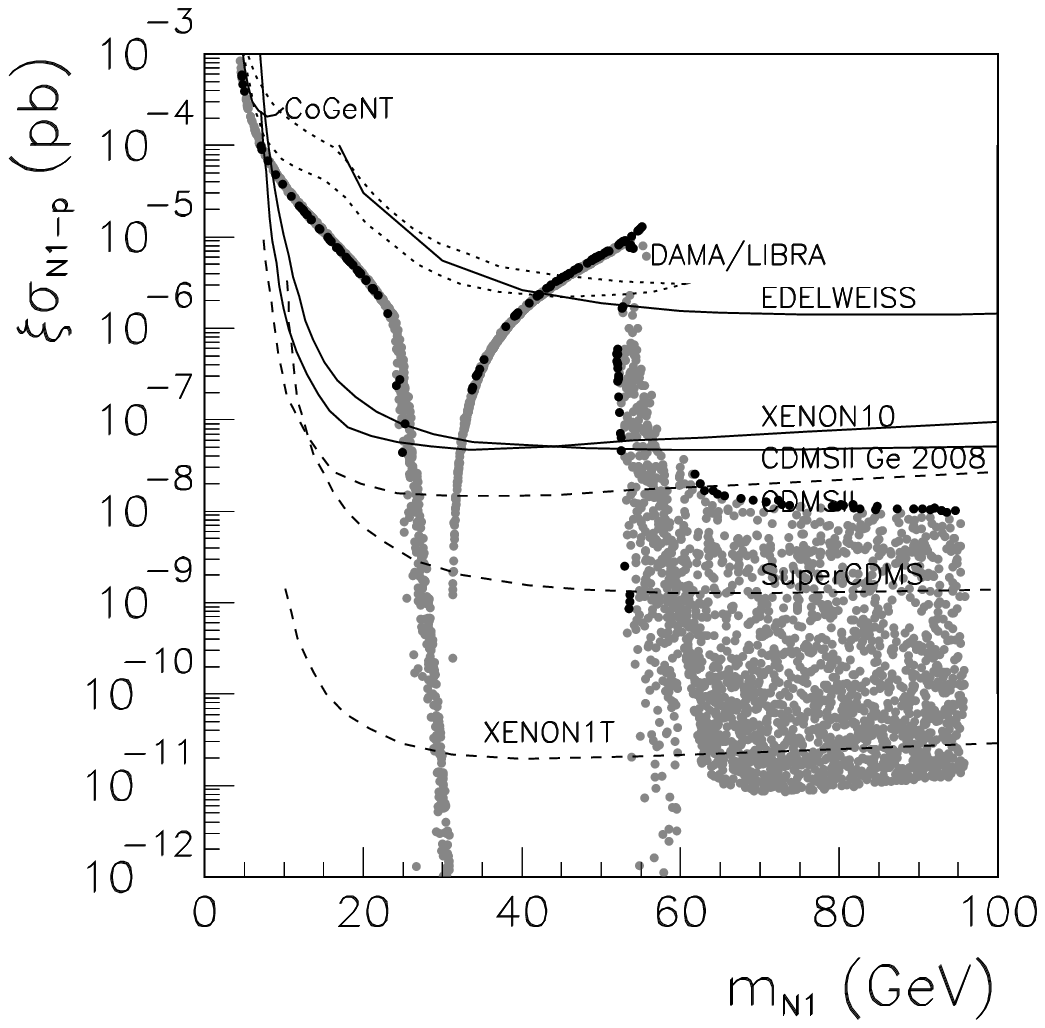,width=9.cm}
  \captions{Theoretical predictions for $\xi\crosssec$, as a function
    of the sneutrino mass for example A) 
    of Table\,\ref{tab:examples-real}. All the points
    represented fulfil all the experimental constraints. Black dots
    correspond to those which reproduce
    the WMAP result for the dark matter relic density whereas grey
    ones represent those with $\snrelic<0.1037$. The
    sensitivities of present and projected experiments are represented
    by means of solid and dashed lines, respectively, in the case of an
    isothermal spherical halo. The large (small) area bounded by
    dotted lines is consistent with the interpretation of DAMA
    experiment in terms of a WIMP. }
  \label{fig:sncross-c}
\end{figure}

We follow this approach and show on the left-hand side of 
Fig.\,\ref{fig:sncross-c} the theoretical predictions for the
spin-independent contribution to 
$\xi\crosssec$ as a function of the sneutrino mass for the example
studied in Section\,\ref{sec:higgs}. This is, case A) with
$\aln=-250$~GeV, 
$\ln\in[0.05,0.1]$ and $\mn\in[0,250]$~GeV.
Black dots correspond to points with a relic density consistent with
the WMAP results, whereas grey dots stand for those with
$\snrelic\le0.1037$.

For light sneutrinos, for which annihilation into Higgses is not
kinematically allowed, the value of $\ln$ in order to have the
correct relic density is large (see
Fig.\,\ref{fig:lnmn-c}). Consequently, the resulting scattering cross
section turns out to be sizable and easily exceeds the sensitivities
of present detectors. On the other hand, when annihilation into
CP-even Higgses is possible, the necessary value of $\ln$ is smaller,
of order $0.1$, resulting in $\xi\crosssec\lsim10^{-8}$~pb. These
results are not yet excluded by present searches, but interestingly,
could be within the reach of future experiments such as SuperCDMS.
Notice also that along the Higgs resonances the sneutrino scattering
cross section decreases considerably since the necessary value of the
$\ln$ 
coupling to obtain the correct relic density becomes much smaller.

On the right-hand side of Fig.\,\ref{fig:sncross-c} we show the
results for case A) but for $\aln=-500$~GeV, which corresponds to the
case analysed in Section\,\ref{sec:bbar}. We remind the reader that in
this example a larger $\ln$ was allowed and we perform a scan in
$\ln\in[0.05,0.35]$ with $\mn\in[0,250]$~GeV. This made it possible to
obtain very light sneutrinos with the correct relic abundance. The
large value of $\ln$ implies that the detection cross section for
these light sneutrinos is also very large. This is clearly evidenced
in the plot, where we find that sneutrinos lighter than $20$~GeV have
a cross section which can be as large as $\xi\crosssec\sim
10^{-4}$~pb.

This situation is very similar to what happens with very light 
neutralinos in the MSSM. 
Very light
neutralinos with masses $\neutmass\gsim7$~GeV 
can be obtained \cite{Bottino:2002ry,Bottino:2003cz}
if the GUT relation for gaugino masses is abandoned.
Although it was argued that these neutralinos could account for the
DAMA/LIBRA annual modulation signal without contradicting the null
results from CDMS and XENON10 \cite{Bottino:2008mf},
this interpretation is now more constrained by the recent results
published by the MAJORANA collaboration using 
results from low-threshold experiments \cite{Aalseth:2008rx}.

\begin{figure}
  \begin{center}
    \epsfig{file=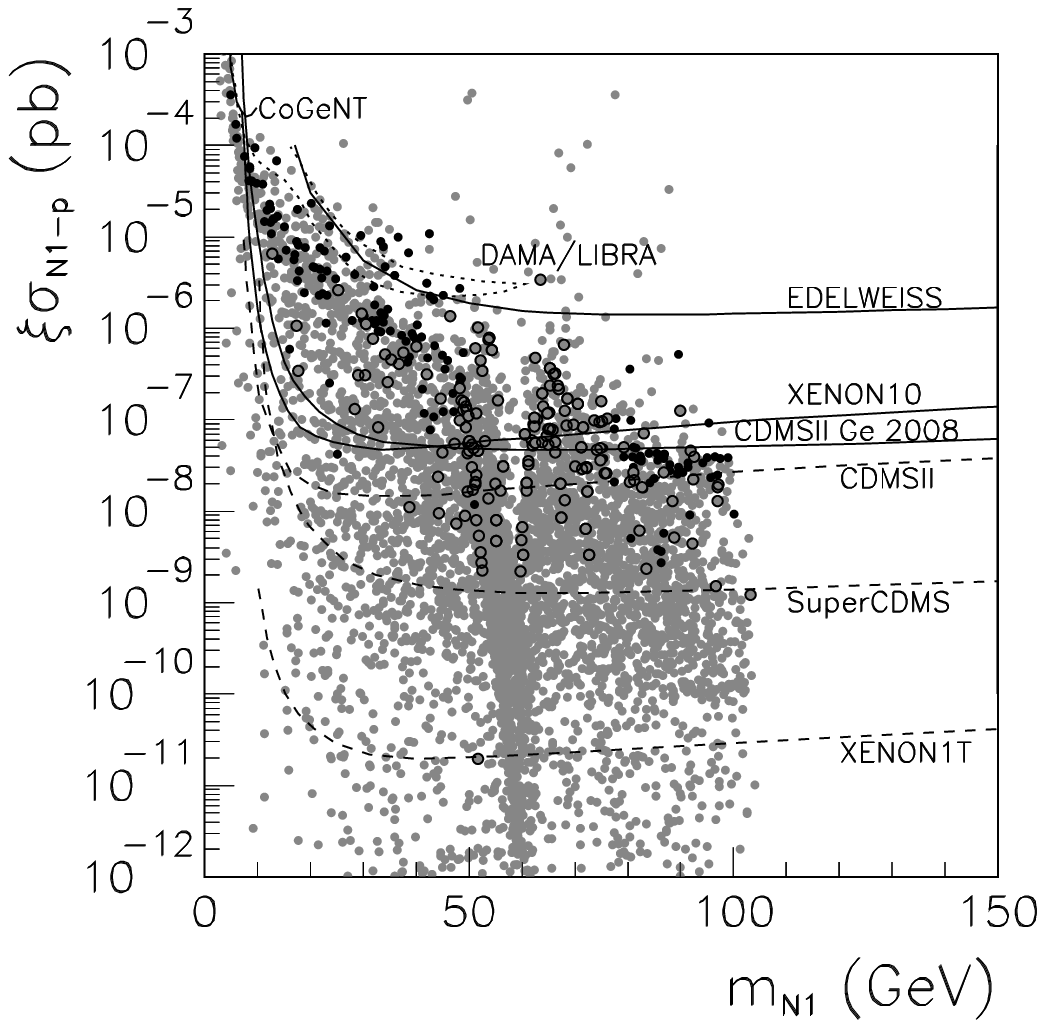,width=9.cm}
  \end{center}
  \captions{The same as in Fig.\,\ref{fig:sncross-c} but for 
    a scan in the $(\l,\k)$ plane. Empty circles correspond to
    points with the correct relic abundance due to resonant
    annihilation effects.
  }
  \label{fig:sncross-scan-c}
\end{figure}

Finally, in order to explore more general variations in the NMSSM
parameters, we have scanned in the allowed $(\l,\k)$ plane, keeping
$\aln=-500$~GeV and the same scan in the $\ln$ and $\mn$ parameters as
in the previous case. As we see from the resulting cross section,
depicted in Fig.\,\ref{fig:sncross-scan-c} where we see again that the
cross section can be rather large (especially in those cases with
light CP-even Higgs bosons). The region with very light sneutrinos,
for with the predicted $\crosssec$ is sizable, 
could be within the reach of low-threshold experiments.

\begin{figure}
  \hspace*{-0.5cm}
  \epsfig{file=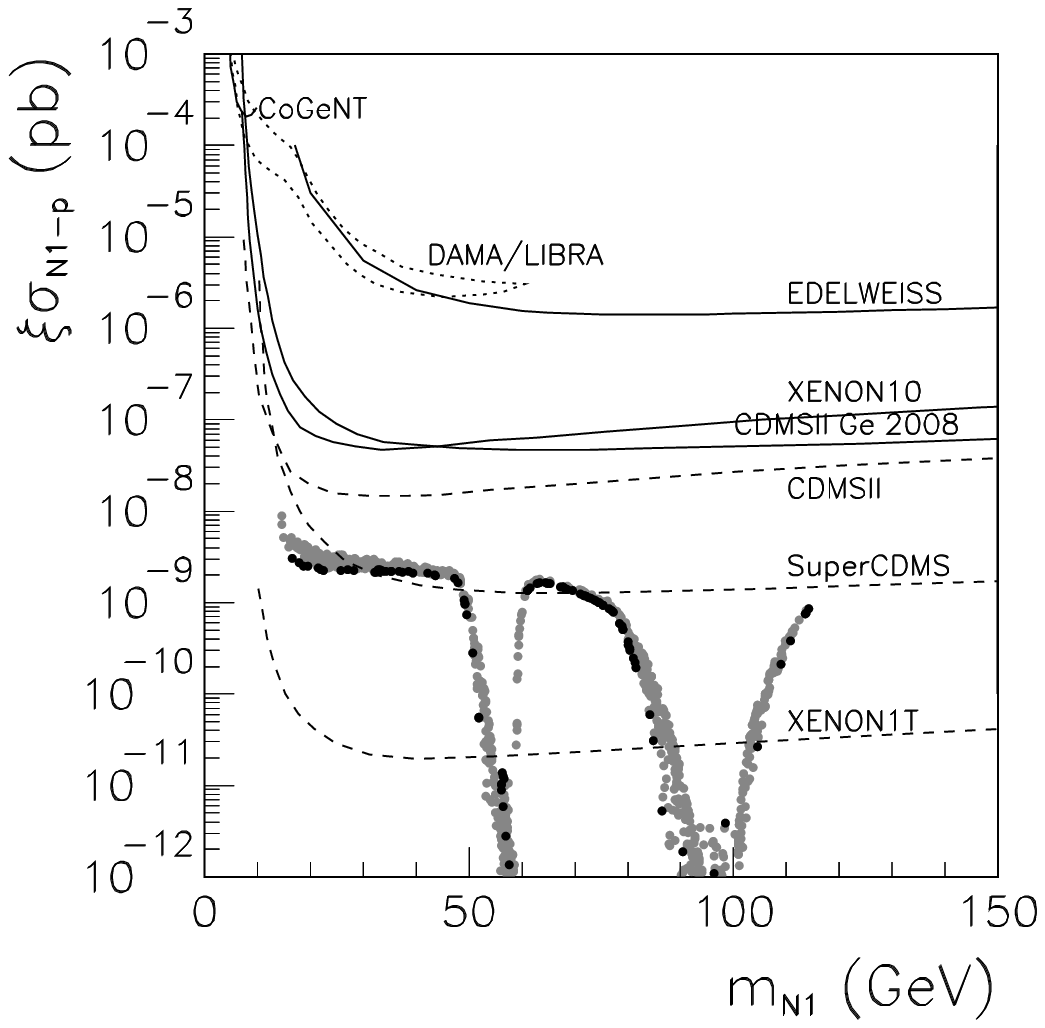,width=9.cm}\hspace*{-1.2cm}
  \epsfig{file=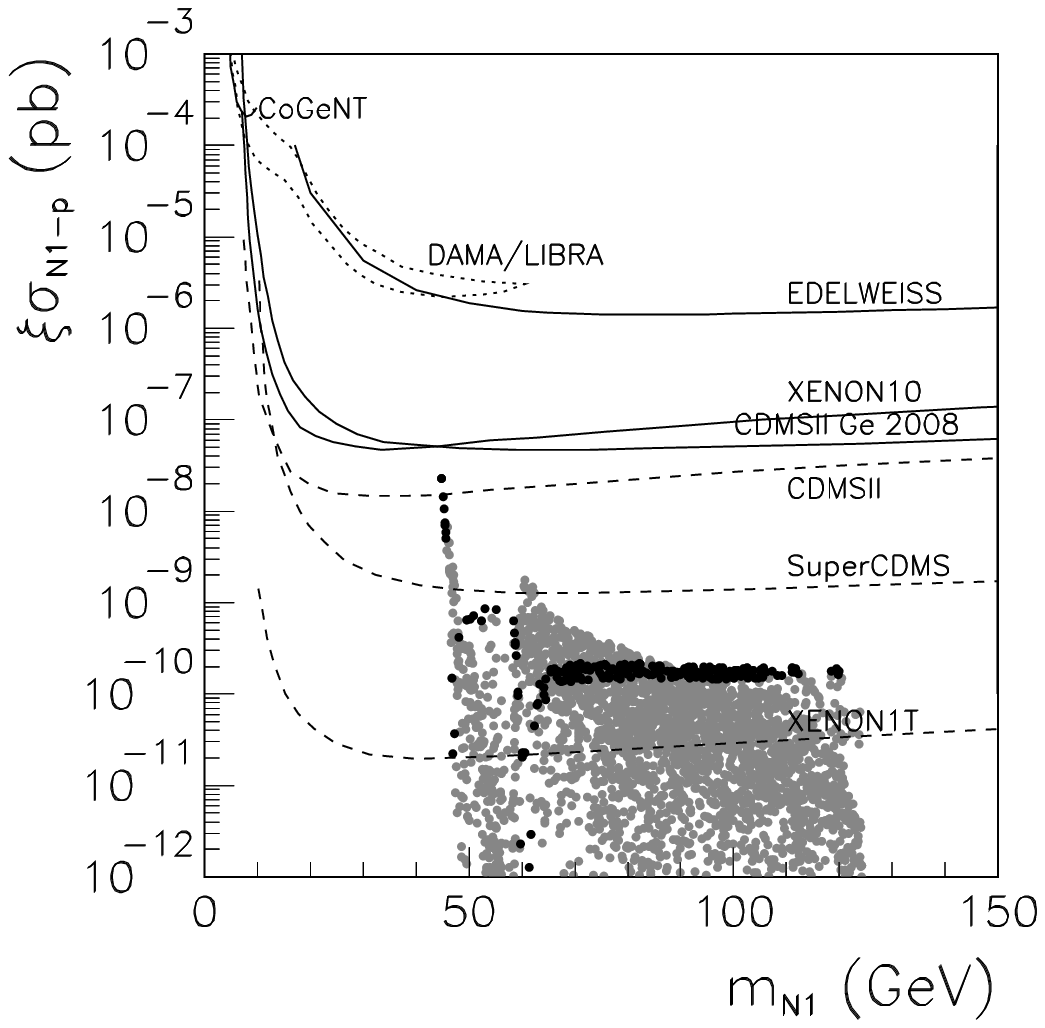,width=9.cm}
  \captions{The same as in Fig.\,\ref{fig:sncross-c} but for 
    examples B1) 
    and B2) of Table\,\ref{tab:examples-real}. 
  }
  \label{fig:sncross-a}
\end{figure}

Next we address the examples B1) and B2) studied in
Section\,\ref{sec:pseudoscalar}, and for which annihilation into
pseudoscalar Higgses was predominant. The results for the sneutrino
scattering cross section are illustrated in
Fig.\,\ref{fig:sncross-a}.

We find a very different behaviour from the previous examples. First,
since light sneutrinos are in this case obtained thanks to their
annihilation into very light pseudoscalars, the value of $\ln$ at
which the WMAP result is reproduced is considerably smaller. Second,
the mass of the CP-even Higgs bosons is larger in this example than in
the previous one. As a
consequence, these very light sneutrinos have a smaller scattering
cross section and can evade the experimental constraints from XENON10
and CDMS.

For example, on the left-hand side of Fig.\,\ref{fig:sncross-a} (which
corresponds to the case analysed in Fig.\,\ref{fig:relic-b-real})
sneutrinos as light as $\snmassr\sim15$~GeV are obtained with the
correct relic abundance and with a scattering cross section of order
$\xi\crosssec\approx2\times10^{-9}$~pb. This is a very interesting
situation, since as we will explain later in more detail, it is a
completely 
different behaviour than the one observed for neutralinos in the
MSSM.

On the right hand-side of Fig.\,\ref{fig:sncross-a} we show the
results corresponding to the case analysed in
Fig.\,\ref{fig:relic-a-real}). In this example the CP-even Higgs is 
heavier than in the previous examples and as a consequence we observe
a decrease in the scattering cross section. In this case
$\xi\crosssec\lsim10^{-9}$~pb and Ton-scale detectors, such as the
projected XENON1T would be necessary to explore them.

\begin{figure}
  \begin{center}
    \epsfig{file=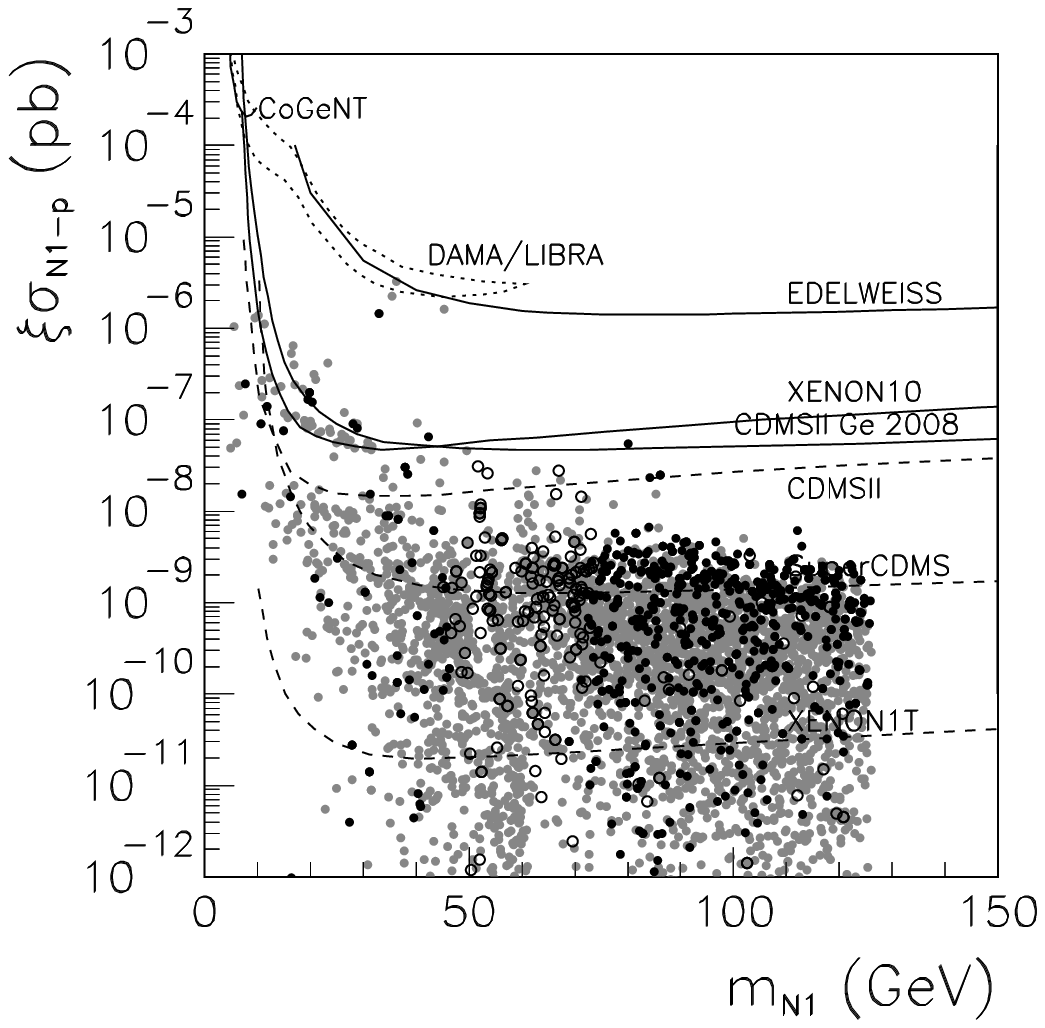,width=9.cm}
  \end{center}
  \captions{The same as in Fig.\,\ref{fig:sncross-a} but for 
    a scan in the $(\l,\k)$ plane. 
  }
  \label{fig:sncross-scan-ab}
\end{figure}

For completeness we have also performed a scan in the $(\l,\k)$ plane,
the results of which are shown on Fig.\,\ref{fig:sncross-scan-ab}. 
Depending on the point of the $(\l,\k)$ plane, the mass and coupling
of the lightest pseudoscalar varies and the predictions for sneutrino
dark matter are seriously affected.
As a consequence, the predicted cross section for very light
sneutrinos spans  
several orders of magnitude, ranging from $10^{-11}$~pb to
$10^{-6}$~pb. Interestingly very light sneutrinos populate an area of
the $\crosssec-\snmassr$ plane which is forbidden to neutralinos in
the MSSM. However,
the predictions for very light neutralinos in the NMSSM are
very similar to those obtained here for sneutrinos
\cite{Aalseth:2008rx}.

\begin{figure}
  \begin{center}
    \epsfig{file=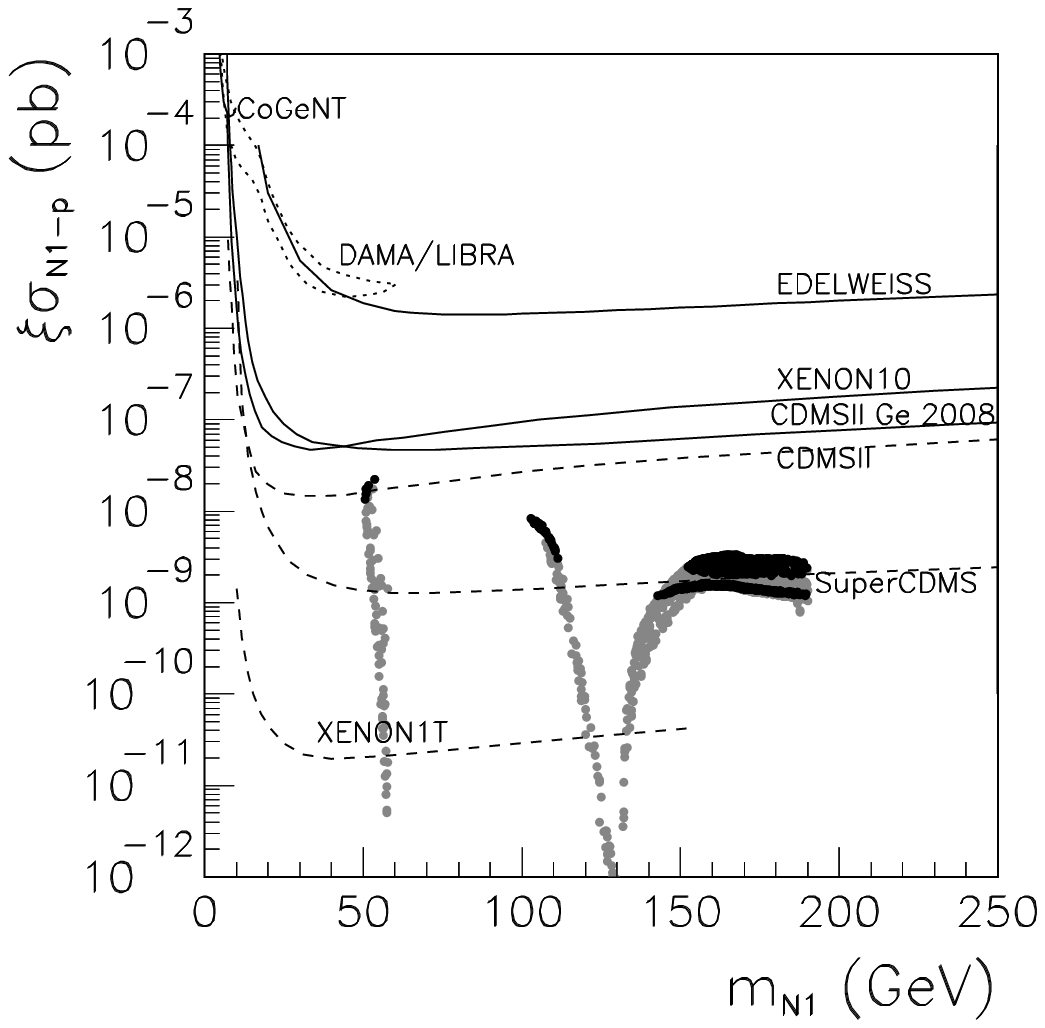,width=9.cm}
  \end{center}
  \captions{The same as in Fig.\,\ref{fig:sncross-c} but for 
    example C) of Table\,\ref{tab:examples-real}. 
  }
  \label{fig:sncross-g}
\end{figure}

Finally let us address an example which we already studied in
Section\,\ref{sec:rhn} and for which annihilation into a pair of
right-handed neutrinos was the predominant sneutrino annihilation
channel. In particular, we take the NMSSM parameters as in case C) of
Table\,\ref{tab:examples-real} with $\aln=-250$ GeV. This correspond
to the example studied in Figs.\,\ref{fig:relic-g-real} and
\ref{fig:lnmn-g}. 
In this example the mass of the scalar Higgses is rather large and
therefore the resulting scattering cross section is reduced. 
The predictions are plotted in 
Fig.\,\ref{fig:sncross-g} as a function of the sneutrino mass and, as
we can see, 
they are of order
$2\times 10^{-9}$~pb. Once more, this result is not excluded by
present dark matter searches and can be within the reach of some of
the future
experiments.

\subsection{Overview and comparison with neutralino dark matter}

We have so far shown that the right-handed sneutrino can be a viable
WIMP in this construction.
The examples throughout this Section illustrate how
sneutrinos within a mass range of $\snmassr\approx5-200$~GeV
and satisfying the WMAP constraint can be obtained for which the
predicted scattering cross section is not excluded by current
experimental searches (unlike in the case of the left-handed sneutrino
in the MSSM). 
Interestingly, the predicted $\crosssec$ is generally within
the reach of future experiments, except along the Higgs resonances in
which case the predicted scattering cross section can even be below
the sensitivity of Ton-scale detectors.

\begin{figure}
  \begin{center}
    \epsfig{file=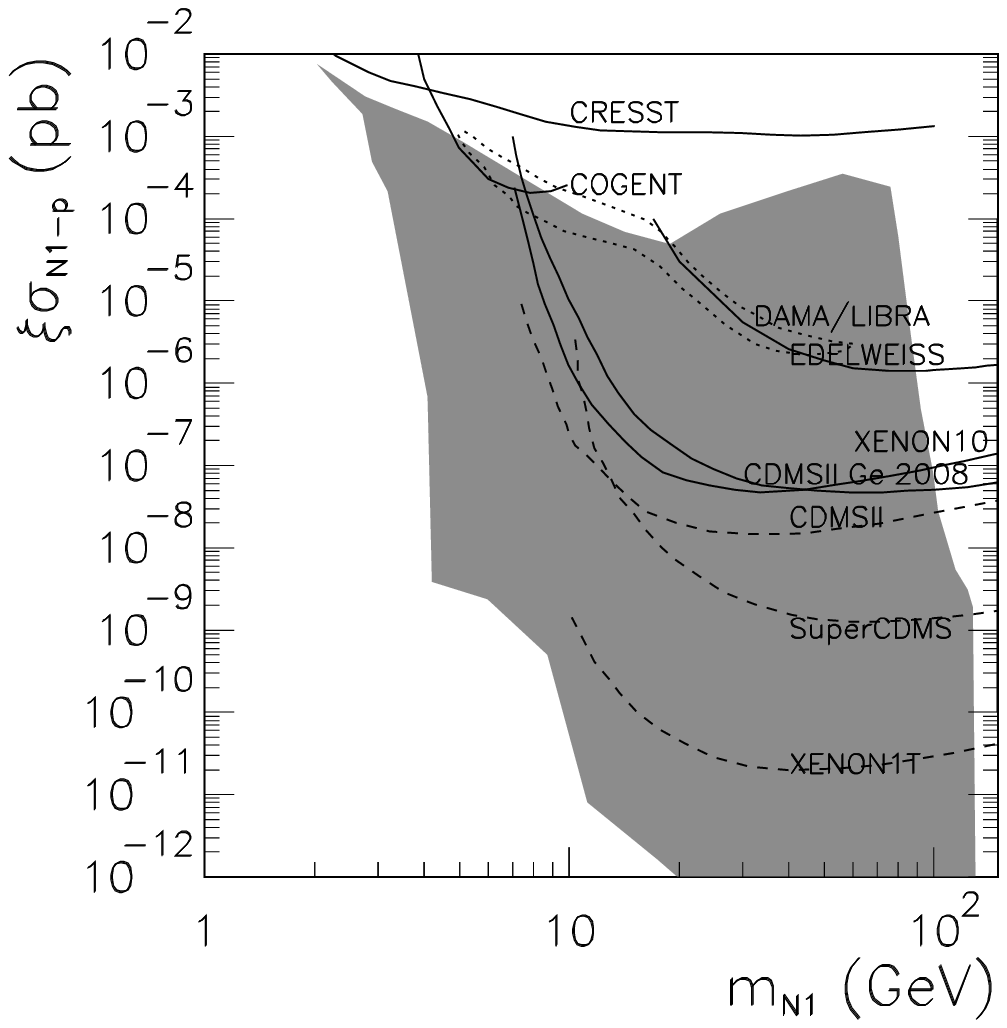,width=9.cm}
  \end{center}
  \captions{Combination of all the theoretical results for the
    sneutrino elastic scattering cross section as a function of the
    sneutrino mass. 
  }
  \label{fig:sncross-total}
\end{figure}

We can combine the results of this Section in order to determine the
region in the $(\crosssec,\snmassr)$ plane that the right-handed 
sneutrino can
cover. This is helpful in order to understand in which regions could
it account for a hypothetical WIMP detection in the future. 
We do this in 
Fig.\,\ref{fig:sncross-total}, which is constructed gathering all the
predictions from this Section,
and which serves as a summary of our results.

This also allows us to compare with the predictions for other existing
WIMPs. In particular, it should be pointed out that the area shown in
Fig.\,\ref{fig:sncross-total} is within the area predicted for
neutralinos in the NMSSM \cite{Cerdeno:2004xw,Aalseth:2008rx}
(as well as for neutralinos in the MSSM except for the region for very
light WIMPs). One may therefore wonder whether these two candidates
can be discriminated in the case of a hypothetical detection of dark
matter. It should be pointed out in this respect that since the
sneutrino has no spin-dependent couplings, the combination of data
from two experiments whose targets are sensitive to spin-dependent and
-independent couplings 
could be used to disentangle between sneutrino and neutralino dark
matter in the same way that it was shown for Kaluza-Klein dark matter
and neutralinos in Ref.\,\cite{Bertone:2007xj}.

\section{Implication for collider physics}
\label{sec:collider}

Finally, we briefly comment on the characteristic signals 
that might be expected from this model at accelerator experiments.

As with other WIMP models, right-handed sneutrinos can be
pair-produced in a collider and subsequently escape undetected. This
would manifest as events with a deficit of transverse momentum.
Notice that one expects the same kind of signals if the neutralino is
the LSP and a dark matter candidate. Thus, the relevant question is
whether or not the identity of the missing particle (sneutrino or
neutralino) can be determined. An obvious difference between both
particles is their spin. In this sense, the study of certain
kinematic variables can give us some insight not only about the mass
of the LSP \cite{Cho:2007dh}, but even about its spin
\cite{Cho:2008tj}. This would also be similar to the way the nature of
the NLSP can be determined in scenarios of gravitino dark matter  
\cite{Okada:2007na}. This, however might not be possible in the 
Large Hadron Collider and one would need to wait for the 
International Linear Collider.
Another crucial difference between the sneutrino and the neutralino is
their lepton number. This was exploited in scenarios of 
sneutrino NLSP with gravitino LSP by studying the
associated lepton production \cite{Covi:2007xj,Ellis:2008as}. However,
this is not directly applicable to our case, since we are dealing with
a pure right-handed sneutrino.

Nevertheless, there exists yet another attractive possibility,
the production of right-handed
neutrinos. 
These can be produced by 
the decay of either a Higgs boson or 
a neutralino (which can be the NLSP and decay into
a sneutrino LSP and a right-handed neutrino).
Once the right-handed neutrino is produced, it would 
be long-lived and decay, through the tiny mixing with the 
left-handed component, 
into a $W$ boson and a charged lepton. This would give rise to a 
displaced vertex which could be observed.

As we have already stressed, 
sneutrino dark mater is very sensitive to the properties of the Higgs
sector. 
When a Higgs boson with significant singlet component decays,
depending on its mass, a pair of LSP right-handed sneutrinos and 
a pair of right-handed neutrinos, 
in other words the missing momentums and long-lived particles, 
are simultaneously produced.
This is also a possible unique signal.

There is another interesting consequence, namely 
the invisible decay of the Higgs boson, 
since the right-handed sneutrino interacts with the CP-even Higgs 
through the coupling $\chisnsn$.
This is a common feature of dark matter particles which predominantly
annihilate through $s$-channel Higgs exchange and can be detected 
through $t$-channel Higgs
exchange~\cite{McDonald:1993ex,Burgess:2000yq}.

In addition, through collider studies, it is possible to discriminate 
our model from other models with thermal right-handed sneutrino dark
matter. 
In particular, there exists a family of models in which thermal
right-handed neutralino dark matter is viable thanks to the inclusion
of a new gauge interaction through a $U(1)'$~\cite{Lee:2007mt} or
$U(1)_{B-L}$~\cite{Allahverdi:2007wt}.
These scenarios present a 
$Z'$ gauge boson which is not present in our construction. The
detection or not of this $Z'$ would discriminate among these models.
Also, another construction in which the Higgs sector is extended was
presented in Ref.\,\cite{Deppisch:2008bp}.
The essential difference with our scenario 
is that the singlet 
Higgs in that model~\cite{pilaftsis,Deppisch:2008bp} is quite heavy and 
exchanging Higgs particles are $SU(2)$ doublet only. 
As a result, the correct relic abundance can only be realized 
with the help of the resonant annihilation. 
Hence, a strong correlation between the sneutrino mass and 
the Higgs mass is present. 
In addition, the right-handed neutrino production is 
only through the neutralino decay because of too heavy singlet Higgs.
Thus, by examining the way of right-handed neutrino production,
we may discriminate between these two models.

\section{Conclusions}
\label{sec:conclusions}

We have investigated 
the properties of the right-handed sneutrino and its
viability as a WIMP dark matter candidate in an extended
version of the NMSSM in which a right-handed neutrino
superfield is included with a coupling to the scalar Higgs in order to 
provide a Majorana right-handed neutrino mass. 
Both the $\mu$
parameter and the neutrino Majorana mass are generated by the vacuum
expectation value of the singlet field after radiative electroweak
symmetry breaking and are therefore naturally of order of the
electroweak scale.
This implies small values for the neutrino Yukawas, of order of the
electron Yukawa coupling and therefore leads to negligible mixing
terms in the sneutrino mass matrix and to pure left- and right-handed
mass eigenstates.
The sneutrino spectrum is then analysed in terms of the three new
parameters, the sneutrino-Higgs coupling $\ln$, the sneutrino soft mass
$\mn$ and the new trilinear term $\aln$.

We compute the thermal relic abundance for the right-handed sneutrino,
including all possible annihilation channels. The annihilation cross
section is extremely dependent on the Higgs sector of the
NMSSM. Interestingly, light CP-even and light CP-odd Higgses are
viable in this model as in the NMSSM (if they have a large singlet
component) and this provides interesting properties for the
sneutrino. 
In particular, annihilation into light Higgses (scalar or
pseudoscalar) makes it possible to reproduce the WMAP result for the
relic abundance for low values of the sneutrino mass. We study the
relevance of each individual annihilation channel and show how
sneutrinos in the mass range 5-200~GeV can reproduce the correct value
for the relic abundance for natural values of the three new
parameters, $\ln={\cal O}(0.1)$, $\mn={\cal O}(100)$~GeV and
$|\aln|={\cal O}(250-500)$~GeV.

Next, the sneutrino-proton elastic scattering cross section is
computed and compared with present experimental sensitivities. We find
that the sneutrino is not ruled out by current searches for dark
matter. Interestingly, many regions of the parameter space could be
within the reach of some of the future dark matter experiments.

Finally, we commented on the possible implications for collider
physics, identifying some potentially interesting signals. We discuss
how, in some cases, our construction can be distinguished
experimentally from other models for sneutrino dark matter as well as
from other dark matter candidates such as the neutralino.

\clearpage

\noindent {\bf Acknowledgments} 

We have greatly benefited from discussions with T.~Asaka, A.~Ibarra
and C.~Mu\~noz.
D.G.C. was supported by the program 
``Juan de la Cierva'' of the Spanish MEC.
and also in part by the Spanish DGI of the
MEC under Proyecto Nacional FPA2006-01105, 
and by the EU 
network
MRTN-CT-2006-035863. 
O.S. was partly supported by the MEC project FPA 2004-02015 and DOE
grant DE-FG02-94ER-40823.
We thank the ENTApP Network of the
ILIAS project RII3-CT-2004-506222 and 
the project HEPHACOS P-ESP-00346 
of the Comunidad de Madrid.


\renewcommand{\thesection}{Appendix~\Alph{section}}
\renewcommand{\thesubsection}{\Alph{section}.\arabic{subsection}}

\setcounter{section}{0}

\clearpage


\renewcommand{\thesection}{\Alph{section}}
\section{Feynman rules}
\renewcommand{\thesection}{\Alph{section}}
\label{sec:feynman}

In this appendix we indicate the relevant vertices for the calculation
of the sneutrino annihilation cross section.
The couplings relevant to our computation and which receive
contributions from the right-handed sneutrino sector we have
introduced read

\begin{tabular}{ll}
 \begin{picture}(150,120)(0,-50)
  \Photon(10,0)(60,0){3}{4}
  \Text(0,0)[c]{$Z^0_{\mu}$}
  \DashArrowLine(110,50)(60,0){5}
  \Text(80,45)[c]{$\tilde{\nu}$}
  \Text(100,25)[c]{$p$}
  \DashArrowLine(60,0)(110,-50){5}
  \Text(80,-45)[l]{$\tilde{\nu}$}
  \Vertex(60,0){2}
  \Text(100,-25)[c]{$q$}
 \end{picture}
 &
 \raisebox{45\unitlength}{
 \begin{minipage}{10cm}
  \lefteqn{
   -\frac{ig}{\cw}\,\left(p+q\right)^\mu\, \left(\sncompl\right)^2
   }
 \end{minipage}
 }
\end{tabular}

\begin{tabular}{ll}
 \begin{picture}(150,120)(0,-50)
  \DashLine(10,0)(60,0){5}
  \Text(0,0)[c]{$H^0_i$}
  \DashLine(110,50)(60,0){5}
  \Text(80,45)[c]{$\tilde{\nu}$}
  \DashLine(60,0)(110,-50){5}
  \Text(80,-45)[l]{$\tilde{\nu}$}
  \Vertex(60,0){2}
 \end{picture}
 &
 \raisebox{45\unitlength}{
 \begin{minipage}{10cm}
  \lefteqn{
   -\frac{ig \mz}{2\cw}\,\left(\cosb\hcompid-\sinb\hcompiu\right)\, 
   \left(\sncompl\right)^2 +
   }
  \lefteqn{i\left[\frac{2\l\ln\mw}{\sqrtwo g}
  \,\left(\sinb\hcompid+\cosb\hcompiu\right)\, + \right.}
  \lefteqn{\left.\left[
  \left(4\ln^2+2\k\ln\right)\vevs+\ln\frac{\aln}{\sqrtwo}
  \right](\hcompis)^2\right]\,\left(\sncompr\right)^2 }
  \lefteqn{\equiv i \chsnsn}
 \end{minipage}
 }
\end{tabular}

\begin{tabular}{ll}
 \begin{picture}(150,120)(0,-50)
  \DashLine(10,0)(60,0){5}
  \Text(0,0)[c]{$H^0_i$}
  \Line(110,50)(60,0)
  \Text(80,45)[c]{$\rhn$}
  \Line(60,0)(110,-50)
  \Text(80,-45)[l]{$\rhn$}
  \Vertex(60,0){2}
 \end{picture}
 &
 \raisebox{45\unitlength}{
 \begin{minipage}{10cm}
  \lefteqn{
    -\frac{i\ln}{\sqrt{2}}\left(\hcompis\right)
    \equiv\cnnhi
   }
 \end{minipage}
 }
\end{tabular}

\begin{tabular}{ll}
 \begin{picture}(150,120)(0,-50)
  \Line(10,0)(60,0)
  \Text(0,0)[c]{$\chi^0_{i}$}
  \DashLine(110,50)(60,0){5}
  \Text(80,45)[c]{$\tilde{\nu}$}
  \Line(60,0)(110,-50)
  \Text(80,-45)[l]{$N$}
  \Vertex(60,0){2}
 \end{picture}
 &
 \raisebox{45\unitlength}{
 \begin{minipage}{10cm}
  \lefteqn{
    -i\sqrt{2}\ln\left(\neuticomps\right)\left(\sncompr\right)
    \equiv\csnnneui
   }
 \end{minipage}
 }
\end{tabular}

\begin{tabular}{ll}
 \begin{picture}(150,120)(0,-50)
  \DashLine(10,50)(60,0){5}
  \Text(0,45)[c]{$H^0_i$}
  \DashLine(10,-50)(60,0){5}
  \Text(0,-45)[c]{$H^0_j$}
  \DashLine(110,50)(60,0){5}
  \Text(80,45)[c]{$\tilde{\nu}$}
  \DashLine(60,0)(110,-50){5}
  \Text(80,-45)[l]{$\tilde{\nu}$}
  \Vertex(60,0){2}
 \end{picture}
 &
 \raisebox{45\unitlength}{
 \begin{minipage}{10cm}
  \lefteqn{i \left(\frac{\l\ln}{2}
  \left(\hcompid\hcompju+\hcompiu\hcompjd\right)
  - \ln(\k+2\ln)\hcompis\hcompjs
  \right) 
  \,\left(\sncompr\right)^2}
  \lefteqn{\equiv i \chhsnsn}
 \end{minipage}
 }
\end{tabular}

\begin{tabular}{ll}
 \begin{picture}(150,120)(0,-50)
  \DashLine(10,50)(60,0){5}
  \Text(0,45)[c]{$\phiggsi$}
  \DashLine(10,-50)(60,0){5}
  \Text(0,-45)[c]{$\phiggsj$}
  \DashLine(110,50)(60,0){5}
  \Text(80,45)[c]{$\tilde{\nu}$}
  \DashLine(60,0)(110,-50){5}
  \Text(80,-45)[l]{$\tilde{\nu}$}
  \Vertex(60,0){2}
 \end{picture}
 &
 \raisebox{45\unitlength}{
 \begin{minipage}{10cm}
  \lefteqn{-i \left(\frac{\l\ln}{2}
  \left(\phcompid\phcompju+\phcompiu\phcompjd\right)
  + \ln(2\ln-\k)\phcompis\phcompjs
  \right) 
  \,\left(\sncompr\right)^2}
  \lefteqn{\equiv i \caasnsn}
 \end{minipage}
 }
\end{tabular}

\begin{tabular}{ll}
 \begin{picture}(150,120)(0,-50)
  \DashLine(10,50)(60,0){5}
  \Text(0,45)[c]{$H^+$}
  \DashLine(10,-50)(60,0){5}
  \Text(0,-45)[c]{$H^-$}
  \DashLine(110,50)(60,0){5}
  \Text(80,45)[c]{$\tilde{\nu}$}
  \DashLine(60,0)(110,-50){5}
  \Text(80,-45)[l]{$\tilde{\nu}$}
  \Vertex(60,0){2}
 \end{picture}
 &
 \raisebox{45\unitlength}{
 \begin{minipage}{10cm}
  \lefteqn{-i 2\l\ln \left(\sncompr\right)^2 }
  \lefteqn{\equiv i \cccsnsn }
 \end{minipage}
 }
\end{tabular}
\vspace*{1cm}

We will be also using the following usual NMSSM Feynman diagrams.

\begin{tabular}{ll}
 \begin{picture}(150,120)(0,-50)
  \DashLine(10,0)(60,0){5}
  \Text(0,0)[c]{$H^0_i$}
  \Photon(60,0)(110,50){3}{4}
  \Text(80,45)[c]{$W^+_{\nu}$}
  \Photon(60,0)(110,-50){3}{4}
  \Text(80,-45)[l]{$W^-_{\mu}$}
  \Vertex(60,0){2}
 \end{picture}
 &
 \raisebox{45\unitlength}{
 \begin{minipage}{10cm}
  \lefteqn{
   ig\mw \left(\cosb\hcompid+\sinb\hcompiu\right)g^{\mu\nu}
   \equiv ig\mw\,  C_{H_i}\, g^{\mu\nu}
  }
 \end{minipage}
 }
\end{tabular}

\begin{tabular}{ll}
 \begin{picture}(150,120)(0,-50)
  \DashLine(10,0)(60,0){5}
  \Text(0,0)[c]{$H^0_i$}
  \Photon(60,0)(110,50){3}{4}
  \Text(80,45)[c]{$Z^0_{\nu}$}
  \Photon(60,0)(110,-50){3}{4}
  \Text(80,-45)[l]{$Z^0_{\mu}$}
  \Vertex(60,0){2}
 \end{picture}
 &
 \raisebox{45\unitlength}{
 \begin{minipage}{10cm}
  \lefteqn{
   \frac{ig\mz}{\cw} 
   \left(\cosb\hcompid+\sinb\hcompiu\right)g^{\mu\nu}
   \equiv \frac{ig\mz}{\cw}\, C_{H_i}\, g^{\mu\nu}
  }
 \end{minipage}
 }
\end{tabular}

\begin{tabular}{ll}
 \begin{picture}(150,120)(0,-50)
  \DashLine(10,0)(60,0){5}
  \Text(0,0)[c]{$H^0_i$}
  \DashLine(110,50)(60,0){5}
  \Text(80,45)[c]{$H^0_j$}
  \DashLine(60,0)(110,-50){5}
  \Text(80,-45)[l]{$H^0_k$}
  \Vertex(60,0){2}
 \end{picture}
 &
 \raisebox{45\unitlength}{
 \begin{minipage}{10cm}
  \lefteqn{\equiv i \chhh}
 \end{minipage}
 }
\end{tabular}

\begin{tabular}{ll}
 \begin{picture}(150,120)(0,-50)
  \DashLine(10,0)(60,0){5}
  \Text(0,0)[c]{$H^0_k$}
  \DashLine(110,50)(60,0){5}
  \Text(80,45)[c]{$\phiggsi$}
  \DashLine(60,0)(110,-50){5}
  \Text(80,-45)[l]{$\phiggsj$}
  \Vertex(60,0){2}
 \end{picture}
 &
 \raisebox{45\unitlength}{
 \begin{minipage}{10cm}
  \lefteqn{\equiv i \chaa}
 \end{minipage}
 }
\end{tabular}

\begin{tabular}{ll}
 \begin{picture}(150,120)(0,-50)
  \DashLine(10,0)(60,0){5}
  \Text(0,0)[c]{$H^0_k$}
  \DashLine(110,50)(60,0){5}
  \Text(80,45)[c]{$H^+$}
  \DashLine(60,0)(110,-50){5}
  \Text(80,-45)[l]{$H^-$}
  \Vertex(60,0){2}
 \end{picture}
 &
 \raisebox{45\unitlength}{
 \begin{minipage}{10cm}
  \lefteqn{\equiv i \chcc}
 \end{minipage}
 }
\end{tabular}

\begin{tabular}{ll}
 \begin{picture}(150,120)(0,-50)
  \DashLine(10,0)(60,0){5}
  \Text(0,0)[c]{$H^0_k$}
  \DashLine(110,50)(60,0){5}
  \Text(80,45)[c]{$\phiggsi$}
  \Photon(60,0)(110,-50){3}{4}
  \Text(80,-45)[l]{$Z$}
  \Vertex(60,0){2}
 \end{picture}
 &
 \raisebox{45\unitlength}{
 \begin{minipage}{10cm}
  \lefteqn{\equiv i \chkaiz}
 \end{minipage}
 }
\end{tabular}

\begin{tabular}{ll}
 \begin{picture}(150,120)(0,-50)
  \DashLine(10,0)(60,0){5}
  \Text(0,0)[c]{$H^0_k$}
  \DashLine(110,50)(60,0){5}
  \Text(80,45)[c]{$H^+$}
  \Photon(60,0)(110,-50){3}{4}
  \Text(80,-45)[l]{$W^-$}
  \Vertex(60,0){2}
 \end{picture}
 &
 \raisebox{45\unitlength}{
 \begin{minipage}{10cm}
  \lefteqn{\equiv i \chkcw}
 \end{minipage}
 }
\end{tabular}

\vspace*{1cm}

The explicit expressions for the couplings $\chhh$, $\chaa$, $\chcc$
$\chkaiz$, and $\chkcw$ can be found in \cite{ff95}, and translated
into our formalism by  
doing the transformations $v_{1,2}\to \frac{v_{1,2}}{\sqrtwo}$ and
$\vevs\to \frac{\vevs}{\sqrtwo}$. 

\clearpage

\renewcommand{\thesection}{\Alph{section}}
\section{Annihilation channels}
\renewcommand{\thesection}{\Alph{section}}
\label{sec:wtilde}

In order to determine the total annihilation cross section we follow
the procedure of \cite{nrr02} and start by
introducing a Lorentz-invariant
function $w(s)$~\cite{swo88}
\begin{equation}
w(s)= \frac{1}{4} \int d\, {\rm LIPS}\, |{\cal A} (\snr\snr
\rightarrow {\rm all})|^2
\label{ws}
\end{equation}
where $|{\cal A} (\snr\snr \rightarrow {\rm all})|^2$
denotes the absolute square of the reduced matrix
element for the annihilation of two sneutrinos, averaged over
initial spins and summed over final spins. The function $w(s)$ is
related to the annihilation cross section $\sigma(s)$ 
via~\cite{lny93}
\begin{equation}
w(s)=  \frac{1}{2} \sqrt{ s (s-4\snmassr^2)}\, \sigma(s).
\label{wtosigma:eq}
\end{equation}

Since $w(s)$ receives contributions from all the kinematically allowed
annihilation process $\snr\snr\rightarrow X_1 X_2$, it can be written as
\begin{eqnarray}
w(s)&=&\frac{1}{32\,\pi}\sum_{X_1 X_2} \bigg[
c \, \theta\left(s-(m_{X_1}+m_{X_2})^2 \right)\,
\beta_X(s,m_{X_1},m_{X_2})\,\widetilde{w}_{X_1X_2}(s)\bigg],
\label{wtowtilde:eq}
\end{eqnarray}
where
the summation extends over all possible two-body final states
$X_1X_2$, $m_{X_1}$ and $m_{X_2}$ denote their respective masses, and
\begin{eqnarray}
\label{color:eq}
c &=& \left\{ \begin{array}{ll}
c_{f}  & \mbox{if $X_{1 (2)}=f(\bar f)$} \\
 1     & \mbox{otherwise,}  \,
\end{array}
      \right.
\end{eqnarray}
where $c_{f}$ is the color factor of SM fermions
($c_{f}=3$ for quarks and $c_{f}=1$ for leptons).
The kinematic factor $\beta_X$ is defined as
\begin{eqnarray}
\beta_X(s,m_{X_1},m_{X_2})\equiv
\left[1-\frac{(m_{X_1}+m_{X_2})^2}{s}\right]^{1/2}
\left[1-\frac{(m_{X_1}-m_{X_2})^2}{s}\right]^{1/2}.
\label{kdef:eq}
\end{eqnarray}
In the CM frame, which we choose for convenience, the function
$\widetilde{w}_{X_1 X_2}(s)$ can be expressed as
\begin{eqnarray}
  \widetilde{w}_{X_1 X_2}(s)\equiv \frac{1}{2}\int_{-1}^{+1} \!d \cos
  \theta_{CM}  
  \,|{\cal A}(\snr \snr \rightarrow X_1 X_2)|^2, \label{wtildedef:eq}
\end{eqnarray}
where $\theta_{CM}$ denotes the scattering angle in the CM frame. In other
words, we write
$|{\cal A}(\snr \snr \rightarrow X_1 X_2)|^2$ as a function of
$s$ and $\cos \theta_{CM}$, which greatly simplifies the computation.

We present here the analytic expressions for the $\tilde w$ functions
for the various annihilation channels. In
doing so, we have assumed the lightest sneutrino to be a pure
right-handed sneutrino, hence $\sncompl=0$ and $\sncompr=1$ with $i = 1$.

In the following expressions we define 
\begin{eqnarray}
\Delta_{i}&\equiv&
(s-\hmassi^2)+\frac{(\hmassi\,\Gamma_i)^2}{s-\hmassi^2}\,, \\
\Delta_{ij}&\equiv&
(s-\hmassi^2)(s-\hmassj^2)+\hmassi\,\hmassj\,\Gamma_i\Gamma_j \nonumber\\   
&&
+\frac{(\hmassi\,\Gamma_i(s-\hmassj^2)-
\hmassj\,\Gamma_j(s-\hmassi^2))^2}
{(s-\hmassi^2)(s-\hmassj^2)+\hmassi\,\hmassj\,\Gamma_i\Gamma_j}
\,,
\end{eqnarray}
for $i,j=1,\,2,\,3$, where $\Gamma_i$ is the decay width of the Higgs
$\higgsi$.

\vspace*{1cm}
$\bullet$ $\to W^+W^-$

The annihilation proceeds through an $s$-channel mediated by a scalar
Higgs, $\higgsi$. 

\begin{equation}
\tilde w_{W^+W^-}=6g^2\mw^2 \sum_{i,j=1}^3 
\frac{\chisnsn\chjsnsn C_{H_i}C_{H_j}}{\Delta_{ij}}
\end{equation}

\vspace*{1cm}
$\bullet$ $\to ZZ$

The annihilation proceeds through an $s$-channel mediated by a scalar
Higgs, $\higgsi$. The $t$-channels where a sneutrino is exchanged do
not contribute since the $Z-\tilde\nu-\tilde\nu$ coupling vanishes
for pure right-handed sneutrinos.

\begin{equation}
\tilde w_{ZZ}=\frac{3g^2\mz^2}{\cw} \sum_{i,j=1}^3 
\frac{\chisnsn\chjsnsn C_{H_i}C_{H_j}}{\Delta_{ij}}
\end{equation}

\vspace*{1cm}
$\bullet$ $\to \higgsi\higgsj$

There are four contributions to this annihilation channel. One comes
from an $s$-channel mediated by a scalar
Higgs, $\higgsk$, the $t,u$-channels where a sneutrino is exchanged,
and the point interaction. 

\begin{eqnarray}
\tilde w_{H_iH_j}&=&
2(\chhsnsn)^2 -
4\sum_{k=1}^3
\frac{\chksnsn\chhhk}{\Delta_{k}}
+
2\sum_{k,l=1}^3
\frac{\chksnsn\chlsnsn\chhhk\chhhl}{\Delta_{kl}}
+ \nonumber\\
&&\frac{2(\chisnsn\chjsnsn)^2(\ahit+\ahiu)^2}{\ahit^2\ahiu^2\Ehi^2s}
\left[\bhit\bhiu
\frac{\Fone{\bhit}+\Fone{\bhiu}}{(\bhit+\bhiu)^3} + \right.\nonumber\\
&&\left.\frac{(\bhit^2+\bhiu^2)-(\bhit\bhiu)^2}
{(1-\bhit^2)(1-\bhiu^2)(\bhit+\bhiu)^2}
\right]+\nonumber\\
&&2\sum_{k=1}^3
\frac{\chhhk\chisnsn\chjsnsn\chksnsn(\ahit+\ahiu)}
{\ahit\ahiu\Ehi\sqrt{s}\Delta_{k}}
\left[\frac{\Fone{\bhit}+\Fone{\bhiu}}{\bhit+\bhiu}\right]-\nonumber\\ 
&&2\frac{\chisnsn\chjsnsn\chhsnsn(\ahit+\ahiu)}
{\ahit\ahiu\Ehi\sqrt{s}}
\left[\frac{\Fone{\bhit}+\Fone{\bhiu}}{\bhit+\bhiu}\right]
\end{eqnarray}
which should be multiplied by a factor $1/2$ in the case of identical
particles in the final state.

We have defined

\begin{equation}
\Fone{x}\equiv \log\left(\frac{1+x}{1-x}\right)
\end{equation}

\begin{eqnarray}
\ahit&\equiv&\frac{\hmassi^2}{\sqrt{s}\Ehi}-1\\
\ahiu&\equiv&\frac{\hmassj^2-s}{\sqrt{s}\Ehi}+1
\end{eqnarray}

and

\begin{eqnarray}
\bhit&\equiv&\frac{\sqrt{\frac{s}{4}-\snmassr^2}
\sqrt{\Ehi^2-\hmassi^2}}{\hmassi^2-\sqrt{s}\Ehi}\\
\bhiu&\equiv&\frac{\sqrt{\frac{s}{4}-\snmassr^2}
\sqrt{\Ehi^2-\hmassi^2}}{\hmassj^2-s+\sqrt{s}\Ehi}
\end{eqnarray}

where

\begin{eqnarray}
\Ehi&=&\frac{s-\left(\hmassj^2-\hmassi^2\right)}{2\sqrt{s}}\\
\Ehj&=&\frac{s+\left(\hmassj^2-\hmassi^2\right)}{2\sqrt{s}}
\end{eqnarray}

\vspace*{1cm}
$\bullet$ $\to \phiggsi\phiggsj$

The annihilation proceeds through an $s$-channel mediated by a scalar
Higgs, $\higgsk$ and a point interaction.

\begin{eqnarray}
\tilde w_{\phiggsi\phiggsj}&=&2(\caasnsn)^2 - 
4\sum_{k=1}^3\frac{\caasnsn\chksnsn\chaak}{\Delta_{k}} +\nonumber\\
&&2\sum_{k,l=1}^3\frac{\chksnsn\chlsnsn\chaak\chaal}
{\Delta_{kl}}
\end{eqnarray}
which is multiplied by a factor $1/2$ in the case of identical
particles in the final state.

\vspace*{1cm}
$\bullet$ $\to f\bar f$

We only include here the 
$s$-channel mediated by a scalar
Higgs, $\higgsk$. 
Once more, the $s$-channel annihilation mediated by the $Z$ boson
vanishes for a pure right-handed sneutrino.
For annihilation into $l\bar l$ ($\nu\bar\nu$) 
one should, in
principle, also consider the $t$-channel mediated by
charginos (neutralinos). However, those diagrams are suppressed by the
very small Yukawa coupling, $Y_N$, and will be here neglected. 

\begin{equation}
\tilde w_{f\bar	f}=2\sum_{i,j=1}^3
\frac{\chisnsn\chjsnsn\cfhi\cfhj}{\Delta_{ij}}
\,(2s-8m_f^2)
\end{equation}

\vspace*{1cm}
$\bullet$ $\to NN$

There are three contributions to this channel, namely the $s$-channel,
mediated by a scalar Higgs, $\higgsk$,
together with 
the $t$- and $u$- channels where (the five) neutralinos
are exchanged. 
%

First, we address the case where the lightest sneutrino is the real 
part of right-handed sneutrino.
\begin{equation}
\tilde w_{NN}= w_{NN}^s + w_{NN}^{st,su} + w_{NN}^{t,u}\, ,
\end{equation}
with    
\begin{eqnarray}
\tilde w_{NN}^s&=&\sum_{k,l=1}^3 
\frac{\chksnsn\chlsnsn\cnnhk\cnnhl}{\Delta_{kl}}
\,(s-4\rhnmass^2)
\end{eqnarray}
\begin{eqnarray}
\tilde w_{NN}^{st,su}&=&\sum_{k=1}^3 \sum_{i=1}^5
\frac{8\chksnsn\cnnhk\left(\csnnneui\right)^2}{\Delta_{k}}
\,\left[-\rhnmass
+\frac{{\cal D}(\neutmassi)}{2A}\log\left(\frac{A_i+A}{A_i-A}\right)
\right]
\end{eqnarray}
\begin{eqnarray}
\tilde w_{NN}^{t,u}&=&\sum_{i=1}^5 
\frac{32\csnnneui^4}{s^2}
\left[-\frac{s^2}{8} +
\frac{{{\cal A}(\neutmassi)}}{\left(A_i^2-A^2\right)} -
\frac{{\cal B}(\neutmassi)}{2A_iA}
\log\left(\frac{A_i-A}{A_i+A}\right)\right]
+\nonumber \\
&&
\sum_{i\ne j=1}^5
\frac{32\csnnneui^2\csnnneuj^2}{s^2}
\left[-\frac{s^2}{8} -
\frac{{\cal C}(\neutmassi,\neutmassj)}
{A\left(A_i^2-A_j^2\right)}
\log\left(\frac{A_i+A}{A_i-A}\right) +\right.\nonumber \\
&&
\left.\frac{{\cal C}(\neutmassj,\neutmassi)}
{A\left(A_i^2-A_j^2\right)}
\log\left(\frac{A_j+A}{A_j-A}\right)\right]
\end{eqnarray}

The following auxiliary functions have been used
\begin{eqnarray}
{\cal A}(\neutmassi)&\equiv&
\left(\frac{s}{4}-m_N^2\right)
\left(m_N+\neutmassi\right)^2+
\frac{s}{4}\left(\frac{s}{4}-\snmassr^2+2m_N^2 A_i+
2m_N\neutmassi A_i\right) -
\nonumber \\
&&
\frac{s^2}{16}A_i^2\, , 
\\
%
{\cal B}(\neutmassi)&\equiv&
\left(\frac{s}{4}-m_N^2\right)
\left(m_N+\neutmassi\right)^2 -
\frac{s}{4}\left(\frac{s}{4}-\snmassr^2+2m_N^2 A_i+
2m_N\neutmassi A_i\right)+\nonumber \\
&&
3\frac{s^2}{16}A_i^2\, , 
\\
%
{\cal C}(\neutmassi,\neutmassj)&\equiv&
\left(\frac{s}{4}-m_N^2\right)
\left(m_N+\neutmassi\right) \left(m_N+\neutmassj\right)A_j+  
\nonumber \\
&&
\frac{s}{4}\left(\frac{s}{4}-\snmassr^2+m_N^2 (A_i+A_j) +
m_N(\neutmassi A_i+\neutmassj A_j)\right)A_i -
\nonumber \\
&&
\frac{s^2}{16}A_i^3 \\
{\cal D}(\neutmassi)&\equiv&\left(1-\frac{4\rhnmass^2}{s}\right)
(\rhnmass+\neutmassi)+\rhnmass A_i
\end{eqnarray}

We have also defined
\begin{eqnarray}
A&\equiv&\frac{1}{s}\sqrt{s-4\snmassr^2\rule{0pt}{1.8ex}}
\sqrt{s-4m_N^2\rule{0pt}{1.8ex}}\, ,
\\
A_i&\equiv&2\ \frac{\snmassr^2+m_N^2-\neutmassi^2}{s}-1
\end{eqnarray}

In the case that the lightest sneutrino is the imaginary part of
right-handed sneutrino, one can obtain the relevant $\tilde w_{NN}$
by replacing 
\begin{equation}
w_{NN}^{st,su} \to -w_{NN}^{st,su},
\end{equation}
and 
\begin{equation}
m_N \to -m_N
\end{equation}
in $w_{NN}^{t,u}$.

\vspace*{1cm}
$\bullet$ $\to H^+ H^-$

As in the case of annihilation into pseudoscalar Higgses, 
the annihilation receives contributions from 
an $s$-channel mediated by a scalar
Higgs, $\higgsk$ and a point interaction.

\begin{eqnarray}
\tilde w_{\chiggsp\chiggsm}&=&2(\cccsnsn)^2 -
4\sum_{k=1}^3\frac{\cccsnsn\chksnsn\chcc}{\Delta_{k}} +\nonumber\\
&&2\sum_{k,l=1}^3\frac{\chksnsn\chlsnsn\chcc^2}
{\Delta_{kl}}
\end{eqnarray}

\vspace*{1cm}
$\bullet$ $\to Z \phiggsi$

This channel receives contribution from $s$-channel scalar Higgs
exchange. Notice that although there would also be an $s$-channel with
$Z$ boson mediation, the smallness of the Yukawa $Y_N$ renders that
contribution negligible.

\begin{eqnarray}
\tilde w_{Z\phiggsi}&=&
\sum_{k,l=1}^3\frac{\chksnsn\chlsnsn\chkaiz\chlaiz}{\Delta_{kl}}\nonumber\\
&&\left(-2s+\mz^2+2\phmassi^2+\frac{s^2+
\phmassi^4-2\phmassi^2 s}{\mz^2}
\right)\ .
\end{eqnarray}

\vspace*{1cm}
\clearpage
$\bullet$ $\to W^- H^+$

As in the case above, 
this channel receives contribution from $s$-channel scalar Higgs exchange.

\begin{eqnarray}
\tilde w_{ W^- H^+}&=&
\sum_{k,l=1}^3\frac{\chksnsn\chlsnsn\chkcw\chlcw}{\Delta_{kl}}\nonumber\\
&&\left(-2s+\mw^2+2\chmass^2+\frac{s^2+
\chmass^4-2\chmass^2 s}{\mw^2}
\right)\ .
\end{eqnarray}

\clearpage

%


\end{document}